\documentclass[a4paper,11pt]{article}
\usepackage{jheppub}

\usepackage{lscape}
\usepackage{array}
\usepackage{natbib}
\usepackage{amsfonts}
\usepackage{amsmath}
\usepackage{booktabs}
\usepackage{amssymb}
\usepackage{bbm} 
\usepackage{amsthm} 
\usepackage{graphicx}
\usepackage{imakeidx}
\usepackage[utf8]{inputenc}
\usepackage[T1]{fontenc}
\usepackage{tikz}
\usepackage{cancel}
\usetikzlibrary{arrows.meta, positioning, decorations.pathreplacing, calligraphy}

\makeindex

\newcommand{\p}{\partial}

\newcommand\bi{\begin{itemize}}
\newcommand\ei{\end{itemize}}

\newcommand\bspl{\begin{split}}
\newcommand\espl{\end{split}}

\newcommand{\susyL}[1]{\underset{\text{SUSY Locus}}{\longrightarrow}}
\newtheorem{theorem}{Observation}[section]

\theoremstyle{definition}

\theoremstyle{definition}

\begin{document}


\title{\boldmath Why Indices Count the Total Number of Black Hole Microstates (at large N) }

\author{Alejandro Cabo-Bizet\,$^a$}

\affiliation[a]{Department of Mathematics, University of Turin, \& I.N.F.N. Turin, Italy}

\emailAdd{acbizet@gmail.com}

\abstract{  
Using supersymmetric localization, we show that the partition function of four-dimensional superconformal gauge theories—computed as a trace over BPS states without the insertion of $(-1)^F$—is perturbatively protected and piecewise independent of the gauge coupling. We derive a matrix-integral representation of this observable at $g_{\text{YM}}=0\,$ for generic four-dimensional superconformal gauge theories. For $U(N)$ maximally supersymmetric Yang--Mills theory we study such a matrix integral and show that, even at finite $N\,$, it localizes to ensembles of superconformal indices near its essential singularities. The latter asymptotic localization projects out any potential discontinuity of the perturbatively protected partition function from zero to strong coupling and explains why single microcanonical indices reproduce the growth of the total number of BPS states in co-dimension one regions of large charges, up to large oscillations due to the insertion of $(-1)^F\,$. To compute quantum corrections to entropy at finite $N$ and small charges, the correct observable is the perturbatively protected partition function, which by definition is a positive quantity.

We propose and test an improvement of the Cardy-like method that allows us to identify and compute perturbatively exact expressions for the leading large-$N$ on-shell action of eigenvalue configurations that we call orbifold, dressed orbifold, and eigenvalue-instanton saddles. These are also saddle points of large-charge expansions at finite $N\,$. We test the conclusions obtained from such large-charge saddle-point analysis at $N=2$ using explicit Cauchy-residue evaluation.
}



\maketitle


\section{Introduction}

The exponential of the thermodynamic entropy~\cite{Bekenstein:1972tm,Hawking:1974rv,Gibbons:1976ue} of supersymmetric and extremal (BPS) black holes in string theory counts BPS configurations in underlying brane descriptions~\cite{Strominger:1996sh}. More precisely, the exponential of the horizon area matches the count of BPS microstates when the latter are counted with a $(-1)^F$ grading—namely when the counting corresponds to a Witten index~\cite{Witten:1982df}, which we denote by $\mathcal{I}_{\text{micro}}$. This identification has been shown to persist in the context of gauge/gravity duality~\cite{Maldacena:1997re,Witten:1998zw,Benini:2015eyy,Cabo-Bizet:2018ehj,Choi:2018hmj,Benini:2018ywd} and is remarkable for several reasons.

For example, from a thermodynamic standpoint, the quantity that would be naturally associated with the gravitational entropy is the logarithm of the total number of BPS states at the relevant charges, rather than the logarithm of the number of bosonic states minus the number of fermionic ones. 

Building on insights from the attractor mechanism~\cite{Ferrara:1995ih} and the AdS/CFT correspondence~\cite{Maldacena:1997re}, Sen proposed in~\cite{Sen:2005wa,Sen:2008vm,Sen:2009vz} that, in the regime where extremal and supersymmetric black holes dominate the gravitational path integral, the microscopic index should asymptote to the index of an emergent effective theory describing fluctuations in the near-horizon $AdS_2$ region. Schematically,
\begin{equation}\label{eq:IndicesEqMicroAdS2}
\mathcal{I}_{\rm micro}
\,\underset{\cdot}{\sim}\,
\mathcal{I}_{AdS_{2}} \,,
\end{equation}
where the left-hand side denotes the index of the UV-complete brane theory, while the right-hand side is a path integral over emergent degrees of freedom localized in the $AdS_2$ throat. The symbol $\underset{\cdot}{\sim}$ indicates the asymptotic expansion in which the black hole geometry dominates the saddle-point approximation to $\mathcal{I}_{\text{micro}}\,$.

A well-defined zero-temperature limit arguably requires~\cite{Iliesiu:2020qvm} the excitations that remain far from the horizon to be gapped relative to the modes localizing near the $AdS_2$ throat~\cite{Sen:2009vz,Iliesiu:2020qvm,Iliesiu:2022onk}. In the presence of supersymmetry, this gap has been convincingly argued to be there~\cite{,Iliesiu:2022onk}. Consequently, the expectation is that the microscopic index factorizes into contributions from states localized far from and near the horizon, with the former becoming irrelevant.

A useful perspective is obtained by embedding BPS black holes into continuous families of non-extremal, non-supersymmetric solutions. Within such families, one may increase the temperature while preserving supersymmetry—there exists a universal prescription for doing so~\cite{Cabo-Bizet:2018ehj,Iliesiu:2021are}. This yields supersymmetric but non-extremal black hole geometries. Since supersymmetry guarantees invariance of the index under variations of the regulator temperature, both the expression for the indices $\mathcal{I}$'s and their asymptotic relation~\eqref{eq:IndicesEqMicroAdS2} remain unchanged as the regulator is removed. Thus, one safely returns to the BPS geometry in the zero-temperature limit.

In these intermediate supersymmetric but non-extremal geometries, the Euclidean continuation retains its cigar topology. Regularity at the tip of the cigar enforces periodic boundary conditions for bosons and antiperiodic ones for fermions along the thermal circle. Importantly, these thermal periodicities survive the limit to extremality~\cite{Cabo-Bizet:2018ehj,Iliesiu:2021are}. This structure implies that, in the expansion $\underset{\cdot}{\sim}\,$, the near-horizon contribution $\mathcal{I}_{AdS_2}$ asymptotically counts the total number of near-horizon excitations (up to an overall fluctuating sign), suggesting that most microstates contributing at fixed charges $P$ (a codimension-one subset of the charges $P_{\text{tot}}$) are either predominantly bosonic or predominantly fermionic.

The zero-temperature condition enforces a nonlinear relation between $P_{\text{tot}}$ and the charges $P$ -- the subset of charges commuting with the supercharges that the index $\mathcal{I}_{\text{micro}}$ counts cohomology elements of with grading $(-1)^F\,$. Schematically, we denote this non-linear relation as
\begin{equation}\label{eq:NonLLocusIntro}
P_{\text{tot}}\,=\, P_{\text{tot}}[P]\,.
\end{equation}
This line of arguments suggest the following relation:
\begin{equation}\label{eq:AdS2IndePartQuantuM}
\mathcal{I}_{\text{AdS}_2}[P]\,:=\,\text{Tr}_{\text{near-hor}, P}(-1)^F \,\underset{\cdot}{\sim}\, s_{P}\, \text{Tr}_{\text{near-hor}, P_{\text{tot}}}(1)\,=:\, s_P Z_{\text{AdS}_2}[P_{\text{tot}}]\,,
\end{equation}
where $s_P$ is a charge-dependent sign. Indeed, using Sen’s entropy functional~\cite{Sen:2005wa,Sen:2008vm}, it has been extensively verified that--e.g. in AdS space~\cite{Morales:2006gm,Suryanarayana:2007rk,Dias:2007dj}
\begin{equation}\label{eq:Zads2BPS}
Z_{AdS_2}[P_{\text{tot}}]\,\underset{\cdot}{\sim}\, e^{\frac{A_{\text{hor}}[P]}{4 G_N}}\,,
\end{equation}
where $A_{\text{hor}}[P]$ is the area of the supersymmetric extremal horizon.

Combining~\eqref{eq:IndicesEqMicroAdS2}, \eqref{eq:AdS2IndePartQuantuM}, and \eqref{eq:Zads2BPS} yields the conjectural relation
\begin{equation}
|\mathcal{I}_{\text{micro}}[P]|\,\underset{\cdot}{\sim}\,e^{\frac{A{\text{hor}}[P]}{4 G_N}}\,,
\end{equation}
a relation that has been confirmed in many explicit microscopic computations.

A further expectation—which has not yet been tested for sufficiently large charges (under the assumption~\eqref{eq:NonLLocusIntro})~\footnote{We assume gravity in a number of dimensions larger than three.}—is that
\begin{equation}\label{eq:AZBPSIntro}
e^{\frac{A_{\text{hor}}[P]}{4 G_N}}\,\underset{\cdot}{\sim}\,\text{Tr}_{BPS, P_{\text{tot}}}(1)\,=:\,Z_{\text{BPS}}[P_{\text{tot}}]\,.
\end{equation}
In other words, the exponential of the black hole area should coincide with the BPS partition function of the microscopic theory, without the $(-1)^F$ insertion. Progress on testing~\eqref{eq:AZBPSIntro} has been limited by the difficulty of computing $Z_{\text{BPS}}$ at strong coupling.

If~\eqref{eq:AZBPSIntro} holds, then at the locus~\eqref{eq:NonLLocusIntro} one expects
\begin{equation}\label{eq:RelationBPSIndexIntro}
Z_{\text{BPS}}[P_{\text{tot}}] \,\underset{\cdot}{\sim}\, |\mathcal{I}_{\text{micro}}[P]|\,,
\end{equation}
where $\underset{\cdot}{\sim}$ reflects the large-$P_{\text{tot}}$ and strong-coupling asymptotic limit in which the black hole geometry dominates the saddle-point expansion.

Although these asymptotic relations follow on the basis of the holographic principle and the near horizon arguments before sketched, the validity of the relations~\eqref{eq:AZBPSIntro} and~\eqref{eq:RelationBPSIndexIntro} remains enigmatic, leaving a relevant conceptual gap in our understanding of the microscopic meaning of (supersymmetric) black hole entropy.

\subsection{Summary of results}

In this paper we derive the two asymptotic relations~\eqref{eq:NonLLocusIntro} and~\eqref{eq:RelationBPSIndexIntro} in AdS$_5$/CFT$_4\,$ duality~\cite{Maldacena:1997re}, concretely, starting from $U(N)$ four-dimensional $\mathcal{N}=4$ super Yang--Mills (SYM) on $\mathbb{R}\times S^3\,$. 

Other interesting results are found, which we summarize as follows:
\begin{itemize}

\item[1.] Our starting point is to show, using supersymmetric localization, that $Z_{\text{BPS}}$ is a perturbatively protected observable for generic four-dimensional superconformal gauge theories in a weak but non vanishing coupling expansion around $g_{\text{YM}}\neq 0$ (Section~\ref{sec:SUSYLoc}). This means that the \emph{total} number of BPS states at fixed charge is piecewise independent of the gauge coupling $g_{\text{YM}}$ and can therefore be computed in the free gauge theory projecting out contributions from BPS states of the free theory that at any order in a weak coupling develop anomalous dimension.~{In forthcoming work we will address the projection to physical BPS states at the level of the integral representation of $Z_{BPS}\,$. For the purposes of this paper such a projection turns out to be dynamically generated for reasons that will be explained next.} 

\item[2.] We compute the matrix–integral representation of $Z_{\text{BPS}}\,$ at zero gauge coupling for generic families of four-dimensional superconformal gauge theories (Section~\ref{sec:SUSYLoc}).~\footnote{At any level of charge, the number of states counted by $Z_{\text{BPS}}$ at zero gauge coupling may overcount or equal the number of states counted by $Z_{\text{BPS}}$ at weak but infinitesimal coupling.  }

For example, for $U(N)$ maximally supersymmetric Yang--Mills theory in the canonical ensemble (Sections~\ref{sec:SUSYLoc} and~\ref{sec:BPSPartHam}) the answer is
\begin{equation}\label{eq:ZBPSNonPerIntro}
\begin{split}
Z_{\text{BPS}}&\,=\,\frac{(Z_0)^N}{N!} 
\int \prod_{i=1}^{N} \frac{\mathrm{d}u_i}{2\pi}\, \prod_{\rho \in \mathrm{Adj}(U(N))\atop \rho\,\neq\,0} \Delta(\rho(u))\,\mathcal{G}_{0}\!\left(\rho(u); \frac{\omega_1}{2\pi i}, \frac{\omega_2}{2\pi i}\right) \\ &
\qquad\qquad\qquad \qquad   \times \prod_{I=1}^3\biggl({ 
\mathcal{G}_{I}\!\left(\rho(u); \frac{\omega_1}{2\pi i}, \frac{\omega_2}{2\pi i}\right)}\biggr)\,,
\end{split}
\end{equation}
where $z=e^{2\pi \text{i}\rho(u)}$, $p=e^{\omega_1}$, $q=e^{\omega_2}\,$, and~$t_{I}=e^{\varphi_I}$, are the gauge, rotational, and $R$–symmetry rapidities, respectively. The $\rho$'s are adjoint weights of $U(N)$:
\begin{equation}
\begin{split}
\Delta(\rho(u))&:=\frac{(1-z)}{(1\,-\,{z\,{\chi^{1/2}})}}\,,
\\
\mathcal{G}_{0}\biggl(\rho(u);\frac{\omega_1}{2\pi \text{i}}, \frac{\omega_2}{2\pi \text{i}}\biggr)&\,:=\,\frac{({z}\,\chi^{1/2};\,p,q)_\infty}{({ \frac{p q}{z}} \, ;\,p,q)_\infty}\,,
\\
\mathcal{G}_{I}\!\left(\rho(u); \frac{\omega_1}{2\pi i}, \frac{\omega_2}{2\pi i}\right)&\,:=\,\frac{(\frac{p q}{z \,t_I}\,\chi^{1/2};\,p,q)_\infty}{({ z\, t_I} \, ;\,p,q)_\infty}\,,
\end{split}
\end{equation}
and
\begin{equation}\label{eq:FinalDefChi}
\chi\,:=\,e^{2\pi \text{i} k}\,\frac{t_1 t_2 t_3}{p q}\,=\,e^{2\pi \text{i} k}\,\frac{t^2}{p q}\,=:\,e^{4\pi \text{i} (\alpha +\tfrac{k}{2})}\,.
\end{equation}
For the choice of branch $k=1\mod 2$ we obtain
\[\chi^{1/2}\,=\, - \frac{t}{\sqrt{p q}}\]
\footnote{Using the distribution rules $(x y)^z =x^z y^z\,$ and $(e^{x})^{y}\,:=\, e^{ x y}\,$ (choice of branch cuts) assumed in this paper.} 
and then~\eqref{eq:ZBPSNonPer} is the BPS partition function whose Taylor coefficients in the expansion around $(p^{\frac{1}{2}},q^{\frac{1}{2}},t_I)=(0,0,0)$ are all positive integers.\footnote{One can move among branches by the shifts $p\to e^{-2\pi \text{i}\widetilde{k}}$, $\widetilde{k}\in\mathbb{Z}$ at fixed $q\,$, $t_{I}\,$.}
Superconformal indices are obtained by imposing $\chi^{1/2}\to 1\,$. The zero–mode contribution $Z_0$ is reported in equation~\eqref{eq:N4SYM}.

\item[4.] Building on our previous work~\cite{Beccaria:2023hip}, we propose an improved method to derive the asymptotic expansion of $Z_{\text{BPS}}$ near its essential singularities. The method can also be applied to $\mathcal{I}_{\text{micro}}\,$ (Section~\ref{sec:LocPartitionF}). 

We implicitly test this method and in a sense the asymptotically generated protectedness of zero-coupling $Z_{\text{BPS}}$,~\footnote{Protectedness at $g_{\text{YM}}\neq 0$.} by reproducing known results for the superconformal index, but starting from the latter which is a different unitary matrix integral. 

For example, we find orbifold saddle configurations of $Z_{\text{BPS}}$ whose contribution at large $N$ is (Section~\ref{sec:Orbifold}):
\begin{equation}\label{eq:ExampleOrbifoldContributionIntro}
e^{-\frac{N^2 (M\varphi _1)^{\pm}(M\varphi_2)^{\pm}(M\varphi_3)^{\pm}}
{2M (M \omega_1+N_1)(M\omega_2+N_2)}}\,.
\end{equation}
These saddles are characterized by a nonzero positive integer $M$ and generic integers $N_1$ and $N_2\,$.
Remarkably, they exist only upon imposing the constraint
\begin{equation}\label{eq:OrbifoldBCIntro}
(M\varphi_3)^{\pm}\,:=\, -(M\varphi_1)^{\pm}-(M\varphi_2)^{\pm} +(M \omega_1+N_1)+ (M\omega_2+N_2)\,\pm 2\pi \text{i}\,.
\end{equation}
(Missing definitions of notation in this equation will be introduced in due time).

\item[5.] A reader experienced in this topic will recognize~\eqref{eq:ExampleOrbifoldContributionIntro} as the contribution of known orbifold solutions to the superconformal index $\mathcal{I}_{\text{micro}}$~\cite{Cabo-Bizet:2019eaf,Cabo-Bizet:2020nkr,Cabo-Bizet:2020ewf,Aharony:2021zkr,ArabiArdehali:2021nsx,Jejjala:2021hlt,Colombo:2021kbb,Cabo-Bizet:2021plf,Cabo-Bizet:2021jar,Jejjala:2022lrm,Mamroud:2022msu,Choi:2023tiq,Aharony:2024ntg}.\footnote{We have also identified these solutions in the superconformal index of ABJM~\cite{BenettiGenolini:2023rkq} and in AdS$_4$ supergravity.} 

This match is part of a broader correspondence arising near the essential singularities of the integral~\eqref{eq:ZBPSNonPerIntro} at large rank $N\,$.  

In analogy with our previous work~\cite{Cabo-Bizet:2021plf,Beccaria:2023hip}, we find that near its exponential singularities in the canonical ensemble, $Z_{\text{BPS}}$ asymptotes to ensembles of superconformal indices.

For example, for a generic expansion of $\omega_{1,2}$ near roots of unity (specified by $M,N_1,N_2\,$), we find that the zero-coupling $Z_{\text{BPS}}$ asymptotes to
\begin{equation}\label{eq:AsymptLocFormulaGeneralIntro}
\begin{split}
&\, \sum_{p\,\in\,\mathbb{Z}}\biggl(e^{-\frac{N^2 (M\varphi _1)^+ (M\varphi_2)^+ (M\varphi_3)^+}{2 M (M \omega_1+N_1)
(M \omega_2+N_2)}} \,+\,\ldots\biggr)\,\delta_{M(\alpha+\tfrac{k}{2}),\,1+3p} \\ &\,\, \,\qquad \qquad +\,\sum_{p\,\in\, \mathbb{Z}}\biggl(e^{-\frac{N^2 (M\varphi _1)^{-} (M\varphi_2)^{-} (M\varphi_3)^-}{2 M (M \omega_1+N_1)
(M \omega_2+N_2)}}\,+\,\ldots\biggr) \delta_{M(\alpha+\tfrac{k}{2}),\,-1+3p}\,,
\end{split}
\end{equation}
where $\ldots$ stands for contributions from other saddle points of the gauge–rapidity integral. All saddles satisfy the same localization condition encoded in the delta functions $\delta_{\ldots}\,$, and their zero locus reproduces~\eqref{eq:OrbifoldBCIntro}. In the bulk we expect each such delta function to correspond to regularity conditions at the tip of the dual background geometry that denotes their horizon~\cite{Cabo-Bizet:2018ehj,Aharony:2021zkr}.

The asymptotic localization~\eqref{eq:AsymptLocFormulaGeneralIntro} means that studying $Z_{\text{BPS}}$ at zero coupling is equivalent to studying $Z_{\text{BPS}}$ at strong coupling in any semiclassical expansion where one of the delta functions dominates the counting, namely, where the saddles dual to gravitational solutions with horizons emerge.~\footnote{As 
$Z_{\text{BPS}}$ at zero gauge coupling localizes to a linear combination of superconformal indices, only genuinely protected BPS states contribute in that region of charges. Effectively, this asymptotic expansion implements the projection to genuinely perturbatively protected states. As a consequence, in this limit the free-coupling value of $Z_{\text{BPS}}$ asymptotes to the strong-coupling value, and vice versa.}

Remarkably, these asymptotic expansions may be simply large charge expansions at finite $N$, e.g., $N=2\,$. Indeed, we test this saddle-point result at $N=2$ in appendix~\ref{sec:App} using explicit Cauchy-residue evaluation of~\eqref{eq:AsymptLocFormulaGeneralIntro}.

\item[6.] This method also allows us to identify new types of saddles and compute their large-$N$ contributions. These include two families that we call \emph{eigenvalue-instanton saddles} (Section~\ref{ref:EigenvalueInstantons}) and \emph{dressed orbifolds} (Section~\ref{eq:DressedOrbifolds}).

Eigenvalue-instantons arise from orbifold solutions by shifting eigenvalues among the equidistant minima of the multiparticle potential.  

For configurations with two stacks of eigenvalues---a fraction $x$ at one vacuum and $1-x$ at another---we show that the integral over the moduli space, $x\,$, localizes to $x=1/2$, i.e.\ to an orbifold configuration. Although we do not provide a general proof, we expect such behavior to hold in broader families.

This simple observation suggests that eigenvalue-instanton saddles are unstable and flow to orbifold saddles after integration over their moduli.

Dressed orbifolds consist of a core orbifold solution with an additional dressing. Evidence suggests they correspond to the large-$N$ asymptotic form of continuous families of Bethe roots at finite $N$ (in regions where the Bethe expansion exists~\cite{Closset:2017bse,ArabiArdehali:2019orz,Lezcano:2021qbj,Benini:2021ano,Cabo-Bizet:2024kfe}).  
Dressing of eigenvalue–instanton type is also possible, but appears to be unstable, just as in the undressed case.

The localization feature~\eqref{eq:AsymptLocFormulaGeneralIntro} implies that all exponentially growing saddle points of $Z_{\text{BPS}}$ for $\mathcal{N}=4$ SYM have on-shell action
\begin{equation}
\frac{P_{3}[\omega_1,\omega_2,\varphi_1,\varphi_2]}{(\omega_1+\tfrac{N_1}{M})(\omega_2+\tfrac{N_2}{M})}\,\pm\, \pi\text{i} N^2\gamma
\end{equation}
at large $N$, where $P_{3}$ is a cubic polynomial in $\omega_1,\omega_2,\varphi_1,\varphi_2\,$ and $\gamma$ is a complex function regular in expansions near roots-of-unity
\[\omega_{a} \to -\frac{N_a}{M}\,.\]
All saddles appear in complex-conjugate or time–reversal–conjugate pairs, corresponding to the two sets of delta functions in~\eqref{eq:AsymptLocFormulaGeneralIntro}.

For dressed orbifolds, the polynomial $P_3$ is identical to that of the core solution. For the cases we study (more general ones exist), the deviation from the core–orbifold action is, up to a $c$-number,
\[N^2\gamma=-\,\frac{\left((M\omega _1+N_1)-(M\omega_2+N_2)\right)^2 N^2}{16
    (M\omega_1+N_1)(M\omega_2+N_2) }\,.\]
Thus these corrections are irrelevant when $\omega_1\equiv \omega_2$, which would also impose $N_1=N_2$.

In the microcanonical ensemble this means that dressed orbifolds only compete with undressed orbifolds in regions of \emph{unequal} angular momentum. In Legendre-dual variables, this corresponds to regions where $\omega_1\neq \omega_2$, consistent with the empirical observation of~\cite{Choi:2025lck,Deddo:2025jrg}.  
The gravitational interpretation of this dressing remains to be understood.

\item[7.] Using these results, we illustrate how, in the microcanonical ensemble and semiclassical expansion~\footnote{To ease the reading, in the discussion we assume $J_1=J_2$ but our analysis covers also $J_1\neq J_2\,$.}
\begin{equation}\label{eq:SemiclassicalExpIntro}
N\to \infty\,, \qquad   \frac{J}{N^2}=j_0=\text{fixed}\neq 0\,,\,\qquad \frac{Q}{N^2}=q_0=\text{fixed}\,,
\end{equation}
the conjectured relations~\eqref{eq:AZBPSIntro} and~\eqref{eq:RelationBPSIndexIntro} follow at infinitely many codimension–1 loci~\eqref{eq:NonLLocusIntro} in the space of spin $J$ and $R$–charges $Q\,$ (Section~\ref{sec:Microcanonical}). In the notation used earlier,
\[
P\,:=\,2J+Q\,,\qquad P_{\text{tot}}:=\{J,Q\}\,.
\]

In field theory these codimension–1 regions determine where a \emph{single} pair of complex-conjugate or time–reversal–conjugate saddles can dominate $Z_{\text{BPS}}$ in the microcanonical ensemble.

Outside these loci, the leading pair of complex saddles would induce large oscillations, incompatible with the fact that $Z_{\text{BPS}}[P]$ must be positive by definition. Therefore, only on loci of the form~\eqref{eq:NonLLocusIntro} can a single pair dominate.  
Elsewhere, different saddle pairs must dominate. We do not attempt a full analysis of these saddle transitions (which are enforced by positivity); we leave this for future work.

\item[8.] We compare these results with the gravitational side of the duality~\cite{Cabo-Bizet:2018ehj} (Section~\ref{sec:BPSCigars}).

\item[9.] In the conclusions (Section~\ref{sec:FinalComments}) we discuss related open problems and directions. We explain how the results of~\cite{Cabo-Bizet:2018ehj} together with the results in this paper further support the conclusions of~\cite{Cabo-Bizet:2024gny} regarding the emergence of non-protected Schwarzian contributions within $\mathcal{N}=4$ SYM on $\mathbb{R}\times S^3\,$ (at large charges of order $N^2$ where saddles dual to black holes dominate) and how such an emergence can be understood with a zero-coupling computation in the gauge-theory side.

\end{itemize}

\subsection{Background information}

\label{sec:BHIntro}

Black holes are central objects in our effort to understand quantum gravity. They are Lorentzian gravitational solutions that, after a Wick rotation
\begin{equation}\label{eq:WickRot}
t_L\to \mp\,\text{i} t\,,
\end{equation}
and a periodic identification of the Euclidean time variable
\begin{equation}\label{eq:ThermalPeriodicity}
t\,\sim\,t +\beta\,, \qquad \beta<\infty
\end{equation}
are mapped into complex and smooth Euclidean classical configurations $g=g_{c}$. These configurations contribute to a hypothetical Euclidean gravitational path integral
\begin{equation}\label{eq:PathIntegral}
Z[\beta]\,=\, \int_{t\,\sim\, t+\beta} [Dg]\, e^{-\frac{S[g]}{G_N}}\,,
\end{equation}
in a particular semiclassical expansion $G_{N}\to 0$. The latter expansion is expected to be an asymptotic expansion in terms of saddle-point contributions
\begin{equation}\label{eq:Saddle-points}
Z[\beta]\,\underset{G_N\to 0}{\sim}\, \sum_{g_c}\,(\text{1-loop Det})\, e^{\mathcal{F}_{g_c}[\beta]}\,.
\end{equation}
The on-shell value of the gravitational action on $g_c$, which is a function of $\beta$,
\begin{equation}
\mathcal{F}_{g_c}[\beta]\,:=\,-\,\frac{S[g_c]}{G_N}
\end{equation}
is called the gravitational free energy, for reasons that will be explained below.

The $g_c$’s are solutions to the Euclidean equations of motion that satisfy the thermal periodicity condition~\eqref{eq:ThermalPeriodicity} and are smooth. Given a Lorentzian solution, there are at least as many $g_c$'s as independent smooth ways of satisfying the thermal periodicity condition~\eqref{eq:ThermalPeriodicity}. The remarkable problem then remains that of classifying all $g_c$’s.\footnote{It is even possible that some of them may not be related to any globally well-defined or real Lorentzian geometries after the inverse Wick rotation~\eqref{eq:WickRot}.}

Understanding the fully quantum regime, $1/G_N=$ finite, of~\eqref{eq:PathIntegral}, namely, a non-perturbative completion of the asymptotic expansion~\eqref{eq:Saddle-points}, remains a formidable challenge in generic theories and space-time dimensions.

AdS$_D$/CFT$_{D-1}$ duality~\cite{Maldacena:1997re} provides a powerful framework to address these questions for gravitational theories in Anti-de Sitter spacetimes (AdS). In natural units, the duality identifies $G_N$ with the inverse of the central charge $c$
\begin{equation}
G_{N}\,\propto\, \frac{1}{c}
\end{equation}
of a dual conformal field theory, which may happen to be a gauge theory of rank $N\,$. In such cases, the central charge $c$, which is a dimensionless parameter, depends on $N\,$. The finiteness of $N$ (i.e. of $c(N)$) corresponds to the quantization of the gravitational theory, $G_{N}\,>\,0$. In the canonical example of $U(N)$ maximally supersymmetric four-dimensional Yang-Mills,
\begin{equation}
G_N\,\equiv\, \frac{\pi}{2N^2}\,.
\end{equation}
The duality also provides a neat state-counting interpretation of~\eqref{eq:PathIntegral}
\begin{equation}\label{eq:StateCounting}
Z\,=\,Z[\beta]:=\text{Tr}_{\mathcal{H}}\, e^{-\beta \widehat{H}}\, =\, \sum_{n} d[E]\, e^{-\beta E}\,,
\end{equation}
where $\widehat{H}$ is the Hamiltonian and $\mathcal{H}$ is the space of states of the dual gauge theory. States in $\mathcal{H}$ are identified as microstates in the dual quantum theory of gravity. There are $d[E]$ of them at energy level $E\,$. The corresponding Boltzmann entropy is defined as
\begin{equation}\label{eq:MicroEntropy}
\mathcal{S}\,=\,\mathcal{S}[E]:=\,\log d[E].
\end{equation}

In virtue of~\eqref{eq:StateCounting}, the degeneracy of microstates $d(E)$ is extracted from the gravitational path integral~\eqref{eq:PathIntegral} by a Laplace transform
\begin{equation}\label{eq:dE}
d[E] \,=\, \oint_{|x|\,=\,1} \frac{dx}{2\pi \text{i} x}\int_{t\,\sim\, t+\beta} [Dg]\,\,e^{-\frac{S[g]}{G_N}}\,x^{-E} \,,\qquad x=e^{-\beta}.
\end{equation}
In the semiclassical expansion
\begin{equation}\label{eq:LargeGNExp}
G_{N}\,\to\,0\,, \,\qquad  G_N E \,=\,e=\text{fixed}\,\neq\, 0 \,,  
\end{equation}
\eqref{eq:dE} can be approximated by a saddle-point expansion. From~\eqref{eq:Saddle-points} it follows that the saddle points of~\eqref{eq:dE} are fixed by the extremization condition
\begin{equation}\label{eq:Ext}
\frac{1}{G_N}\,\text{ext}_{\beta}\biggl(-S[g_c]+\beta e\biggr)\,,
\end{equation}
where in principle $g_c$ ranges over all smooth-enough Euclidean saddle points in the asymptotic expansion~\eqref{eq:Saddle-points}. One possibility is that, at very leading order in~\eqref{eq:Saddle-points}, all dominating $g_c$'s come from Wick rotations of Lorentzian black hole solutions constrained by regularity conditions at the tips of the corresponding cigar. Examples of such regularity conditions that will be relevant to our discussion were put forward in~\cite{Cabo-Bizet:2018ehj} and~\cite{Aharony:2021zkr}.

If one only cares about the leading asymptotic behavior in the expansion~\eqref{eq:PathIntegral}, then one only needs to focus on the $g_c$'s that maximize the real part of~\eqref{eq:Ext}; we will denote those as~$g^\star\,$
\begin{equation}\label{eq:CriterumDominance}
g^\star= \qquad  \text{$g_c$'s that maximize} \qquad \text{Re}\biggl(\text{ext}_{\beta}\biggl(\mathcal{F}_{g_c}[\beta]+\beta E\biggr)\biggr)\,.
\end{equation}
In the semiclassical expansion~\eqref{eq:LargeGNExp}, the degeneracy of states is given by
\begin{equation}\label{eq:Averaging}
d[E]\,\sim\, \sum_{g^\star} (\text{1-loop Det})\,e^{\biggl(-\frac{S[g^\star]}{G_N}\,+\,\beta E\biggr)_{\beta=\beta^\star}}
\end{equation}
\footnote{From now on, we ignore subleading corrections in asymptotic relations $\sim\,$ under the expansion~\eqref{eq:LargeGNExp}.} where
\begin{equation}
\beta^\star =\beta^\star[E]\,, \qquad E\,=\,\frac{1}{G_N}\, \partial_{\beta} S[g^\star]\bigg|_{\beta=\beta^\star}\,.
\end{equation}

Some of the dominating solutions $g^\star$ are expected to come from Wick rotations of the \emph{larger} Lorentzian black hole in the original gravitational theory, for given values of $E\,$. By larger we mean, roughly speaking, the one with the largest horizon area.

The simplest possible geometries $g^\star$ can be represented as two-dimensional cigar geometries with a puncture at the tip of the cigar. The tip denotes the horizon of the solution ($r=r_+$). The radial direction out of the tip corresponds to the radial direction of the geometry $g^\star$, $r\,$. Each point in the cigar represents the codimension-two space defined by fixing $(t,r)$ coordinates in the geometry. The boundaries of $g^\star$ are represented by the puncture $r=r_+$, which is the horizon, and the codimension-one locus at $r=\infty\,$, which is the conformal boundary where the CFT$_{D-1}$ lives.

\subsection{Black hole thermodynamics} \label{sec:BHIntro2}
Focusing on a single gravitational cigar $g^\star\,$, its gravitational entropy is defined as
\begin{equation}\begin{split}
\mathcal{S}_{g^\star}[E]& := \text{ext}_\beta \biggl( \mathcal{F}_{g^\star}[\beta]+\beta E\biggr)\,,
\end{split}
\end{equation}
which is the Legendre transform of the free energy $\mathcal{F}_{g^\star}[\beta]$, i.e.
\begin{equation}
\mathcal{S}_{g^\star}\,=\,(1-\beta \partial_{\beta})\,\mathcal{F}_{g^\star}\,,
\end{equation}
where $\beta$ and $E$ are Legendre dual variables
\begin{equation}
 E \,=\,-\partial_\beta \mathcal{F}_{g^\star}[\beta]\,.
\end{equation}

The quantity $\mathcal{S}_{g^\star}$ can be computed solely with data encoded in the tip of the cigar geometry $g^\star$ via the formula~\cite{Gibbons:1976ue}
\begin{equation}\label{eq:AreaHorDef}
A_{\,g^\star}\,=\,\int_{\text{tip of cigar}} d x^{D-2} \sqrt{g^\star}_{r=r_+}
\end{equation}
where $\sqrt{\det g^\star_{r=r_+}}$ is the volume form of the Wick-rotated metric $g^{\star}$ restricted to the tip of the cigar geometry. At that point, the thermal cycle collapses and the corresponding hypersurface has codimension $D-2\,$.

Equation~\eqref{eq:AreaHorDef} follows from the fact that the on-shell action $\mathcal{F}_{g^\star}$, which is the integral of a total derivative, receives contributions only from the boundary of the cigar. The contribution coming from the codimension-one boundary is cancelled by the $+\beta E$ term. The contribution from the tip of the cigar is simply the integral~\eqref{eq:AreaHorDef}.

In static Lorentzian solutions, the Wick rotation~\eqref{eq:ThermalPeriodicity} keeps the metric real. In those cases, for real mass $E$, the quantity
\begin{equation}
A_{g^\star}\,,
\end{equation}
computed by~\eqref{eq:AreaHorDef}, is real and positive, and by definition matches the area of the horizon of the Lorentzian solution,
\begin{equation}\label{eq:BekensteinHawking}
\mathcal{S}_{g^\star}\,=\,\frac{A_{\,g^\star}}{4 G_N}\,.
\end{equation}

In more general examples, for instance in the presence of rotation, the Wick rotation~\eqref{eq:ThermalPeriodicity} renders the metric $g^\star$ complex, and thus $A_{g^\star}$, as computed by~\eqref{eq:AreaHorDef}, may become complex in certain regions of charges. Whenever complex cigars emerge, they appear in complex-conjugate pairs of solutions. These solutions map to each other under the time-reversal symmetry $t\to-t\,$ (which corresponds to the exchange of signs in~\eqref{eq:WickRot}). This is the case for the rotating black holes we will study here~\cite{Chong:2005da,Gutowski:2004ez,Cabo-Bizet:2018ehj}.

We will call the classical or non-CTC locus of a dominating cigar $g^\star\,$ with identification~\[t\sim t+\beta\,,\](resp. of its complex-conjugate dual), the codimension-one region in its space of charges for which they share the same $\mathcal{S}_{g^\star}$, i.e.,
\begin{equation}\label{rm:NonLinearConstraintOfCharges}
\text{Im}(\mathcal{S}_{g^\star})\,=\,0\,.
\end{equation}
It is only in the classical or non-CTC locus that the entropy $\mathcal{S}_{g^\star}$ of a single semiclassical solution $g^\star$ can be associated with the horizon area of the parent Lorentzian solution with the same charges. Equation~\eqref{rm:NonLinearConstraintOfCharges} will also be called the non-linear constraint of charges.

As we will explain in due time, in examples with rotation there are also orbifold time-periodicity conditions~\cite{Aharony:2021zkr}. Our field theory results, including past results in~\cite{Cabo-Bizet:2019eaf},  predict that the condition of reality for entropy and charges at the tip will give codimension-one loci different from~\eqref{rm:NonLinearConstraintOfCharges}. Consequently, our field-theory results predict that such cigars correspond to supersymmetric Lorentzian solutions with naked CTC.

\section{Index and BPS partition function}

Computing the degeneracy of all states in the field theory $d[E;\ldots]$ at generic energies $E$ in the semiclassical expansion~\eqref{eq:LargeGNExp} would require dealing with coupling corrections. A much simpler problem is to deal with protected quantities such as Witten indices, or superconformal indices~\cite{Romelsberger:2005eg,Kinney:2005ej}.

At the microscopic level we will focus on four-dimensional $\mathcal{N}=1$ superconformal theories on $S^{3}$ space. These theories have a $U(1)$ R-symmetry represented by the Cartan charge operator $Q\,$, and a pair of complex-conjugate supercharges $\mathcal{Q}$ and $\mathcal{Q}^\dagger$ such that
\begin{equation}
2\{\mathcal{Q},\mathcal{Q}^\dagger\}\,=\,E-2J-\frac{3}{2}Q \,\geq \, 0\,, \qquad 2J\,:=\, J_1\,+\,J_2\,,
\end{equation}
and
\begin{equation}
\left[\mathcal{Q},J_a+\tfrac{Q}{2}\right]\,=0\,,\qquad \left[\mathcal{Q},Q\right]\,\neq\, 0\,.
\end{equation}
For them, the superconformal index can be defined as
\begin{equation}\label{eq:Index}
\mathcal{I}[\omega_1,\omega_2]\,=\,\text{Tr}_\mathcal{H}\biggl((-1)^F x^{E-{2J}-\frac{3}{2}Q} p^{J_1+\tfrac{Q}{2}}q^{J_2+\tfrac{Q}{2}} \biggr)\,,\qquad  p=e^{\omega_1}\,, \, q:= e^{\omega_2}\,,
\end{equation}
following the conventions of~\cite{Cabo-Bizet:2018ehj}. This object is protected against coupling corrections and it can then be computed at zero gauge coupling. It depends only on the rapidities $p$ and $q$, not on the temperature.

As for the generic partition function $Z[\beta]$ before, indices can also be defined as Euclidean path integrals upon imposition of periodic boundary conditions for bosons and fermion fields $X$ on the thermal circle
\begin{equation}\label{eq:PeriodicBC}
X[t+\beta]\,=\,+\,X[t]\,.
\end{equation}

To compute the index~\eqref{eq:Index} upon imposing the thermal periodicities~\eqref{eq:PeriodicBC} requires turning on angular velocities 
\begin{equation}\label{eq:DefOmega12}
\Omega_{1,2} \,=\,1\,+\, \frac{\omega_{1,2}}{\beta}
\end{equation}
for the space $S^3\,$, and turning on a background gauge potential $A^{Q}_0$ for the $U(1)$ R-symmetry proportional to
\begin{equation}\label{eq:DefPhi}
\Phi=\frac{3}{2}\,+\,\frac{\varphi}{\beta}\,, \qquad \varphi\,=\,\tfrac{1}{2}(\omega_{1}+\omega_2)\,.
\end{equation}
This change in the background geometry produces the effect that the twisted weights
\begin{equation}
x^{-{2J}-Q} p^{J_1+\tfrac{Q}{2}}q^{J_2+\tfrac{Q}{2}}\,
\end{equation}
produce in the trace representation.
One can use the same trick, but the other way around: we keep $\Omega_{1,2}$ as before and redefine the background R-symmetry potential by exchanging 
\begin{equation}
\Phi\to \Phi- 2\pi \text{i}\alpha/\beta\,.
\end{equation}
In the new background the relation between $\Phi$ and $\Omega$ reads
\begin{equation}\label{eq:GeneralDeformations}
2\Phi \,-\,\Omega_1\,-\,\Omega_2\, -1\,=\, 4 \pi \text{i} \,\frac{\alpha}{\beta}\,,
\end{equation}
or equivalently
\begin{equation}\label{eq:DefAlphaVarphi}
2\varphi\,-\,\omega_1 \,-\omega_2\,=\,4\pi \text{i} \alpha\,.
\end{equation}
In this new background the path integral formulation of the index $\mathcal{I}_{\partial}$ is computed with the following twisted thermal periodicity conditions, instead of~\eqref{eq:PeriodicBC},
\begin{equation}\label{eq:Ambiguity}
X[t+\beta] \,=\, + e^{2\pi \text{i} \alpha r} X[t]\,,
\end{equation}
which guarantees its independence of $\alpha$. Here $r$ is the R-charge $Q$ of the field $X\,$. In a gauge theory with spectrum of $Q$ such that (for some fractional c-numbers $\kappa_a$)
\begin{equation}\label{eq:KappaShifts}
e^{\pi\text{i}Q}\,=\, e^{\pi \text{i} \sum_{a=1}^{2}\kappa_a (J_a+\tfrac{Q}{2})} (-1)^F
\end{equation}
\footnote{For example, fixing $\kappa_a=1$ one obtains the spin statistics identity~$e^{-\pi \text{i}\, (2J)}=(-1)^F\,$.} then fixing
\begin{equation}\label{eq:PeriodOfAlpha}
\alpha= \tfrac{1}{2} \bmod 1
\end{equation}
after implementing appropriate shifts of $\omega_1$ and $\omega_2\,$, can be understood as imposing \emph{periodicity conditions} consistent with both supersymmetry and thermality
\begin{equation}
X[t+\beta] \,=\,(-1)^F X[t]\,,
\end{equation}
the very same boundary conditions that define the usual partition function $
{Z}\,$.

Concretely, this only means that the index $\mathcal{I}$ can be obtained from the partition function upon imposition of the constraint~\eqref{eq:PeriodOfAlpha}. More precisely, that
\begin{equation}
\begin{split}
\mathcal{I}[\omega_1+\pi\text{i}\kappa_1,\omega_2+\pi\text{i}\kappa_2]=Z[\beta,\omega,\varphi]&\,=\,\text{Tr}_{\mathcal{H}} x^{E} e^{\beta\Omega_1 J_1}e^{\beta\Omega_2 J_2} e^{\beta \Phi Q}\, \\
&\,=\,\text{Tr}_{\mathcal{H}} x^{E-2J-3/2 Q} e^{ \omega_1 J_1}e^{\omega_2 J_2} e^{\varphi Q}
\end{split}
\end{equation}
for arbitrary temperature
\begin{equation}
\beta \text{ arbitrary}
\end{equation}
if the linear and complex constraints 
\begin{equation}\label{eq:ConstraintSect2Canonical}
2\Phi \,-\,\Omega_1-\Omega_2 -1\,=\, 4 \pi \text{i} \,\frac{\alpha}{\beta}\,, \qquad \alpha\,=\,\frac{1}{2}\,\bmod 1,
\end{equation}
or equivalently
\begin{equation}\label{eq:ConstraintSect2}
2\varphi \,-\,\omega_1-\omega_2 \,=\, 4\pi \text{i}\,\alpha, \qquad \alpha\,=\,\frac{1}{2}\bmod1\,,
\end{equation}
are imposed on the \emph{BPS chemical potentials}
\begin{equation}\label{eq:BPSChemcialPotentials}
\omega_a\,=\, \beta (\Omega_a -1)\,, \qquad \varphi \,=\,\beta (\Phi-\frac{3}{2})\,.
\end{equation}

\paragraph{Adding flavour potentials} We can also insert flavor rapidities $\underline{\nu}=\{\nu_j\}$ dual to Cartan flavor or R-symmetry generators,~\footnote{In the gravity side this would correspond to temporal components of gauge field potentials.} $\underline{Y}=\{Y_j\}$, including $Q\,$. In the microscopic theory we will always assume that the spectrum of $\underline{Y}$ is quantized in integer units.

Then we can define the following refined or flavored partition function
\begin{equation}\label{eq:PhysicalPartitionFunction}
Z[\beta,\omega,\varphi]\,=\,\text{Tr}_{\mathcal{H}} x^{E-2J-3/2 Q} e^{ \omega_1 J_1}e^{2 \omega_2 J_2} e^{\varphi Q} e^{\underline{\nu}\cdot \underline{Y}}
\end{equation}
For technical reasons, we will be interested in the choices
\begin{equation}\label{eq:ChoicesOfFlavorRapidities}
n_j:=e^{\nu_j}\,=\bigl(\chi^{1/2}\bigr)^{c_j}\,,\qquad \chi\,:=\,e^{2\pi \text{i} k}\frac{t^2}{p q}\,=\,e^{2\pi \text{i} k}e^{4\pi \text{i}\alpha }\,,
\end{equation}
with $c_j$ some real number and $k$ some integer number. This guarantees that on the BPS locus~\eqref{eq:ConstraintSect2}~\footnote{As already said, in this paper we use the rule $(e^{X})^{y} \,:=\,e^{X y}$\,.}
\begin{equation}
Z[\beta,\omega,\varphi]\,=\,\mathcal{I}[\omega]\,.
\end{equation}

Instead, we define the refined BPS partition function of a theory with enhanced supersymmetry and three independent R-symmetries $Q_{a}$, $a=1,2,3$ as
\begin{equation}\label{eq:ZBPSdefinition}
Z_{\text{BPS}}[\omega,t_1,t_2,t_3]\,:=\, \text{Tr}_{\mathcal{H}} x^{E-2J-\sum_{I} Q_I} p^{J_1}q^{ J_2} t_1^{Q_1} t_2 ^{Q_2}t_3^{Q_3}\biggl|_{x=0}\,.
\end{equation}
If we define~\footnote{\label{ftn:ChoiceFlavourRapiditiesConst}We will at some point make use of a co-dimension one choice of flavour rapidities, for which $\prod_a n_a=1\,$. }
\begin{equation}
\begin{split}
&\,\,\,\,\,\,\,t_a\,=\, t^{2/3}\, n_a=e^{\varphi_a}\,,\\  &\frac{3}{2}Q\,:=\,Q_1+Q_2+Q_3\,,\quad Y_a\,=\, Q_a\,,
\end{split}
\end{equation}
then we can recast this partition function in a series of equivalent representations where $Q$ is defined as the $U(1)$ R-charge of reference
\begin{equation}
\begin{split}
Z_{\text{BPS}}[\omega,\varphi_1,\varphi_2,\varphi_3] 
&\,=\,\text{Tr}_{\text{BPS}}\,\,  p^{J_1}q^{J_2}t^{2(Q_1+Q_2+Q_3)/3} n_1^{Y_1} n^{Y_2}_2 n^{Y_3}_3 \\
&\,=:\,\text{Tr}_{\text{BPS}}\,\,  p^{J_1}q^{J_2}  (t)^{Q} n^{Y_1}_1 n^{Y_2}_2n^{Y_3}_3 \\
&\,=\,\text{Tr}_{\text{BPS}}\,\,  p^{J_1+Q/2}q^{J_2+Q/2}  (\chi)^{Q/2} n^{Y_1}_1 n^{Y_2}_2 n^{Y_3}_3 \\
&\,=:\, \text{Tr}_{\text{BPS}}\,\,  p^{J_1+Q/2}q^{J_2+Q/2}  (\chi)^{Q/2} n^{Y_1}_1 n^{Y_2}_2 n^{Y_3}_3 \\
&\,=\, \text{Tr}_{\text{BPS}}\,\,  p^{J_1+Q/2}q^{J_2+Q/2}  e^{2\pi\text{i}\alpha Q} e^{\underline{\nu}\cdot \underline{Y}}\,.
\end{split}
\end{equation}
The eigenvalues of $Q$ are denoted as $r\,$, and are by definition
\begin{equation}
\frac{3}{2}r \,=\, \sum_{a} r_a\,,
\end{equation}
where $r_a$ denote the eigenvalues of the $Q_a$'s.
For later use, we note that a simultaneous shift
\begin{equation}\label{eq:SymmetryShift}
\omega_{a}\to \omega_{a}\mp 2\pi \text{i}
\end{equation}
 at fixed $\alpha$ and $\nu$, can be rephrased as a shift
\begin{equation}
\alpha\to \alpha=\alpha\mp 2
\end{equation}
at fixed $\omega_a$ and $\nu$ (provided $2J \,\in\, \mathbb{Z}$ as we know it is the case).

\paragraph{Refined path integral from unrefined path integral}
Let us assume
\[
Z_{{\text{BPS}}}[\omega,\varphi_1,\varphi_2,\varphi_3]\]
is a path integral computed only at the sublocus of background potentials
\begin{equation}
\varphi_a\,=\,\varphi^{(0)}_a\,=\, \frac{2\varphi}{3}\,.
\end{equation}
Then the answer in generic background potentials $\varphi_{a}$
follows from a spectrum of eigenvalues $\Lambda_{f}\,$ that is related to the spectrum of eigenvalues of the former path integral, $\Lambda_0\,$, via the substitution rule
\begin{equation}\label{eq:EqChangeBCEigenvalue}
\Lambda_f =\Lambda_0\,+\,2\pi\text{i} \,\widetilde{r}\, \alpha\,.
\end{equation}
with
\begin{equation}
\widetilde{r}\,:=\,\sum_{a} c_a r_a\,.
\end{equation}
For us
\begin{equation}\label{eq:ChoiceOfMIxing}
c_{a}\,=\,\frac{2}{3} \implies \widetilde{r}=r\,.
\end{equation}
\eqref{eq:EqChangeBCEigenvalue} is simply the change induced by adding the corresponding temporal components of the background gauge potential to space-time covariant derivatives. 
Alternatively, at $\beta\to\infty$ one can interpret these deformations as deforming whichever were the periodicity conditions for fields $X$ defining $Z_{\text{BPS}}[\omega,\varphi^{(0)}_{1},\varphi^{(0)}_{3},\varphi^{(0)}_{3}]$ into the twisted boundary conditions
\begin{equation}
X[t+\beta]\,=\, X[t]\,\underbrace{\ldots}_{\text{previous twists}}\, \chi^{\frac{\sum_{a}c_a R_a}{2}}\,,
\end{equation}
without need of deforming the covariant derivatives.

This implies that the minimally refined BPS partition function encodes the fully refined BPS partition function provided the spectrum of R-charges of BPS states in the theory, $\{r_I\}$, is known.

\subsection{$Z_{\text{BPS}}$ from supersymmetric localization}\label{sec:SUSYLoc}
Next, borrowing the conventions from section 4 of~\cite{Cabo-Bizet:2018ehj}, we define a path integral
\[ Z^{(L)}_{\text{BPS}} \,=\, \int_{L} [DX]\, e^{-S_{\text{phys}}[X]}\,. \]
The physical action $S_{\text{phys}}[X]$, the matter content $\{X=\phi_{\text{there}}\}$, and the supersymmetry algebra are all the ones used in section 4 of~\cite{Cabo-Bizet:2018ehj}. This time, the functional integral is computed over a space of fields $\{X\}$ on an interval of length \(L\), and not over an $S^1$, which was the case in~\cite{Cabo-Bizet:2018ehj}.

Let us define the following combination of chemical potentials
\begin{equation}\label{eq:BPSChemcialPotentialsInterval}\begin{split}
\omega^{(L)}_{a}&\,:=\, L (\Omega_a -1)\,, \\  \varphi^{(L)} &\,:=\,L (\Phi-\frac{3}{2})\,,\\\
4\pi \text{i} \alpha^{(L)}&\,:=\,2\varphi^{(L)}\,-\,\omega^{(L)}_1 \,-\omega^{(L)}_2\,.
\end{split}
\end{equation}
In the limit \( L \to \infty \) at fixed values of the potentials~\eqref{eq:BPSChemcialPotentialsInterval}, by definition, $Z^{(L)}_{\text{BPS}}$ should become the path-integral representation of the BPS partition function \( Z_{\text{BPS}} \) defined in~\eqref{eq:ZBPSdefinition}:
\begin{equation}\label{eq:ZLZBPS}
Z^{(L\to \infty)}_{\text{BPS}} = Z_{\text{BPS}}\,.
\end{equation}
(We will show this a posteriori in a concrete example.)
From now on, to ease the reading, we remove the superindex $(L)\,$, but the reader should recall that it is $L$ (the length of the interval) -- and not $\beta$ (the period of $S^1$ in the previous section), the implicit parameter in the definition of BPS chemical potentials which we will use in this section.

The relevant path integral is over the functional space of fields \(X\) -- the cohomologically unpaired variables defined in section 4 of~\cite{Cabo-Bizet:2018ehj} -- but this time imposing upon them the following supersymmetric Neumann-like boundary conditions at the endpoints of the time interval:
\begin{equation}\label{eq:ConstrainsPathIntegral}
\widehat{D}_t X = 0\,, \qquad 2\{\mathcal{Q},\mathcal{Q}^\dagger\} = \widehat{D}_t\,,
\end{equation}
instead of periodicity constraints. These boundary conditions also annihilate potential boundary-term variations of the classical action $S_{\text{phys}}$, making the semiclassical variational principle well-defined on the interval $L$.~\footnote{In the case of gauge-fields one needs to be more careful to see this, but the same conclusion holds. For example, in the Landau gauge $D^\mu A_\mu=D^\mu \delta A_\mu=0$ the boundary variation of the Yang-Mills kinetic term is the integral over $S^3$ of a c-number times~$[D^{t},D^\mu] \delta A_{\mu}\,$ which obviously vanishes upon imposition of the Neumann-like boundary conditions~\eqref{eq:ConstrainsPathIntegral}. } \( \widehat{D}_t \) is (up to a c-number proportionality factor) the differential operator \(\widehat{H}\) defined in equation (4.28) of~\cite{Cabo-Bizet:2018ehj}.

By looking at the supersymmetry transformations~(4.26) and (4.27) in~\cite{Cabo-Bizet:2018ehj}, it is clear that the boundary conditions~\eqref{eq:ConstrainsPathIntegral} are consistent with supersymmetry. The relevant Killing spinors are the ones we reported in equation (4.6) of~\cite{Cabo-Bizet:2018ehj}, e.g.,
\[
\begin{pmatrix}
e^{\tfrac{t}{2}(1 - 2\Phi + \Omega_1 + \Omega_2)} \\
0 \\
0 \\
e^{\tfrac{t}{2}(-1 + 2\Phi - \Omega_1 - \Omega_2)}
\end{pmatrix}.
\]
In the interval, though, we do not need to impose the global constraint~\eqref{eq:ConstraintSect2Canonical}. The existence of these Killing spinors in the interval means that the path integral $Z_{\text{BPS}}^{(L)}$ can be computed using supersymmetric localization at generic values of the gauge coupling $g_{\text{YM}}\,$. This is because the gauge-coupling dependence in the Lagrangian is $\mathcal{Q}$-exact, and the corresponding path integral is invariant under the addition of $\mathcal{Q}$-exact terms~\cite{Cabo-Bizet:2018ehj}.~This means that the corresponding path integral is piecewise independent of the gauge coupling. 

The only dependence of $Z_{\text{BPS}}$ on $g_{\text{YM}}$ can arise at the position of discontinuities that occur at values of the coupling where the functional space over which the path integral is computed over changes. This functional space is determined by the asymptotic form of the physical potential in the Lagrangian density of the corresponding theory. Accordingly, $Z_{\text{BPS}}$ can jump only at values of the coupling where the asymptotics of the physical potential in the relevant theory change.~\footnote{This is because such a change may modify the basis of Gaussian massless fluctuations, or more precisely, BPS classical fluctuations around the relevant vacua with classical charge $\widehat{H}=0$.} 

For example, in $U(N)$ $\mathcal{N}=4$ SYM, such a potential is obtained after integrating out auxiliary fields, and it contains terms of the form $g_{\text{YM}} \bigl|[\phi_a,\phi_b]\bigr|^2\,$. The fact that the asymptotic functional form of the potential, at least in $\mathcal{N}=4$ SYM, changes only at $g_{\text{YM}}=0$ implies that a discontinuity may occur only when one moves from $g_{\text{YM}}=0$ to $g_{\text{YM}}\neq 0\,$. However, as long as $g_{\text{YM}}\neq 0$ supersymmetric localization predicts $Z_{\text{BPS}}$ to remain constant all the way up to strong coupling.~{In this sense, supersymmetric localization implies the non-renormalization conjecture of~\cite{Grant:2008sk}.}

The Neumann-like boundary condition~\eqref{eq:ConstrainsPathIntegral}
annihilates all Fourier modes in the interval that are not in the kernel of $\widehat{D}_t$. This is because these modes are eigenfunctions of a one-dimensional Laplacian (in a Kaluza-Klein reduction to $S^1$); thus, imposing two independent conditions completely fixes the solution to modes in the kernel of $\widehat{D}_t$ all along the time direction~\cite{Cabo-Bizet:2016ars} and not just at the extrema. In virtue of~\eqref{eq:ConstrainsPathIntegral}, these are, by definition, the BPS states annihilated by $\mathcal{Q}$ and $\mathcal{Q}^\dagger\,$.

The localizing $\mathcal{Q}$-exact action is the physical action at zero coupling. The spectrum of eigenvalues that determines the localizing action one-loop determinant is that of the free theory. Using equation~\eqref{eq:EqChangeBCEigenvalue}, we recover the eigenvalues in the presence of minimal refinement $\varphi$ from the ones obtained in the unrefined case studied in~\cite{Cabo-Bizet:2018ehj}, i.e., from those computed under the implicit assumption
\begin{equation}\label{eq:ConstraintFieldTheory}
2\varphi_{\text{there}}=2\varphi^{(0)}\,:
=\,{\omega_1+\omega_2+2\pi \text{i} k}\,, \qquad k\in \mathbb{Z}\,.
\end{equation}
For the moment we drop $k$ from the equations ($k=0$). This dependence can be, and it will be, recovered eventually by shifting either $\omega_{1}$ or $\omega_2\,$ at fixed $\varphi\,$. The interval path integral $Z_{\text{BPS}}$ (not the index) computed at the choice of branch $k=0$ will always be denoted with the supraindex $(k=0)$.

The counting partition function, for which all Taylor coefficients are positive integers, is obtained from the latter by shifting $\omega_{a}\to\omega_a+\pi\text{i} k\,$, or $\omega_1\to \omega_1+2\pi \text{i} k\,$, or $\omega_2\to \omega_2+2\pi \text{i} k\,$ at fixed $\varphi$, for some integer $k\,$.

As explained, the minimally refined eigenvalues follow from the unrefined ones,~$\Lambda\,$, computed in
\cite{Cabo-Bizet:2018ehj}, by replacing
\begin{equation}\label{EigenVChange}
\Lambda\,\to\,\Lambda +2\pi  r\alpha
\end{equation}
where $r$ is the R-charge of a generic (cohomologically) unmatched mode necessary to compute one-loop determinants. Due to supersymmetric localization, we can work with the spectrum of R-charges of the free theory.

For the vector modes, the unrefined one-loop computation is~\cite{Cabo-Bizet:2018ehj}
\begin{equation}
\small
Z^{\text{vector}, \rho}_{\text{1-loop}}(u)
= \prod_{n_0\,\in\,\mathbb{Z}}^{\infty}\frac{1}{2\pi n_0\,+\,2\pi\rho(u)}\,\cdot\,\prod_{n_0 \in \mathbb{Z}} \prod_{n_1, n_2 \ge 0}
\frac{2\pi n_0 +\rho \cdot u -\text{i} n_1 \omega_1 - \text{i} n_2 \omega_2}
{2\pi n_0 + \rho \cdot u  - \text{i} (n_1+1) \omega_1 - \text{i} (n_2+1) \omega_2}\,.
\end{equation}
This first factor ($\rho\neq0$), which usually cancels the Vandermonde contribution after regularization, comes from adding a missing zero mode ($n_1=n_2=0$) in the numerator for the gaugino contribution~\cite{Assel:2014paa,Alday2013a}
\begin{equation}
\prod_{n_0\,\in\,\mathbb{Z}}^{\infty}\frac{1}{2\pi n_0\,+\,2\pi\rho(u)}\,.
\end{equation}
and in the present case becomes
\begin{equation}\label{eq:DeltaChiDivergent}
\prod_{n_0\,\in\,\mathbb{Z}}^{\infty}\frac{1}{2\pi n_0\,+\,2\pi\rho(u)\,+\,2\pi \alpha}
\end{equation}
after using~\eqref{EigenVChange}, because the gaugino has $R$-charge $r=+1\,$.
The unique meromorphic function in the complex variable
\begin{equation}
z=e^{2\pi \text{i}(\rho(u))}\,,
\end{equation}
up to a constant in $z\,$, that has poles at the position in the complex $z$-plane indicated by the denominator in~\eqref{eq:DeltaChiDivergent} is
\begin{equation}
\frac{\Delta(\rho(u))}{1-z}:=\frac{1}{1-z \,\chi^{1/2}}\,=\, \text{exp}\biggl(+\sum_{n=1}^\infty \frac{1}{n} z^n\, \chi^{n/2}\biggr)
\,.
\end{equation}
The ambiguous regularization constant is fixed by demanding that the regularized form has a counting interpretation, with a single entity counted at $z\,=\,0\,$. 

The other contributions from the gaugino, the ones in the numerator, transform into
\begin{equation}
\prod_{n_0\,\in\,\mathbb{Z}}^{\infty}\prod_{n_1, n_2 \ge 0}{2\pi n_0\,+\,2\pi\rho(u)\,+\,2\pi \alpha -\text{i} n_1 \omega_1 -\text{i}n_2 \omega_2}
\end{equation}
whose regularized version is
\begin{equation}
(z \chi^{1/2};p,q)_\infty\,=\, \text{exp}\biggl(-\sum_{n=1}^\infty \frac{z^n \chi^{n/2}}{n(1-p^n)(1-q^ n)}\biggr)\,.
\end{equation}
The contribution from the vector field remains the same, because it has zero R-charge
\begin{equation}
\prod_{n_0\,\in\,\mathbb{Z}}^{\infty}\prod_{n_1, n_2 \ge 0} \frac{1}{2\pi n_0\,+\,2\pi\rho(u) -\text{i} (n_1+1) \omega_1 -\text{i}(n_2+1) \omega_2}\,.
\end{equation}
Its regularized version is
\begin{equation}
\frac{1}{(z p q; p,q)_\infty}\,=\,\text{exp}\biggl(+\sum_{n=1}^\infty \frac{(z p q)^{n}}{n(1-p^n)(1-q^ n)}\biggr)\,.
\end{equation}
The unregularized contribution of a chiral multiplet with R-charge \[Q=r=r_B\] is~\cite{Cabo-Bizet:2018ehj}
\begin{equation}\label{eq:Z1Loop}
\small
Z^{\text{chiral}, \rho}_{\text{1-loop}}(u)
= \prod_{n_0 \in \mathbb{Z}} \prod_{n_1, n_2 \ge 0}
\frac{2\pi n_0 +\rho \cdot u\,+\,2\pi \,{r}_{F}\, \alpha -\text{i} (r_F-1) \,\omega_+ +\text{i} n_1 \omega_1 + \text{i} n_2 \omega_2}
{2\pi n_0 + \rho \cdot u \,+\,2\pi \,{r}_B\, \alpha - \text{i} r \,\omega_+ - \text{i} n_1 \omega_1 - \text{i} n_2 \omega_2}\,,
\tag{D.1}
\end{equation}
$r_{B}$ is the R-charge $Q$ of the boson in the corresponding multiplet. $r_F$ is the R-charge $Q$ of the fermion in the corresponding multiplet.~\footnote{In the conventions of~\cite{Cabo-Bizet:2018ehj}, for chiral multiplets. $r_F=r_B-1$ or equivalently~$r_F=r-1\,   $. This, however, does not need to be the case, as we will illustrate below. } The numerator in~\eqref{eq:Z1Loop} is regularized as
\begin{equation}
\biggl(s \frac{1}{(p q)^{( 1+\Delta r)/2}}\bigl({\chi}\bigr)^{\frac{r_F}{2}};\frac{1}{p},\frac{1}{q}\biggr)\,=\,\biggl(s (p q)^{(1-\Delta r)/2}\bigl({\chi}\bigr)^{\frac{r_F}{2}};{p},{q}\biggr)\,,
\end{equation}
where
\begin{equation}
 \Delta r:= r_B-r_F\,.
\end{equation}
The denominator in~\eqref{eq:Z1Loop} is regularized as
\begin{equation}
\biggl(s\chi^{\frac{r_B}{2}};{p},{q}\biggr)\,
\end{equation}
The complete interval BPS path integral for generic four-dimensional $\mathcal{N}=1$ superconformal gauge theories, can be packaged as follows:
\begin{equation}\label{eq:ZBPS}
\begin{split}
^{(k=0)}Z_{\text{BPS}}[\omega,\varphi]
&\,=\,\frac{\mathcal{N}_0}{|\mathcal{W}|} 
\int \prod_{i=1}^{\mathrm{rk}(G)} \frac{\mathrm{d}u_i}{2\pi}\, \prod_{\rho \in \mathrm{Adj}(G)\atop\rho\,\neq\,0} \Delta(\rho(u))\,\cdot\,\prod_{\rho \in \mathrm{Adj}(G)\atop\rho\,\neq\,0}\mathcal{G}_{V}\!\left(\rho(u); \frac{\omega_1}{2\pi i}, \frac{\omega_2}{2\pi i}\right) \\ &
\qquad\qquad \qquad\qquad \times { \prod_I \prod_{\rho \in Rep_I \atop\rho\,\neq\,0} 
\mathcal{G}_{\Delta r_I}\!\left(v(\rho, r_I); \frac{\omega_1}{2\pi i}, \frac{\omega_2}{2\pi i}\right)}{}\,.
\end{split}
\end{equation}
The~$|\mathcal{W}|$ is the dimension of the Weyl group. The~$\rho$'s are weight elements of a representation of the gauge group $G\,$. The label $I$ counts chiral multiplets in representations $Rep_I$ of $G\,$.

An important component in the definition~\eqref{eq:ZBPS} is the contribution from zero gauge-charge modes
\begin{equation}
\mathcal{N}_0\,:=\,
\biggl(\frac{\mathcal{G}_{V}\!\left(0; \frac{\omega_1}{2\pi i}, \frac{\omega_2}{2\pi i}\right)}{(1-\chi^{1/2})}\biggr)^{\text{rk}(G)}\times \text{(chiral multiplet $\rho=0$ contributions)}\,.
\end{equation}
For $\rho\neq0$
\begin{equation}
\Delta(\rho(u))\,=\, \frac{(1-z)}{(1\,-\,{z\,{\chi^{1/2}})}}
\end{equation}
is a contribution that reduces to $1$ at the locus
\[
\chi^{1/2}\,=\, 1\,.
\]
The contributions from vector and chiral multiplets are as follows:
\begin{equation}\label{eq:Gdefs}
\begin{split}
\mathcal{G}_{V}\biggl(\rho(u);\frac{\omega_1}{2\pi \text{i}}, \frac{\omega_2}{2\pi \text{i}}\biggr)&\,:=\,\frac{\biggl({z}\,{\chi^{1/2}}\,;p,q\biggr)_\infty}{\biggl({z}p q\,;\, p,q\biggr)_\infty}\,=\,\exp\biggl(\sum_{n=1}^{\infty} {\frac{(z p q)^n  \,-\,{z^n}  \, {\chi^{n/2}}}{n(1-p^{n})(1-q^{n})}}\biggr)\,,
\\
\mathcal{G}_{r}\biggl(v;\frac{\omega_1}{2\pi \text{i}}, \frac{\omega_2}{2\pi \text{i}}\biggr)&\,:=\,\frac{\biggl( {s}\,(p q)^{(1-\Delta r)/2}\,{\chi^{r_F/2}}\,;p,q\biggr)_\infty}{\biggl({s}\,{\chi^{r_B/2}}\,;\, p,q\biggr)_\infty}\\&\,=\,\exp\biggl(\sum_{n=1}^{\infty} {\frac{s^n \, {\chi^{n r_B/2}} \,-\,{s^n} \,(p q)^{n(1-\Delta r)/2} \, {\chi^{n r_F/2}}}{n\,(1-p^{n})(1-q^{n})}}\biggr)\,.
\end{split}
\end{equation}
We define rapidities as
\[\,s\,:=\,e^{2\pi \text{i}v}\,, t=e^{\varphi}= (\underbrace{e^{-2\pi \text{i} k}}_{\underset{k=0}{=}1}\,\chi p q)^{\frac{1}{2}}\,,\,p=e^{\omega_1}\,, \, q=e^{\omega_2}\,,\, n_j=e^{\nu_j}\,,\]\footnote{This factor of $e^{-2\pi \text{i} k}$ is the inverse of the one in~\eqref{eq:ChoicesOfFlavorRapidities}; however, at this point we are implicitly assuming the trivial branch $k=0\,$, so we can substitute it by $1\,.$}
and
\[
v={v}(\rho,r) := \frac{1}{2\pi} \left( \rho \cdot u - \,\text{i} r\,\frac{\omega_1+\omega_2}{2} -\text{i}\nu_j y_j  \right).
\]
In contrast to the analysis in~\cite{Cabo-Bizet:2018ehj}, in this section the variable $\varphi$ is not constrained
\begin{equation}
2\varphi= \omega_1+\omega_2+4\pi \text{i}\,\alpha\,.
\end{equation}
The label~$y_{j}$ is the charge of the corresponding multiplet with respect to the $j$-th flavor Cartan charge $Y_j\,$.

The interval path integral $Z_{\text{BPS}}$ at a generic branch $k$ is obtained from the right-hand side of~\eqref{eq:ZBPS} by introducing the redefinition (when $Z_{\text{BPS}}$ is written as a function of $p$, $q$ and $t$)
\[p^{x}\to e^{2\pi \text{i} k x} p^x\,, \qquad t=\text{fixed}\, \]
for any arbitrary power $x\,$. We will come back to elaborate on this below.

For the choice of flavor rapidities~\eqref{eq:ChoicesOfFlavorRapidities}, then for the choice of BPS locus (at generic branch $k=1\bmod 2$)
\[
\chi^{1/2}\to 1\,,
\]
and assuming real representations, namely representations with symmetry $\rho\to-\rho\,$, then
\[{Z_{\text{BPS}}[\omega,\varphi]}\,\to\, \mathcal{I}(\omega_1,\omega_2,\varphi=\varphi_0)\,\] where $\mathcal{I}(\omega_1,\omega_2,\varphi)$ is the superconformal index, for example, as reported in equation (4.57) of~\cite{Cabo-Bizet:2018ehj} (in that case $\Delta r=1$).

The canonical example of an $\mathcal{N}=1$ superconformal gauge theory is $U(N)$ $\mathcal{N}=4$ SYM, that is, a vector multiplet and three matter multiplets $I=1,2,3$ in the adjoint of $U(N)\,$. In that case, \[\rho_{i,j}(u)=u_{i,j}=u_i-u_j\] and \[\prod_{\rho}=\prod_{i,j=1}^N\,.\] We will come back to this example below.

At fixed BPS chemical potentials~\eqref{eq:BPSChemcialPotentialsInterval}, the BPS partition function~\eqref{eq:ZBPS} on the interval $\mathbb{R}_L\,$ is independent of $L\,$. As at $\beta= +\infty$, it is bound to equal the $\beta=+\infty$ value of the (path integral representation of the) thermal partition function then
\begin{equation}\label{eq:BPSPartitionFunction}
Z^{(L)}_{\text{BPS}}[\omega,\varphi]\,=\, Z[\beta=\infty,\omega,\varphi,\nu]\,,
\end{equation}
for all $L$ and for a particular choice of $\nu\,$. 

\eqref{eq:BPSPartitionFunction} means that the supersymmetric path integral~$Z^{(L)}_{\text{BPS}}$ has a trace representation as well:
\begin{equation}\label{eq:CountingInterpretationZBPS}
Z_{\text{BPS}}[\omega,\varphi]\,=\, \text{Tr}_{\text{BPS}} \,q^{2J+Q} e^{2\pi \text{i}\alpha Q}\,\,e^{\underline{\nu}\,\cdot\,\underline{Y}}\,.
\end{equation}
In this expression, Tr$_{\text{BPS}}$ means a trace over states satisfying
\begin{equation}\label{eq:BPSEq1}
E-2J-\tfrac{3}{2}Q\,=\,0\,.
\end{equation}
We will explicitly verify this claim for $\mathcal{N}=4$ SYM next (at zero coupling).

\subsection{The Hamiltonian perspective on $Z_{\text{BPS}}$}\label{sec:BPSPartHam}

As mentioned above, the path integral $Z^{(L)}_{\text{BPS}}$ has a counting interpretation~\eqref{eq:CountingInterpretationZBPS}. The single-letter operators of free $U(N)$ $\mathcal{N}=4$ SYM satisfying the BPS condition
\begin{equation}
E-J_1-J_2-Q_1-Q_2-Q_3\,=\,E-2J-3/2 Q\,=\,0\,,
\end{equation}
are summarized in table~\eqref{tab:Table1} of~\cite{Kinney:2005ej}.
\begin{table}[h]\label{tab:Table1}
\centering
\caption{Letters with $E-2J-3/2 Q=0$ (upper/lower signs are correlated with upper/lower signs.)}
\setlength{\tabcolsep}{8pt}
\renewcommand{\arraystretch}{1.2}
\begin{tabular}{lccc}
\toprule
\textbf{Letter} & $(-1)^F\,[J_1,J_2]$ & $[Q_1,Q_2,Q_3]$ & \\
\midrule
$X,Y,Z$ &
$[0,0]$ &
$[1,0,0]+\text{cyclic}$ &
\\
$\psi_{+,0;\,-+++}+\text{cyc}$ &
$-\big[\tfrac{1}{2},\tfrac{1}{2}\big]$ &
$\big[-\tfrac{1}{2},\tfrac{1}{2},\tfrac{1}{2}\big]+\text{cyc}$ &
\\
$\psi_{0,\pm,\ +++}$ &
$-\big[\pm\tfrac{1}{2},\mp\tfrac{1}{2}\big]$ &
$\big[\tfrac{1}{2},\tfrac{1}{2},\tfrac{1}{2}\big]$ &
\\
$F_{++}$ &
$[1,1]$ &
$[0,0,0]$ &
\\
$\partial_{++}\psi_{0,-;\ +++}+\partial_{+-}\psi_{0,+;\ +++}=0$ &
$\big[\tfrac{1}{2},\tfrac{1}{2}\big]$ &
$\big[\tfrac{1}{2},\tfrac{1}{2},\tfrac{1}{2}\big]$ &
\\
$\partial_{\pm\pm}$ &
$\big[\tfrac{1\pm 1}{2},\tfrac{1\mp 1}{2}\big]$ &
$[0,0,0]$ &
\\
\bottomrule
\end{tabular}
\end{table}
The translation of charges to our conventions is as follows:
\begin{equation}
\begin{split}
Q_{1,2,3,\text{here}}&\,=\,q_{1,2,3,\text{there}}\,, \\
J_{1,\text{here}}&\,=\, \frac{j_{1,\text{there}}+j_{2,\text{there}}}{2}\,,\\
J_{2,\text{here}}&\,=\,\frac{j_{1,\text{there}}-j_{2,\text{there}}}{2}
\,.
\end{split}
\end{equation}
In the decompactification limit $\beta\to\infty$ the fully refined partition function
\[Z[\beta,\omega,\varphi_1,\varphi_2,\varphi_3]\,=\,\text{Tr}_{\mathcal{H}} x^{E-J_1-J_2-Q_1-Q_2-Q_3} p^{J_1}q^{J_2} t_1^{Q_1} t^{Q_2}_2 t^{Q_3}_3\]
equals
\[Z[\beta=+\infty,\omega,\varphi_1,\varphi_2,\varphi_3]\,=\,\text{Tr}_{\text{BPS}}\, p^{J_1}q^{J_2} t_1^{Q_1} t^{Q_2}_2 t^{Q_3}_3\,.\]
We focus on the minimally refined case
\[t_a= t^{2/3}= e^{\frac{2 \varphi}{3}}\]
denoted as
\begin{equation}
Z_{\text{BPS}}[\omega,\varphi]\,=\, \text{Tr}_{\text{BPS}}\,\,  p^{J_1+Q/2}q^{J_2+Q/2}  e^{2\pi\text{i}\alpha Q}\,,
\end{equation}
where
\[\frac{3}{2}Q\,:=\,Q_1+Q_2+Q_3\,.\]
The bosonic single-letter partition function $f_B$, computed by summing
\begin{equation}
p^{J_1+Q/2}q^{J_2+Q/2}  e^{2\pi\text{i}\alpha Q}
\end{equation}
over bosonic single letters is
\begin{equation}
\begin{split}
f_B[p,q, \chi]&\,=\,\frac{3 \,{(p q)^{1/3} (\chi^2)^{1/6}
   }}{(1-p) (1-q)}+\frac{p q}{(1-p) (1-q)}\,.
\end{split}
\end{equation}
In $\mathcal{N}=1$ language, the first contribution comes from three scalar letters $X,\, Y$ and $Z$, which can be understood as scalars in a four-dimensional $\mathcal{N}=1$ chiral multiplet. The second contribution comes from the vector modes.

The fermionic single-letter partition function is
\begin{equation}
\begin{split}
f_F[p,q, \chi]&\,=\,\frac{3 (p q
   )^{2/3} \chi^{1/6}}{(1-p) (1-q)}\,+\,\frac{p \sqrt{\chi }}{(1-p) (1-q)}+\frac{q \sqrt{\chi
   }}{(1-p) (1-q)}-\frac{p q \sqrt{\chi }}{(1-p) (1-q)}\,.
\end{split}
\end{equation}
The first contribution comes from three fermionic letters in an $\mathcal{N}=1$ chiral multiplet with R-charge $Q=\tfrac{2}{3}\sum_a Q_a$
\begin{equation}
r_{B}=r=\frac{2}{3}\,\,, \qquad r_F=\frac{1}{3}\,, \qquad \Delta r=\frac{1}{3}\,.
\end{equation}
The second and third contributions come from gaugino components (third row in table~\ref{tab:Table1}). The fourth contribution comes from the constraint imposed by the EoMs of the gaugino components.

It is convenient to note the following identity:
\begin{equation}\label{eq:Identity}
\frac{p \sqrt{\chi
   }}{(1-p) (1-q)}+\frac{q \sqrt{\chi }}{(1-p) (1-q)}\,-\frac{p q \sqrt{\chi }}{(1-p) (1-q)}\,=\,\frac{\sqrt{\chi }}{(p-1) (q-1)}-\sqrt{\chi }\,,
\end{equation}
which combines the second, third, and fourth contributions in $f_F\,$.

Given a single gauge mode $z=e^{\rho(u)}\,$, the quantity
\begin{equation}\label{eq:PFInitial}
\exp{\biggl(\biggl(\sum_{n=1}^{\infty}\frac{1}{n}(-1\,+\,f_B[p^n,q^n,\chi^n]+(-1)^{n+1}f_F[p^n,q^n,\chi^n]\biggr) z^n\biggr)}
\end{equation}
corresponds to computing a path integral with periodic boundary conditions for bosons and antiperiodic for fermions (before the decompactification limit $\beta\to\infty$). The $-1$ accounts for the Vandermonde contribution.

Using~\eqref{eq:Identity} we can recast~\eqref{eq:PFInitial} as
\begin{equation}\label{eq:BPSZCounting}
\exp{\biggl(\biggl(\sum_{n=1}^{\infty}\frac{1}{n}(-1\,+\,\widetilde{f}_B[p^n,q^n,\chi^n]+(-1)^{n+1}\widetilde{f}_F[p^n,q^n,\chi^n]\biggr) z^n\biggr)}
\end{equation}
where
\begin{equation}\label{eq:Feffect}
\begin{split}
\widetilde{f}_B[p,q,\chi]&\,:=\,\frac{3 \,{(p q \chi)^{1/3}
   }}{(1-p) (1-q)}+\frac{p q}{(1-p) (1-q)}\\ &=\frac{3 \,{t^{2/3}
   }}{(1-p) (1-q)}+\frac{p q}{(1-p) (1-q)}\,,\\
   \widetilde{f}_F[p,q,\chi]&\,:=\, \frac{3 (p q \chi
   )^{2/3} \chi^{-1/2}}{(1-p) (1-q)}\,+\,\frac{{\chi }^{1/2}}{(1-p) (1-q)}-{\chi }^{1/2}\,\\&=\,\frac{3\, t
   ^{4/3} \chi^{-1/2}}{(1-p) (1-q)}\,+\,\frac{{\chi }^{1/2}}{(1-p) (1-q)}-{\chi }^{1/2}.
   \end{split}
\end{equation}
The \[(-1)^{n}\] in~\eqref{eq:BPSZCounting}, which assigns fermionic states grading $+1\,$ instead of $-1\,$, can be exchanged by the simultaneous redefinitions (demanding $k=1\bmod 2$)
\[p^{x}\to e^{2\pi \text{i} k x} p^x\, , \quad \chi^x \to e^{-2\pi \text{i} k x} \chi^x\,, \]
or equivalently
    \[p^{x}\to e^{2\pi \text{i} k x} p^x\,, \qquad t=\text{fixed}\,. \]
This redefinition shows how the BPS partition function on the interval can be equated to (given an ambiguous choice of branch $k=1\bmod 2$)
\begin{equation}\label{eq:ZInterm}
\exp{\biggl(\sum_{n=1}^{\infty}\biggl(\frac{1}{n}(-1\,+\,\widetilde{f}_B[e^{-2\pi \text{i} k n}p^n,q^n, e^{2\pi \text{i} k n}\chi^n]-\widetilde{f}_F[e^{-2\pi \text{i} k n}p^n,q^n,e^{2\pi \text{i} k n}\chi^n]\biggr) z^n\biggr)}\,.
\end{equation}
More generally, for a generic choice of branch $k$,~\eqref{eq:ZInterm} matches the answer obtained from the Lagrangian perspective for the interval path integral~\eqref{eq:ZBPS} at generic $k$ in~\eqref{eq:ConstraintFieldTheory} (after some algebra that involves assembling contributions from all gauge charge vectors $\rho$)
\begin{equation}
\begin{split}
Z_{\text{BPS}}[\omega,\varphi]
&\,=\,\frac{(Z_0)^N}{N!} 
\int \prod_{i=1}^{N} \frac{\mathrm{d}u_i}{2\pi}\, \prod_{\rho \in \mathrm{Adj}(U(N))\atop \rho\,\neq\,0} \Delta(\rho(u))\,\mathcal{G}_{V}\!\left(\rho(u); \frac{\omega_1}{2\pi i}, \frac{\omega_2}{2\pi i}\right) \\ &
\qquad\qquad \qquad\qquad\qquad \qquad \qquad \qquad  \times \biggl({  
\mathcal{G}_{\frac{1}{3}}\!\left(v(\rho, 2/3); \frac{\omega_1}{2\pi i}, \frac{\omega_2}{2\pi i}\right)}\biggr)^3\,,
\end{split}
\end{equation}
where as defined in~\eqref{eq:Gdefs}
\begin{equation}
\begin{split}
\mathcal{G}_{1/3}\biggl(v;\frac{\omega_1}{2\pi \text{i}}, \frac{\omega_2}{2\pi \text{i}}\biggr)&\,:=\,\frac{\biggl( {s}\,(p q)^{1/3}\,{\chi^{1/6}}\,;p,q\biggr)_\infty}{\biggl({s}\,{\chi^{1/3}}\,;\, p,q\biggr)_\infty}\\&\,=\,\exp\biggl(\sum_{n=1}^{\infty} {\frac{s^n \, {\chi^{n/3}} \,-\,{s^n} \,(p q)^{n/3} \, {\chi^{n/6}}}{n\,(1-p^{n})(1-q^{n})}}\biggr)\,.
\end{split}
\end{equation}
An analogous computation, with the very same letters quoted in table~\ref{tab:Table1}, and again using~\eqref{eq:Identity} and the symmetry $\rho\to -\rho\,$, gives us the fully refined BPS partition function
\begin{equation}\label{eq:ZBPSNonPer}
\begin{split}
Z_{\text{BPS}}[\omega,\varphi_{1},\varphi_{2},\varphi_{3}]&\,=\,\frac{(Z_0)^N}{N!} 
\int \prod_{i=1}^{N} \frac{\mathrm{d}u_i}{2\pi}\, \prod_{\rho \in \mathrm{Adj}(U(N))\atop \rho\,\neq\,0} \Delta(\rho(u))\,\mathcal{G}_{0}\!\left(\rho(u); \frac{\omega_1}{2\pi i}, \frac{\omega_2}{2\pi i}\right) \\ &
\qquad\qquad\qquad \qquad   \times \prod_{I=1}^3\biggl({ 
\mathcal{G}_{I}\!\left(\rho(u); \frac{\omega_1}{2\pi i}, \frac{\omega_2}{2\pi i}\right)}\biggr)\,,
\end{split}
\end{equation}
where ($z=e^{2\pi \text{i}\rho(u)}$)
\begin{equation}
\begin{split}
\mathcal{G}_{0}\biggl(\rho(u);\frac{\omega_1}{2\pi \text{i}}, \frac{\omega_2}{2\pi \text{i}}\biggr)&\,:=\,\frac{({z}\,\chi^{1/2};\,p,q)_\infty}{({ \frac{p q}{z}} \, ;\,p,q)_\infty}\,,
\\
\mathcal{G}_{I}\!\left(\rho(u); \frac{\omega_1}{2\pi i}, \frac{\omega_2}{2\pi i}\right)&\,:=\,\frac{(\frac{p q}{z \,t_I}\,\chi^{1/2};\,p,q)_\infty}{({ z\, t_I} \, ;\,p,q)_\infty}\,,
\end{split}
\end{equation}
and
\begin{equation}\label{eq:FinalDefChi}
\chi\,:=\,e^{2\pi \text{i} k}\,\frac{t_1 t_2 t_3}{p q}\,=\,e^{2\pi \text{i} k}\,\frac{t^2}{p q}\,.
\end{equation}
For the choice of branch $k=1\mod 2$ we obtain
\[\chi^{1/2}\,=\, - \frac{t}{\sqrt{p q}}\]
~\footnote{and for the choice of distribution rules $(x y)^z =x^z y^z\,$, and $(e^{x})^{y}\,:=\, e^{ x y}\,$ (choice of branch cuts) that we are assuming in this paper.} and then~\eqref{eq:ZBPSNonPer} is the BPS partition function whose Taylor coefficients in the expansion around $(p^{\frac{1}{2}},q^{\frac{1}{2}},t)=(0,0,0)$ are all positive integers.~\footnote{We reiterate, one can move among branches by the shifts $p\to e^{-2\pi \text{i}\widetilde{k}}$, $\widetilde{k}\in\mathbb{Z}$ at fixed $q\,$, $t_{I}\,$.}

Concretely, at $k=1\bmod 2$ the interval path integral $Z_{\text{BPS}}\,$, thought of as a function of the rapidities $p$, $q$, and $t_I$, is the generating function counting the BPS states without grading $(-1)^F$. The BPS locus is then necessarily located at $\alpha= \frac{1}{2}\mod 1$ if $k=1\mod 2\,$
\[ \chi^{1/2}\,=\,e^{\pi \text{i}k}e^{2\pi \text{i}\alpha} \,=1. \]
For these values of $(\alpha,k)\,$, $Z_{\text{BPS}}$ reduces to the known expression of the fully refined superconformal index of $U(N)$ $\mathcal{N}=4$ SYM on $S^3\,$ in terms of elliptic Gamma functions (up to potential shifts in angular velocities induced by the constant $\kappa_a$ defined in equation~\ref{eq:KappaShifts}).

\subsection{Index vs BPS partition function: microcanonical ensemble}
Let us assume $p=q$ ($J_a=J$), with a single $R$-symmetry chemical potential $\varphi$ turned on, for simplicity of presentation but without loss of generality.

The superconformal index in canonical ensemble~\footnote{For simplicity of presentation, we assume in the following discussion that $\kappa_1=\kappa_2=0$. This need not be the case in general. However, since we are interested in Laplace transforming the $\omega$'s, the dependence on $\kappa_1$ and $\kappa_2$ is spurious (as such a Laplace transform involves an averaging over $\omega$'s along their periods). Thus, in order to keep the presentation as simple as possible, we will assume $\kappa_1=\kappa_2=0$ here. The discussion for the general case $\kappa_1=\kappa_2\neq 0$ is analogous, though. We note that this latter is the case of $\mathcal{N}=4$ SYM where the eigenvalues of $Q$ are discretized in units of $\frac{1}{3}\,$. }
\begin{equation}
\begin{split}
\mathcal{I}[\omega]\,\underset{\eqref{eq:ConstraintSect2}}{=}\,Z[\beta,\omega,\varphi]&\,\,\,\,=\,\,\,\sum_{J,Q} d[J,Q] \,e^{2\omega J}e^{\varphi Q} \\
&\,\underset{\eqref{eq:ConstraintSect2}}{=}\,\sum_{y\,=\,2J +Q} \widetilde{d}[\,\mathfrak{j}\,]\, e^{\omega \,\mathfrak{j}\,}
\end{split}
\end{equation}
counts states with a certain grading. On the other hand, $d[J,Q]$ is a positive integer that counts physical states in the BPS sector of the CFT with charges $J$ and $Q$ at the locus~\eqref{eq:BPSEq1}. Instead, the index counts an average
\begin{equation}\label{eq:IndexAveragePhysStates}
\widetilde{d}[\,\mathfrak{j}\,]\,:=\, \sum_{Q\,:\, 2J= n-Q } d[J,Q]\,e^{\pm \pi \text{i} Q}\,,
\end{equation}
which can be a positive or a negative integer, depending on whether there are more bosonic or fermionic states at the level $\,\mathfrak{j}\,:=2J+Q\,$. The $(-1)^F$ grading comes from the $e^{\pm \pi \text{i} Q}$ grading.

To understand how these two observables compare to each other, we study the leading order in the natural extension of the semiclassical expansion~\eqref{eq:LargeGNExp}
\begin{equation}\label{eq:SemiclassicalExp2}
G_{N}\,\to\, 0\,, \quad G_N J_{a}=j_{a,0}=\text{fixed}\,\neq 0, \quad G_N Q=q_0=\text{fixed}\,\neq 0
\end{equation}
of the integral definitions of the total number of BPS states $d[J,Q]\,$,
\begin{equation}\label{eq:DTotal}
d[J,Q]\,=\,\oint_{|q|=1} \frac{dq}{2\pi \text{i}q}\oint^{(3)}_{|t|=1} \frac{dt}{6\pi \text{i} t} \,Z[\beta=+\infty,\omega,\varphi]\, q^{-2 J} t^{-Q}\,,
\end{equation}
and the total number of operators counted with a $(-1)^F$ grading $\widetilde{d}\,$,
\begin{equation}\label{eq:dn}
\begin{split}
\widetilde{d}[\,\mathfrak{j}\,]&\,=\,(-1)^Q\oint^{(3)}_{|q|=1} \frac{dq}{6\pi \text{i}q}\oint_{t \,\circlearrowleft\, -q} \frac{dt}{2\pi \text{i} t(1+\tfrac{q}{t})} \,Z[\beta{=+\infty},\omega,\varphi]\, q^{-2 J} t^{-Q}\,,
\\
&=(-1)^Q\oint^{(3)}_{|q|=1} \frac{dq}{6\pi \text{i}q} \,Z[\beta{=+\infty},\omega,\varphi]\, q^{-2 J} t^{-Q}\,\biggl|_{t\,\to\,-q}
\\
&=(-1)^Q\oint^{(3)}_{|q|=1} \frac{dq}{6\pi \text{i}q} \,\mathcal{I}[\omega]\, q^{-(2 J+Q)} (-1)^{-Q}\,\\
&=\oint^{(3)}_{|q|=1} \frac{dq}{6\pi \text{i}q} \,\mathcal{I}[\omega]\, q^{-(2 J+Q)}\,.
\end{split}
\end{equation}
\footnote{Recall that $Z[\beta=+\infty,\omega,\varphi]$ equals $\mathcal{I}[\omega]$ at $t=-q$\,. Thus, it does not have poles at $t=-q\,$.}
The supraindex $(3)$ denotes an integral over a triple cover of the corresponding unit circle $|\xi|=1$, with $\xi=\{q,t\}\,$. This is necessary to integrate to zero the non-integer powers of $\xi\,$. If charges dual to $\xi$ are multiples of $\frac{1}{3}$, an integral over the triple cover of the unit circle $|\xi|=1$ would annihilate them unless the power is $\xi^{-1}$, in which case it contributes. If the eigenvalues of $Q$ were integers, as are those of $2J$, then only integrals over a single cover of the unit circle $|\xi|=1$ would be necessary to project to the microcanonical index.

The prefactor $(-1)^Q$ in the first line is added to cancel the opposite contribution $(-1)^{-Q}$ that comes from the evaluation of the residue in the second line. The latter term is to be canceled because the definition of the microcanonical index is as follows:
\begin{eqnarray}\label{eq:MicrocanonicalIndexDef}
\widetilde{d}[\,\mathfrak{j}\,]\,:=\,\oint^{(3)}_{|q|=1} \frac{dq}{6\pi \text{i}q}\oint_{t \,\circlearrowleft\, -q} \frac{dt}{2\pi \text{i} t(1+\tfrac{q}{t})} \,Z[\beta{=+\infty},\omega,\varphi]\, q^{-(2 J+Q)}\,.
\end{eqnarray}

The integral representation of the index $\widetilde{d}[\,\mathfrak{j}\,]$ in the first line of~\eqref{eq:dn} shares the same large-$N$ eigenvalues as the integral representation of the microcanonical partition function $d[J,Q]\,$. The only differences come from the extra factor
\begin{equation}
\frac{1}{1+\tfrac{q}{t}}
\end{equation}
and the contour of integration over the variable $q\,$. The latter difference is radical because it makes the two integration contours homologically inequivalent~\cite{Witten:2010cx}. One of them localizes the integral over $q$ to the specific value $q=-t$, while the other one does not. That said, in the semiclassical large-$N$ expansion~\eqref{eq:SemiclassicalExp2}, something interesting will happen.

Computations will be performed for $U(N)$ $\mathcal{N}=4$ SYM from now on, but we expect our conclusions to generalize beyond that example.

\section{The BPS partition function localizes to ensembles of indices}\label{sec:LocPartitionF}

The leading singularities of the integrand of~$Z_{\text{BPS}}$ determine the saddle points that contribute to the semiclassical expansion~\eqref{eq:SemiclassicalExp2} of $d[J,Q]$ that we are after~\cite{Beccaria:2023hip}.

We start by exploring the space of essential singularities of the integrand of $Z_{\text{BPS}}$
\begin{equation}
\frac{\mathcal{N}_0}{N!}\prod_{i\,<\, j=1}^{N} I[U_{ij}]I[{U_{ji}}]\,=\,\frac{(Z_0)^N}{N!}\,e^{\sum_{i,j=1 \atop i\,\neq\, j}^{N}V_{\text{eff}}[U_{ij}]} 
\end{equation}
where
\begin{equation}\label{eq:N4SYM}
\begin{split}
Z_0&\,:=\,\frac{1}{(1-{{\chi^{1/2}})}}\, \frac{(\chi^{1/2};\,p,q)_\infty}{({p q} \, ;\,p,q)_\infty}\,\prod_{I=1}^{3}\frac{( \tfrac{p q}{t_I }\,\chi^{1/2};\,p,q)_\infty}{({t_I} \, ;\,p,q)_\infty}\,,\\
I[U]&\,:= \, \frac{(1-U)}{(1\,-\,{{U}{\chi^{1/2}})}}\, \frac{({U}\,\chi^{1/2};\,p,q)_\infty}{({U}{p q} \, ;\,p,q)_\infty}\,\prod_{I=1}^{3}\frac{( U\tfrac{p q}{t_I }\,\chi^{1/2};\,p,q)_\infty}{({ U\, t_I} \, ;\,p,q)_\infty}\,.
\end{split}
\end{equation}
For technical reasons, it is convenient to define the effective potential as a function invariant under inversion of the variables $U$
\begin{equation}
e^{2V_{\text{eff}}[U]}\,:=\,I[U] \,I[\tfrac{1}{U}]\,.
\end{equation}
Let us start with the essential singularities
\begin{equation}\label{eq:LimitTOIntegers}
\delta\omega_{a}\,=\,\,\omega_a-2\pi \text{i}n_a\,=\, 0\,,\qquad n_a\in\mathbb{Z}\,.
\end{equation}
of the effective potential
\begin{equation}
\begin{split}
V_{\text{eff}}[U]&\,=\, \frac{\sum_{n,m=0} V^{(m,n)}_{\text{eff}}[{U;\chi}]\, \delta\omega^{n}_1 \,\delta\omega^m_2}{\delta \omega_1 \delta \omega_2} \,+\, (\text{non-pert. suppressed as $\delta\omega_a\to0$})\,.
\end{split}
\end{equation}
We derive these expansions using the definition
\begin{equation}
(z;p,q)_\infty\,:=\,\exp\biggl(-\sum_{n=1}^{\infty}\frac{1}{n}\,\frac{z^n}{(1-p^n)(1-q^n)}\biggr)\,.
\end{equation}
For technical advantage, in intermediate computations we introduce a cutoff $\Lambda$ in the sum
\begin{equation}
\sum_{n=1}^{\infty} \,\to\, \sum_{n=1}^{\Lambda}\,. 
\end{equation}
\begin{table}[h]\label{tab:Table2}
\centering
\caption{The truncated perturbative effective potential for $t_a=\nu_a e^{4\pi \text{i}\alpha/3}\,p^{1/3}q^{1/3}$, $\prod_{a=1}^3\nu_a=1\,$. We define truncated polylogs as $\text{Li}_{s\Lambda}(z):=\sum_{n=1}^{\Lambda}\tfrac{z^n}{n^s}$. Examples of coefficients. }
\setlength{\tabcolsep}{8pt}
\renewcommand{\arraystretch}{1.2}
\begin{tabular}{lccc}
\toprule
\textbf{(m,n)} & $V^{(m,n)}_{\text{eff},\Lambda}[U,\chi]$ &  \\
\midrule
$(0,0)$ &\begin{tabular}{c}
$\displaystyle \sum_{i,j=1\atop i\,\neq\, j}^{N}\biggl(-\sum_{a=1}^3 \text{Li}_{3\Lambda}\left(X_a
   U_{{ij}}\right)+\sum_{a=1}^3 \text{Li}_{3\Lambda}\left(Y_a U_{{ij}}\right)-\ldots$ \\[2pt]
$\displaystyle \qquad \qquad\qquad\qquad-\,\text{Li}_{3\Lambda}\left(Z U_{{ij}}\right)+\text{Li}_{3\Lambda}\left(U_{{ij}}\right)\Bigr)$
\end{tabular}
&
\\
$(1,0)=(0,1)$ &\begin{tabular}{c}
$\displaystyle \sum_{i,j=1\atop i\,\neq\, j}^{N}\tfrac{1}{2}\biggl(-\sum_{a=1}^3 \text{Li}_{2\Lambda}\left(X_a
   U_{{ij}}\right)- \sum_{a=1}^{3}\text{Li}_{2\Lambda}\left(Y_a U_{{ij}}\right)+\ldots$ \\[2pt]
$\displaystyle \qquad \qquad\qquad\qquad\,+\text{Li}_{2\Lambda}\left(Z   U_{{ij}}\right)+\text{Li}_{2\Lambda}\left(U_{{ij}}\right)\Bigr)$
\end{tabular}
   &
\\
$(1,1)$ &\begin{tabular}{c}
$\displaystyle \sum_{i,j=1\atop i\,\neq\, j}^{N}\tfrac{1}{4}\biggl(-\sum_{a=1}^3\text{Li}_{1\Lambda}\left(X_a
   U_{{ij}}\right)+ \sum_{a=1}^3\text{Li}_{1\Lambda}\left(Y_a U_{{ij}}\right)+\ldots$ \\[2pt]
$\displaystyle \qquad \qquad\qquad\qquad\,+3\text{Li}_{1\Lambda}\left(Z
   U_{{ij}}\right)-3\text{Li}_{1\Lambda}\left(U_{{ij}}\right)\Bigr)$
\end{tabular}
   &
\\
$(2,1)=(1,2)$ &\begin{tabular}{c}
$\displaystyle \sum_{i,j=1\atop i\,\neq\, j}^{N}\tfrac{1}{24}\biggl(- \sum_{a=1}^3\text{Li}_{0\Lambda}\left(X_a
   U_{{ij}}\right)-\sum_{a=1}^3 \text{Li}_{0\Lambda}\left(Y_a U_{{ij}}\right)+\ldots$ \\[2pt]
$\displaystyle \qquad \qquad\qquad\qquad\,+\,\text{Li}_{0\Lambda}\left(Z
   U_{{ij}}\right)\,+\,\text{Li}_{0\Lambda}\left(U_{{ij}}\right)\Bigr)$
\end{tabular}
   &
\\
$(2,2)$ &\begin{tabular}{c}
$\displaystyle \sum_{i,j=1\atop i\,\neq\, j}^{N}\tfrac{1}{144}\biggl(- \sum_{a=1}^3\text{Li}_{-1\Lambda}\left(X_a
   U_{{ij}}\right)+\sum_{a=1}^3 \text{Li}_{-1\Lambda}\left(Y_a U_{{ij}}\right)-\ldots$ \\[2pt]
$\displaystyle \qquad \qquad\qquad\qquad\,-\,\text{Li}_{-1\Lambda}\left(Z
   U_{{ij}}\right)+\text{Li}_{-1\Lambda}\left(U_{{ij}}\right)\Bigr)$
\end{tabular}
   &
\\
$(X_a,Y_a,Z)$ &
$(\tfrac{1}{\nu_a} e^{\frac{2  \pi \text{i} (\alpha+k/2) }{3}},{\nu_a} e^{\frac{4 \pi \text{i}
   (\alpha+k/2) }{3}},e^{2 
   \pi \text{i} (\alpha+k/2) })$ &
\\
\bottomrule
\end{tabular}
\end{table}

\begin{table}[h]\label{tab:Table3}
\centering
\caption{The regularized perturbative effective potential in the improved Cardy-like expansion proposed here. This table reports only the particular case $t_1=t_2=t_3=t^{2/3}=e^{4\pi \text{i}\alpha/3}p^{1/3}q^{1/3}$. We define truncated polylogs as $\text{Li}_{s\Lambda}[z]:=\sum_{n=1}^{\Lambda}\tfrac{z^n}{n^s}\,$. $N^2\gg1$.}
\setlength{\tabcolsep}{8pt}
\renewcommand{\arraystretch}{1.2}
\begin{tabular}{lccc}
\toprule
\textbf{(m,n)} & $\biggl(V^{(m,n)}_{\text{eff},\Lambda}[U,\chi]\biggr|_{\alpha+\frac{k}{2}\,=\, \pm 1\mod 3,\,\,U_{ij}=1}\biggr)\biggl|_{\Lambda\to\infty}$ &  \\
\midrule
$(0,0)$ & $\mp\frac{4 N^2 \text{i} \pi ^3}{27}$
&
\\
$(1,0)=(0,1)$ & $\frac{2N^2 \pi ^2}{9}$
   &
\\
$(1,1)$ &
$\mp\frac{N^2\text{i} \pi }{36}$ &
\\
$(2,1)=(1,2)$ &
$\pm\frac{N^2}{18}$ &
\\
$(2,2)$ &
$0$ &
\\
$(3,0)=(0,3)$ &
$-\tfrac{N^2}{54}$ &
\\
$(X,Y,Z)$ &
$(e^{\frac{2  \pi \text{i} (\alpha+k/2) }{3}},e^{\frac{4 \pi \text{i}
   (\alpha+k/2) }{3}},e^{2 
   \pi \text{i} (\alpha+k/2) })$ &
\\
\bottomrule
\end{tabular}
\end{table}
\noindent This introduces a cutoff $\Lambda$ in the polylogarithms that define $V^{(m,n)}_{\text{eff}}$
\[V^{(m,n)}_{\text{eff}} \,=\,V^{(m,n)}_{\text{eff},\Lambda}\biggr|_{\Lambda\to\infty}.\]
Some examples of the truncated $V^{(m,n)}_{\text{eff},\Lambda}$ are reported in table~\ref{tab:Table3}.
In the presence of the cutoff $\Lambda$ there is a finite number---bounded by a power of $N$\---of saddle points for the eigenvalues of the unitaries
\[U_{ij}=e^{2\pi \text{i}u_{ij}}=e^{2\pi \text{i}(u_{i}-u_{j})}\,, \qquad |U_{ij}|=1\,.\]
These are saddles of all the $V_{\text{eff},\Lambda}^{(m,n)}\,$, independently, and for all values of $\chi=e^{4\pi \text{i}(\alpha+\tfrac{k}{2})}\,$. The simplest example is the one in which all eigenvalues collapse to
\begin{equation}\label{eq:SaddlesU}
U_{ij}\,=\,U^\star_{ij}\,=\,1\,.
\end{equation}
{{Quite remarkably, due to contributions with sufficiently large values of $m$ and $n\,$, 
\begin{itemize}
\item[1.] and upon the assumptions on the potentials $\alpha$
\begin{equation}\label{eq:NonBalancingCondition}
\alpha+\frac{k}{2}\,\neq\, \pm 1\mod 3\,,
\end{equation}
\item[2.] and $\delta\omega_a$
\begin{equation}\label{eq:DomainOmega}
\delta \omega_{a}\,<\,0\,, \qquad \tfrac{\delta \omega_1}{\delta \omega_2}\,=\,\zeta\,\,\, (\text{fixed and close enough to $1$})
\end{equation}
\end{itemize}
it follows that for all values of $\chi\,$ and $\nu$'s,
\begin{equation}\label{eq:NonChi1}
e^{\sum_{m,n=0}\frac{V_{\text{eff},\,\Lambda}^{(m,n)}[1+\text{i}0\,,\,\chi]}{\delta \omega_1 \delta \omega_2}}\biggl|_{\Lambda\to+\infty}\,=\,e^{-\infty}\,=\, 0\,.
\end{equation}
}}
Just to illustrate this, an example of such contributions is:
\begin{equation}\label{eq:ExampleDivergent}
V_{\text{eff},\,\Lambda}^{(2,2)}[U=1+\text{i}0\,,\,\chi]\,\underset{\Lambda\to \infty}{=}\, 
\begin{cases}\tfrac{1}{144} \frac{1}{(\text{i}0)^{2}}\,+\,(\text{sub}) =-\infty  & \alpha+\frac{k}{2}\,\neq\, \pm 1\mod 3\,, \\ 0 & \alpha+\frac{k}{2}\,=\, \pm 1\mod 3\,.
\end{cases}
\end{equation}
The same conclusion~\eqref{eq:NonChi1} holds for every divergent term labeled $(m,n=m)$ with $2m>3\,$. The $\Lambda\to\infty$ value of such terms is a function of $U$ (invariant under inversion $U\to\tfrac{1}{U}$) with poles precisely at the position of the finite-$\Lambda$ saddle points $U=1\,$. After expanding these functions around such poles, the coefficient of the leading divergent term happens to be a $c$-number independent of $\varphi_{1,2,3}\,$, and in particular of $\alpha\neq \frac{1}{2}\mod 1\,$. In all such examples that we have checked, the $c$-number multiplied by the leading negative power of $\tfrac{1}{(U\mp1)}$ is such that, as $U\to 1+\text{i} 0\,$, the corresponding~$\text{Re}(V^{(m,n=m)}_{\text{eff}})\to -\infty$.~\footnote{We have not proved this feature, but we have checked that it holds in all the many cases that we have tried, conjecturing its generality and postponing an analytic proof of it for future work.}

More generally, we can collect all leading singularities coming from coefficients of the same order $\gamma$ in the small $\omega_a$ expansion at fixed and finite $\zeta$ (close enough to $1$)
\[
\delta \omega_1^n \omega_2^{\gamma-n}\,, \qquad n\,=\,-1,\ldots, \gamma+1\,.
\]
All these terms have a leading singularity as $U\to 1$ of the form
$\frac{1}{(U-1)^{\gamma}}\,$. Concretely,
\begin{equation}
\frac{\sum_{n=-1}^{\gamma+1} c_{n}\delta\omega_1^{n} \delta\omega_2^{\gamma-n}}{(U-1)^\gamma}\,.
\end{equation}
The $c$-numbers $c_n$ are always such that in the small $\omega_{a}$ expansion at fixed $\zeta$ within the domain~\eqref{eq:DomainOmega}
\begin{equation}\label{eq:ConditionWeak}
\text{Re}\biggl(\frac{\sum_{n=-1}^{\gamma+1} c_{n}\delta\omega_1^{n} \delta\omega_2^{\gamma-n}}{(U-1)^\gamma} \biggr)\,\underset{U\to 1+\text{i}0}{\to}\, - \infty\,.
\end{equation}
For example, at order $\gamma=6\,$, the leading contributions add to
\begin{equation}
-\tfrac{\delta\omega _1^7}{10080 (U-1)^6 \delta\omega _2}+\tfrac{\delta\omega _2
   \delta\omega _1^5}{3024 (U-1)^6}+\tfrac{\delta\omega _2^3 \delta\omega
   _1^3}{4320 (U-1)^6}+\tfrac{\delta\omega _2^5 \delta\omega _1}{3024
   (U-1)^6}-\tfrac{\delta\omega _2^7}{10080 (U-1)^6 \delta\omega _1}
\end{equation}
which, in the region~\eqref{eq:DomainOmega}, satisfies~\eqref{eq:ConditionWeak}. Again, we have checked this feature at several values of $\gamma$. We conjecture that it holds at arbitrary values of $\gamma$.

We then conclude that if~\eqref{eq:NonBalancingCondition} holds, then for a small-$\omega_a$ expansion within the domain~\eqref{eq:DomainOmega}, the vanishing condition~\eqref{eq:NonChi1} holds as well. Notice that condition~\eqref{eq:ConditionWeak} continues to hold if we substitute simultaneously $\delta\omega_a \to e^{\pi\text{i}\kappa}\delta\omega_a$ and $U-1\to e^{\pi\text{i}\kappa} (U-1)\,$. This means that we can also take the limit of small $\delta\omega_a$ along the angle $e^{\pi\text{i}\kappa}$ if $U\to1$ along the axis $e^{\pi\text{i}(\kappa\pm \tfrac{1}{2})}$.

On the other hand, as illustrated in the second line of example~\eqref{eq:ExampleDivergent}, all potentially divergent contributions at order $m+n>3$ vanish if the following balancing condition is imposed
\[
\alpha+\frac{k}{2}\,=\, \pm 1\mod 3\,,
\]
and we obtain the large asymptotic growth given by the terms in table~\ref{tab:Table3}, which are summarized in the following formula at leading order in the large-$N$ expansion (in the minimally refined case)~\footnote{Up to a pure imaginary $c$-number that exponentiates to an $O(N^0)$ contribution. Since we are not looking at one-loop determinant contributions we will simply ignore these subleading corrections. That said, the method we propose can be used to straightforwardly compute those corrections as well. Computing them lies beyond the scope of this paper. As an aside comment we note that this method can be also applied at finite-$N$ in large-charge expansions, as the saddle-points discussed before are also saddle points in such expansion. We will comeback to this point in Appendix~\ref{sec:App}.}
\begin{equation}\label{eq:NonChi1General}
\sum_{m,n=0}\sum_{i,j}\frac{V_{\text{eff},\,\Lambda}^{(m,n)}[1+\text{i}0\,,\,\chi]}{\delta\omega_1\delta\omega_2}\,=\,\begin{cases}-\frac{N^2 \left(-2 \text{i} \pi+\delta\omega _1+\delta\omega _2
   \right){}^3}{54 \delta\omega _1 \delta\omega _2}\, & \qquad \alpha+\frac{k}{2}\,=\, \quad 1\mod 3\,,\\ - \frac{N^2 \left(+2 \text{i} \pi+\delta\omega _1+\delta\omega _2
   \right){}^3}{54 
   \delta\omega_1 \delta\omega _2} &\qquad \alpha+\frac{k}{2}\,=\, - 1\mod 3\,.
   \end{cases}
\end{equation}

{{
\eqref{eq:NonChi1General} means that if $d[J,Q]$ is dominated by these saddles in some region of $(j_0,q_0)$ in the semiclassical expansion~\eqref{eq:SemiclassicalExp2}, then in such a region of charges the integral over $t$ in~\eqref{eq:DTotal}, when translated to variables $\alpha$, localizes to the infinitesimal vicinities of
\begin{equation}\label{eq:VicinitiesAlpha}
\alpha\,+\frac{k}{2}\,=\, \pm 1\mod 3\,,\qquad k=1\bmod2\,
\end{equation}
that are intersected by the integration contour of $\alpha\,$. Due to the shift symmetry~\eqref{eq:SymmetryShift} the positions $\alpha=\pm\tfrac{3}{2}$ are, in a sense, isomorphic to the positions $\alpha=\mp\tfrac{1}{2}\,$. That is, one maps to the other by changing the representatives $n_a$ in~\eqref{eq:LimitTOIntegers}. 

In the language of distributions, we obtain (in the fully refined case)
\begin{equation}\label{eq:AsymptLocFormula}
\begin{split}
e^{\sum_{m,n=0}\sum_{i,j}\frac{V_{\text{eff},\,\Lambda}^{(m,n)}[1+\text{i}0\,,\,\chi]}{\delta\omega_1\delta\omega_2}}\biggl|_{\Lambda\to+\infty} &\,\sim\, \sum_{p\,\in\,\mathbb{Z}}\biggl(e^{-\frac{N^2 (\varphi_1)^- (\varphi _2)^- (\varphi_3)^-}{2 \delta\omega_1
\delta \omega_2}}\,+\,\ldots\biggr) \,\delta_{\alpha+\tfrac{k}{2}, 1+3p} \\ &\qquad \,+\,\sum_{p\,\in\, \mathbb{Z}}\biggl(e^{-\frac{N^2 (\varphi_1)^+ (\varphi_2)^+ (\varphi_3)^+}{2 \delta\omega_1
\delta\omega_2}}\,+\,\ldots\biggr)\,\delta_{\alpha+\tfrac{k}{2},\, 2+3p}\,,
\end{split}
\end{equation}
where we define
\begin{equation}\label{eq:BC}
(\varphi_3)^{\pm}\,:=\, -(\varphi_1)^{\pm}-(\varphi_2)^{\pm} +(\delta\omega_1)+ (\delta\omega_2)\,\pm 2\pi \text{i}\,,
\end{equation}
\begin{equation}
(\varphi_a)^\pm\,:=\, \varphi_a -2\pi \text{i}  p_a\,,\qquad a=1,2,
\end{equation}
and $p_a$ are integer-valued discontinuous functions defined from the condition
\[0\leq\text{sign}(\pm 1)\,\text{Im}\biggl((\varphi_a)^\pm\biggr)<2 \pi\,\text{sign}(\pm 1)\,.\]
}} 
The $\ldots$ stand for contributions coming from eigenvalue-instanton saddle contributions that are also attached to each one of the Dirac delta's above. These will be discussed below.

The Dirac delta's~\footnote{We expect these Dirac delta's to have a similar origin as the bits contributions of~\cite{Cabo-Bizet:2021jar}. These will be analyzed elsewhere.}
\[\delta_{\alpha+\tfrac{k}{2},\pm 1+3p}
\]
indicate that the only non-vanishing contributions from the contour integral over $\alpha\,$ come from the vicinities~\eqref{eq:VicinitiesAlpha}. The potential $\alpha$ is integrated along the period~$(\alpha\sim\alpha+{3},\,p\sim p,\,q\sim q)\,$, which is equivalent to the triple-cover period $(t\sim e^{6\pi \text{i} }t,\,p\sim p, \,q\sim q)\,$ specified by the integral over $t$ in the definition~\eqref{eq:DTotal}. Thus, only one delta function in each of the two groups in~\eqref{eq:AsymptLocFormula} is intersected by the corresponding contour integral.

We also note that using Poisson resummation identity in the sum over delta functions, e.g., on
\begin{equation}
\sum_{p\in \mathbb{Z}} \delta_{\alpha+\frac{k-2}{2},\, 3 p} \sim \sum_{p\in \mathbb{Z}} e^{2\pi \text{i}(\alpha\,+\, \frac{k-2}{2}) (\frac{p}{3})}\,.
\end{equation}
we naturally recover the fact that the R-charge $Q$ dual to $\alpha$ is quantized in units of $\frac{1}{3}\,$. This is an interesting feature. Multiplying such series by a source term $ t^{-Q}\,=\,p^{-Q/2} q^{-Q/2} e^{-2\pi \text{i} \alpha  Q}$ (e.g. the one appearing in~\eqref{eq:DTotal}) and integrating over $\alpha$ selects only the mode 
\[\frac{p}{3}=Q\,.\]
It remains then the complex phase
\[e^{\pi \text{i} (k-2) Q}\,,\] which, as we will explain below, it has physical relevance. 

\subsection{Orbifold saddle points of $Z_{\text{BPS}}$}
\label{sec:Orbifold}

There are many other saddle points of the complete potential $V[U,\chi]\,$
\[u_{i} \,\equiv\, g_c\,.\] 
They are localized around the singularities of $V[U;\chi]$~\cite{Beccaria:2023hip}. 
$Z_{\text{BPS}}$ also has singularities located at rational values
\begin{equation}\label{eq:General Expansions}
\delta\omega_a:=\omega_a -2\pi\text{i}\frac{n_a}{m_a}\,=\,0\,,\quad m_a, n_a\in \mathbb{Z}\,, \quad   \text{gcd}(m_a,n_a)=1\,
\end{equation}
which correspond to orbifold solutions. 
For the index, the on-shell action of some of these configurations has been found in the literature~\cite{Cabo-Bizet:2019eaf,Aharony:2021zkr,Cabo-Bizet:2020nkr} using various methods. We proceed to compute the contribution of these saddles of $Z_{\text{BPS}}\,$ using the truncation method we have just introduced. 

Let us define the least common multiple of $m_1$ and $m_2$ as
\[M=\text{lcm}(m_1,m_2)\,.\] 
Then a computation shows that in the expansions~\eqref{eq:General Expansions}
\begin{equation}
\begin{split}
V_{\text{eff}}[U]&\,=\, \frac{\sum_{n,m=0} \mathcal{V}^{(m,n)}_{\text{eff}}[{U;\chi}]\, \delta\omega_1^n \delta \omega_2^{m}}{\delta \omega_1 \delta \omega_2} \,+\, (\text{non-pert. suppressed as $\delta\omega_a \to0$})\,,
\end{split}
\end{equation}
where, except for the pure phase contributions $(m,n)=(1,1)$~\footnote{...which can be computed using this method as well...}, the truncated effective potential coefficients
\[\mathcal{V}^{(m,n)}_{\text{eff},\Lambda}[{U;\chi}]\]
are obtained from ${V}^{(m,n)}_{\text{eff},\Lambda}[{U;\chi}]$, that is, from the values reported in Table~\ref{tab:Table2}, by replacing
\[
\text{Li}_{s\Lambda}[z] \longrightarrow \tfrac{1}{M^s} \text{Li}_{s\Lambda}[z^M]\,.
\]
This means that~\eqref{eq:SaddlesU}
\begin{equation}\label{eq:SaddlesUPlus}
U_{i j}=U^\star_{i j}\,=\,1\,\quad \longleftrightarrow \quad u_{ij}:=u_{i}-u_j=0=u^\star_{ij}\,.
\end{equation}
remains a saddle point of the truncated effective potentials. This also means that the expansion in small $\delta \omega_a$ truncates exactly at the same order as before. 

In summary, in the language of distributions, we obtain the following asymptotic expansion~\footnote{In this formula, we ignore the one-loop determinant contributions and the contributions from the pure imaginary $c$-number shift of the free energy that depends on $M\,$. Such contributions, which are independent of the $\omega_{a}$'s and $\varphi_{I}$, will be studied elsewhere.}
\begin{equation}\label{eq:AsymptLocFormulaGeneral}
\begin{split}
e^{\sum_{m,n=0}\sum_{i,j}\frac{\mathcal{V}_{\text{eff},\,\Lambda=\infty}^{(m,n)}[1+\text{i}(0)\,,\,\chi]}{\delta\omega_1\delta\omega_2}} &\,\sim\, \sum_{p\,\in\,\mathbb{Z}}\biggl(e^{-\frac{N^2 (M\varphi _1)^+ (M\varphi_2)^+ (M\varphi_3)^+}{2 M (M\delta \omega_1)
(M\delta\omega_2)}} \,+\,\ldots\biggr)\,\delta_{M(\alpha+\tfrac{k}{2}),\,1+3p} \\ 
&\,\, \,+\,\sum_{p\,\in\, \mathbb{Z}}\biggl(e^{-\frac{N^2 (M\varphi _1)^{-} (M\varphi_2)^{-} (M\varphi_3)^-}{2 M (M\delta\omega_1)
(M\delta\omega_2)}}\,+\,\ldots\biggr) \delta_{M(\alpha+\tfrac{k}{2}),\,-1+3p}\,,
\end{split}
\end{equation}
where we define
\begin{equation}\label{eq:OrbifoldBC}
(M\varphi_3)^{\pm}\,:=\, -(M\varphi_1)^{\pm}-(M\varphi_2)^{\pm} +(M \delta\omega_1)+ (M\delta\omega_2)\,\pm 2\pi \text{i}\,.
\end{equation}
The $\ldots$ on the right-hand side represent extra eigenvalue-instanton saddles, which will be studied below. Again, these are solutions for which at least one of the $N^2-N$ eigenvalues $U_{ij}$ equals $1+\text{i}0\,$. 

For the particular case $M=m_a$,~\eqref{eq:OrbifoldBC} coincides with the BPS constraints obeyed by the gravitational orbifold cigars of~\cite{Aharony:2021zkr}. In particular, for $M=1$ and in the minimally refined case, it coincides with the BPS constraint of the cigar geometries studied in~\cite{Cabo-Bizet:2018ehj}. As announced, the~\emph{ orbifold contributions} to $Z_{\text{BPS}}$
\begin{equation}\label{eq:ExampleOrbifoldContribution}
e^{-\frac{N^2 (M\varphi _1)^{\pm}(M\varphi_2)^{\pm}(M\varphi_3)^{\pm}}
{2M (M\delta \omega_1)(M\delta \omega_2)}}
\end{equation}
matches the semiclassical contributions of the known family of orbifold saddle points~\cite{Cabo-Bizet:2019eaf} 
associated with the two independent canonical superconformal indices 
$\mathcal{I}_{\pm}$, obtained by fixing $\alpha = \pm\tfrac{1}{2} \bmod 2$ (assuming $k=1$) in $Z_{\text{BPS}}\,$. It also contains additional solutions, since $m_1$ and $m_2$ are totally unconstrained in our approach. 

These undressed saddle points organize into pairs whose contributions to the asymptotic expansion of $d[J,Q]$ are complex conjugates of each other. For example, for $M=1$ and $\omega_a=\omega$, one such pair is associated with the infinitesimal vicinities 
\begin{equation}\label{eq:OmegaCC}
\omega\,=\, \{0\,,\, \mp 4\pi \text{i}\}\,.
\end{equation}
As the contour of integration over the variables $(q)$ (at fixed $\alpha$) is a triple cover of the unit circle,~\eqref{eq:MicrocanonicalIndexDef}, we have to pick both of them. We remark that these two complex conjugate saddle points appear as two different saddle points of the integral over the gauge potentials $u_i$ and the integral over the angular velocities $\omega$'s at fixed contributing values of $\alpha\,$. 

Naively,~\footnote{This is because commuting the integral with the sum over different replica delta functions may bring problems. The safer procedure is to resum the series over delta functions before integration over $\alpha$ is performed. We explain this procedure in the appendix. } in the microcanonical partition function~$d[J,Q]\,$,~\eqref{eq:DTotal}, the contour of integration over $\alpha$ (at fixed $\omega$), e.g.\ the period $\alpha\in[-\tfrac{3}{2},\tfrac{3}{2})\,$, also picks a single delta function from each of the two independent groups in~\eqref{eq:AsymptLocFormulaGeneral} ($M=1$) labelled by the sign choice in the exponential prefactor $\pm$. However, in order to avoid sign oscillations we have to choose $k=1\bmod 2$, and as a consequence, these two contributions cannot be identified as complex conjugates of each other (at fixed $\omega$). 

{{A choice of a single superconformal index~$\mathcal{I}_{\pm}[\omega]$ by imposing the constraint~$\alpha=\pm\tfrac{1}{2}\mod 2$ in $Z_{\text{BPS}}[\omega,\varphi]$ corresponds to picking a single delta function in one of the two complex conjugate subgroups in~\eqref{eq:AsymptLocFormulaGeneral} (and not in the other).}}

{{\label{eq:IndexVSPartitionFunction} The dynamically generated balancing condition~\eqref{eq:BalancingCondition3} implies that every $g_c$ of the partition function $d[J,Q]$ is a $g_c$ of a superconformal index~$\widetilde{d}[2J+Q]\,$ in the semiclassical expansion~\eqref{eq:SemiclassicalExp2}. This follows from the asymptotic formulae~\eqref{eq:AsymptLocFormula} and~\eqref{eq:AsymptLocFormulaGeneral} (as each delta function in $\alpha$ localizes to an independent superconformal index) and from the fact that these asymptotic singularities dominate the semiclassical expansion~\eqref{eq:SemiclassicalExp2} of $d[J,Q]$ and $\widetilde{d}[2J+Q]$~\cite{Beccaria:2023hip}. }}

\subsection{Eigenvalue-instanton saddle points of $Z_{\text{BPS}}$} \label{ref:EigenvalueInstantons}

{{ The $g_c$'s also include eigenvalue-instanton saddle points. We define these as solutions which are not invariant under any subgroup of $\mathbb{Z}_N$ and which are obtained by shifting the positions of eigenvalues relative to orbifold solutions.
}}

Let us focus on a generic expansion of the $\omega_a$'s near rational numbers~\eqref{eq:General Expansions} from below. 
The $(m,n)$ components of the single-particle potential $V^{(m,n)}_{\text{eff},\Lambda}[U]$ have two vacua at $U = \pm 1$. 
These vacua are fixed points of a~$\mathbb{Z}_2$ symmetry corresponding to the inversion transformation $U \to \tfrac{1}{U}$.

The orbifold solutions discussed before are all located in one of these two vacua. As mentioned before, there are more general solutions.

For example, the $\mathbb{Z}_2$ eigenvalue-instanton solutions form a continuous family of solutions that interpolate between orbifold ones. These are configurations in which a fraction $x\in [0,1)$ of the $N$ eigenvalues is located at one vacuum of the single-particle potential, while the remaining fraction $(1-x)$ is located at the other.
Concretely, the configuration in which only one holonomy $U_{i_0}$ (corresponding to $x = \tfrac{1}{N}$) is shifted from one vacuum to the other can be written more precisely as
\begin{equation}\label{eq:OneEigevalueAside}
U_{i \neq i_0} = e^{2\pi \mathrm{i} u_{i \neq i_0}} = U_0, 
\qquad \text{and} \qquad 
U_{i_0} = e^{2\pi \mathrm{i} u_{i_0}} = -\,U_0\,,
\end{equation}
which is a saddle point of ${V}_{\text{eff},\Lambda}[U]$ for any value of $N$, $U_0$, $\chi$ or $\varphi_a$'s.

A computation then shows that the on-shell action of the undressed $\mathbb{Z}_2$ eigenvalue-instanton solutions can be recovered from the previous results by using the formula
\begin{equation}\label{eq:xDependence}
\begin{split}
\sum_{i,j }{V}_{\text{eff},\Lambda} [U^\star_{x}] \,=\,  &2{N}^2 \left(x-\frac{1}{2}\right)^2
   \biggl({V}_{\text{eff},\Lambda} [1] -{V}_{\text{eff},\Lambda} [-1]\biggr) \\ &\qquad 
   \qquad \qquad\qquad +\frac{1}{2} {N}^2 \Biggl({V}_{\text{eff},\Lambda} [1]+{V}_{\text{eff},\Lambda} [-1]\Biggr)\,,
   \end{split}
\end{equation}
at relevant orders $(m,n)\,$ in the small-$\delta\omega_a$ expansions at fixed $\zeta$ close enough to one.~For example, at order $(m,n)=(1,1)$, which corresponds to the pure-phase contribution, this gives ($s_{g_c}=\pm 1$))
\[-s_{g_c}\frac{\pi\text{i} N^2}{72}   \left(1-12
   \left(x-\frac{1}{2}\right)^2\right)\,.\]

Eigenvalue-instanton solutions have been studied in the context of the superconformal index.~\footnote{In work to appear by Aharony, O.; Benini, F.; Mamroud, O.;\ldots.} Our approach shows that the free energy of these configurations is bound to be the same for both 
$Z_{\text{BPS}}$ and $\mathcal{I}_{\pm}\,$, as the dynamically generated localization of $Z_{\text{BPS}}$ to an ensemble of superconformal indices 
also applies to these solutions.

We reiterate that such a match follows from the fact that, for the existence of exponentially leading saddle points $U_i$ of the $(m,n)$ components of the truncated effective potentials 
$\mathcal{V}^{(m,n)}_{\text{eff},\Lambda}[U;\chi]$ in the expansions to rational numbers~\eqref{eq:General Expansions}, the dynamically generated balancing constraint
\begin{equation}\label{eq:BalancingCondition3}
Z^M = e^{2M\pi \mathrm{i} (\alpha + k/2)} = 1\,,
\end{equation}
is a necessary condition.

Without this condition, and for sufficiently large values of $(m,n)$, 
the real part of the on-shell potential $\text{Re}\big(\mathcal{V}^{(m,n)}_{\text{eff},\Lambda}\big)$ 
evaluated at the finite-$\Lambda$ saddle points evaluates to $-\infty$ 
at $\Lambda = +\infty$. 
Thus, for values of $\alpha$ that do not meet the balancing condition above, 
there is no asymptotic contribution to the integral $Z_{\text{BPS}}$ 
coming from the expansions around roots of unity~\eqref{eq:General Expansions}.~\footnote{At this point, exponentially subleading power-like singularities should become leading, but studying this is beyond the scope of this paper.} 
Upon the balancing condition~\eqref{eq:BalancingCondition3}, the series of coefficient functions 
$\mathcal{V}^{(m,n)}_{\text{eff},\Lambda}$, when evaluated at the finite-$\Lambda$ saddle points, vanishes for $m+n>3$ after taking the limit $\Lambda \to \infty$. This 
produces a cubic polynomial truncation for $V_{\text{eff}}[U;\chi]$ around $U = U^\star\,$.

More general solutions can be constructed by placing eigenvalues at positions that are fixed points of the $\mathbb{Z}_N$ symmetry of the $N$-particle potential \[\sum_{i,j}\mathcal{V}^{(m,n)}_{\text{eff},\Lambda}[U_{i,j};\chi]\,.\] For example, equidistant packs of distributions of eigenvalues. Not all packs need to have the same number of eigenvalues.

As in the case of orbifold saddles, there are more general $\mathbb{Z}_{k \leq N}$ eigenvalue-instanton solutions which also come in complex conjugate pairs. 

{{\label{rem:LocalizEigenvalueInstantons} At large-$N$ these configurations are classified by a set of discrete parameters that become continuous $N=\infty\,$, e.g., for $\mathbb{Z}_2$ ones, by a filling fraction $x\,$. At large-$N$ we need to integrate over these continuous moduli. In the semiclassical expansion~\eqref{eq:SemiclassicalExp2} such an integral may be approximated by the saddle-point method. 

For example, for $\mathbb{Z}_2$ eigenvalue-instantons, the naive saddle-point evaluation localizes~$x$ to the value $x=\frac{1}{2}\,$, which corresponds to the configuration where $N/2$ eigenvalues are in one vacuum and the other $N/2$ are in the other $\mathbb{Z}_2$-dual one. This is also an orbifold configuration (with parameter $M$).

In this sense, we say that the $\mathbb{Z}_2$ eigenvalue-instanton configurations flow or localize to an undressed orbifold configuration in the semiclassical expansion~\eqref{eq:SemiclassicalExp2}. If this mechanism extends to other $\mathbb{Z}_N$ eigenvalue-instanton sectors, as we suspect is the case, then it would follow that eigenvalue-instanton saddles would not dominate any region of charges in the semiclassical expansion~\eqref{eq:SemiclassicalExp2}. Thus, in a sense, they would be unstable saddle points that flow between stable orbifolds. It would be interesting to check more complicated examples to see if this observation extends universally.
}}

\subsection{Dressed orbifold saddle points  of $Z_{\text{BPS}}$}\label{eq:DressedOrbifolds}

{{There are more general saddle points, which we will call \emph{dressed orbifold} saddles.~\footnote{Although the focus in this section will be on a large-$N$ expansion to simplify their expression. There are also generalizations of the solutions presented in this subsection at finite-$N$ and large charges (small $\delta{\omega}_a$.). Those come come the fact that there exist solutions to the saddle point-equations of~\eqref{eq:EffPotentialUT} at finite $N\,$. We leave the study of those solutions for the future. } 

In the context of individual superconformal indices we expect these to correspond to continuous families of Bethe ansatz solutions.~\footnote{At finite $N=2,3$, those solutions have been shown to be attached to topological or Hong-Liu Bethe roots in the appropriate expansion, in this case by the limit $\delta\to0\,$~\cite{Cabo-Bizet:2024kfe}. We recall that now we are not talking about a supercoformal index but about the BPS partition function, which we have shown to localize to an ensemble of superconformal indices in the semiclassical expansion~\eqref{eq:SemiclassicalExp2}.} Their ansatze depend on continuous moduli, e.g., $\delta$. An example of these ansatze is
\begin{equation}\label{eq:AnsatzDressed}
U_{ij}\,=\,e^{\delta (\widetilde{u}_i-\widetilde{u}_j) }\,.
\end{equation}
The parameter $\delta$
is an infinitesimal parameter-function as $\delta \omega_a\to0\,$. We define the ratios
\[{\delta}_a\,=:\,\frac{\delta}{\delta\omega_a}\,=\,\text{fixed}\,+\,O(\omega_a)\,, \qquad \frac{\delta_2}{\delta_1}\,=\,\frac{\delta\omega_1}{\delta\omega_2}\,=\,\zeta\,,\]
which remain constant as $\omega_a\to0\,$ and which may depend on the flavour rapidities $\varphi_{1,2}$ (always within the balanced locus~\eqref{eq:BalancingCondition3}).

A computation shows that the effective action for the dressed eigenvalue potentials~$\widetilde{u}_{ij}$ truncates to (initially assuming the $u_{ij}$ belong to the original contour of integration $u_{ij}\in \mathbb{R}$)
\begin{equation}\label{eq:EffPotentialUT}
\sum_{i\neq j=1}^{N}\pi \text{i} s_{g_c} \biggl( \frac{\delta_1 \delta_2}{2}\,\widetilde{u}_{ij}^2\, -\, \frac{1}{2}( \delta_1+\delta_2) \,|\widetilde{u}_{ij}|\biggr) \,+\, (\widetilde{u}_{ij}\text{-independent})\,.
\end{equation}
The $s_{g_c}=\pm 1$ is correlated with the sign in the parent undressed orbifold contributions~\eqref{eq:ExampleOrbifoldContribution}, and comes from the two independent ways to solve the balancing condition~\eqref{eq:BalancingCondition3}, i.e., from the two independent groups of periodic delta functions in~\eqref{eq:AsymptLocFormulaGeneral}. The
\begin{equation}
\delta_1 \delta_2 =f(\omega_a)\,=\,f_0+\text{i} f_1
\end{equation}
where $f_{0,1} (\omega_a)\,\in\, \mathbb{R}\,$ are the real and imaginary parts of $\delta_1 \delta_2\,$, which we are free to take as Taylor polynomials~\footnote{Polynomial truncations of Taylor expansions.} of generic real holomorphic and smooth functions at $\delta\omega_a=0\,$.

To obtain the effective potential~\eqref{eq:EffPotentialUT} we have taken the limit $\Lambda\to\infty$ of the $(m,n)$ coefficients of the truncated effective potentials in the small-$\omega_a$ expansion at fixed $U_{ij}\,$ (examples of such coefficients were reported in Table~\ref{tab:Table2}), reaching an expression in terms of polylogarithms in $U_{ij}\,$.~\footnote{All finite expressions are regular, in part due to the symmetry $U\to \tfrac{1}{U}$ of the $N$-particle potential.} Then we have substituted the ansatz~\eqref{eq:AnsatzDressed} into such expressions and, at last, expanded the answer at $\delta=0\,$. At the balancing locus~\eqref{eq:BalancingCondition3} we obtained the truncated effective potential~\eqref{eq:EffPotentialUT}. 

Away from the balancing condition~\eqref{eq:BalancingCondition3}, the $\widetilde{u}_{ij}$-independent part of the effective potential, which equals the free energy of the core orbifold, still blows up with the required signature, assuming $\delta$ is a pure imaginary quantity with $\widetilde{u}_{ij}\in\mathbb{R}$ and~\eqref{eq:DomainOmega}, for example. Such a localization condition for the $\Lambda\to\infty$ expansion implies the constraint
\begin{equation}
{f_{1}}\underset{\delta\omega_a\to0}{=}0\,.
\end{equation}
at every order in the small-$\delta\omega_a$ perturbations in the domain~\eqref{eq:DomainOmega}. This constraint implies $\delta_1 \delta_2$ to be real-holomorphic in $\delta \omega_a$ at leading order in the small $\delta \omega_a$ expansion.~\footnote{The ambiguity in the choice of the continuous single modulus $\delta$ disappears in the final answer for the on-shell action. }

At this point, we look for complex saddle points of the truncated effective potential~\eqref{eq:EffPotentialUT} aside from the orbifold ones we have already found at $u_{i,j}\,=\, 0\,$. To compute the asymptotic form of these corrections we use the original analytic extension of the effective potential which is obtained by replacing
\[-\pi \text{i}|u|\to u\log u - u \log(-u)\]
Now we assume an ansatz in which the eigenvalues group into two groups $a=1,2$, corresponding to two subsets of $\tfrac{N}{2}$ labels $\{i_a\}\,$. The distance among eigenvalues in the same pack being
\[u_{i_a,j_a}=\frac{v_{i_aj_a}}{N}=O(1/N)\,.\]
And the distance between eigenvalues in two distinct packs being (at leading order in the large-$N$ expansion)
\[u_{i_1,j_2}= \Delta u+ O(1/N)\,.\]
The saddle-point equations take the following form at leading order in the large-$N$ expansion in question:
\begin{equation}
\underbrace{\frac{1}{2} \left(-\delta _1-\delta _2\right)}_{\text{constant force coming from derivatives over logs}}+\delta _1 \delta
   _2 \Delta u\,=\,0\,+\, O(1/N)\,.
\end{equation}
After evaluating the on-shell potential at the solution to this equation, we obtain the contribution of the dressed saddles (with the contribution of the core orbifold incorporated):
\begin{equation}\label{eq:DressedContributions}
e^{-\frac{N^2 (M\varphi _1)^{\pm}(M\varphi_2)^{\pm}(M\varphi_3)^{\pm}}
{2M (M\delta \omega_1)(M\delta \omega_2)}\,+\, \pi \text{i} s_{g_c} N^2 \gamma}\,.
\end{equation}
Again, this large-$N$ result is exact at all orders in perturbations around $\delta \omega_a\,=\,0\,$. In that expansion, corrections to this result are non-perturbatively suppressed in the region of chemical potentials that we have assumed. To compute the corrections of each one of these saddle points to the microcanonical ensemble at leading order in the semiclassical expansion~\eqref{eq:SemiclassicalExp2}
we can extend the domain of the $\delta \omega_a$'s to their entire complex plane, the missing contributions will be exponentially suppressed in that expansion.

The corrections introduced by the dressing to the free energy of its background undressed orbifold solutions, which we will call \emph{the core orbifold solutions}, are
\begin{equation}\label{eq:gammaDEIS}
\begin{split}
N^2 \gamma&\,:=\,-\frac{\left(\delta _1+\delta _2\right)^2 N^2}{16
   \delta_1\delta_2 }\,=\,-\frac{\left(\delta \omega _1+\delta \omega_2\right)^2 N^2}{16
   \delta \omega_1\delta \omega_2 }\\
   &\,=\,-\,\frac{\left(\delta \omega _1-\delta \omega_2\right)^2 N^2}{16
   \delta \omega_1\delta \omega_2 }+c\text{-constant}
   =\,-\frac{\left(\frac{1}{\zeta^{1/2}}+\zeta^{1/2}\right)^2 N^2}{16}\,.
   \end{split}
\end{equation}
Note that the dressing function $\gamma$ can still become complex for complex values of $\delta\omega_a\,$. Moreover, consistently with finite-$N$ field theory expectations~\cite{Cabo-Bizet:2024kfe} regarding continuous families of Bethe roots contributions which we expect to correspond to dressed eigenvalue configurations at large-$N\,$, the dependence on the continuous modulus $\delta$ disappears from the on-shell action.

Note also that $\gamma$ is a $c$-number if $\delta\omega_{1}= \delta\omega_2\,$. This means that at large $N$ these dressed orbifold saddle point solutions are irrelevant, in comparison with undressed orbifolds, in regions of charges in the microcanonical ensemble that via the Legendre transform map to the region of chemical potentials 
\begin{equation}\label{eq:HalfCardyExp}
\delta \omega_{1 (\text{resp.}2)} =\text{fixed}\neq 0\,,\qquad \delta\omega_{2(\text{resp.}1)}\to 0\,.
\end{equation}
Or equivalently 
\begin{equation}
\delta \omega_{1 (\text{resp.}2)} =\text{fixed}\neq 0\,,\qquad \zeta\, (\text{resp $\frac{1}{\zeta}$})\to \infty\,.
\end{equation}
We will come back to comment on this below.

There are more dressed orbifold saddle point solutions. Notice that the solutions $\widetilde{u}^\star$ have the form of $\mathbb{Z}_2$ eigenvalue-instanton saddles as well, as they correspond to two packs of almost coinciding saddles.

More generally, we obtain new solutions by distributing a fraction \[x\neq \tfrac{1}{2} \in (0,1)\] of eigenvalues in one of the two packs and $(1-x)$ in the other. In that case, the final answer for their correction to the on-shell action of the core orbifold solution that we have obtained can be recovered by replacing the $\gamma$ defined in~\eqref{eq:gammaDEIS} as follows (as we have already explained for undressed eigenvalue-instantons)
\begin{equation}
\gamma\to 4x(1-x) \gamma\,.
\end{equation}
Again, at $N=\infty$ the filling fraction $x$ is continuous. We call these solutions dressed eigenvalue-instanton saddle points. They flow between orbifold solutions $(x=0)$ and dressed orbifold solutions $(x=\frac{1}{2})\,$. Even more general dressed $\mathbb{Z}_{k\leq N}$ eigenvalue-instanton saddles exist, but those will be studied elsewhere.

The asymptotic expansion of $Z_{\text{BPS}}$ in the small-$\delta \omega_a$ expansion also receives contributions from dressed saddle points of the gauge eigenvalues $U_{ij}$. Such contributions have the form:
\begin{equation}\label{eq:DressedAsymptLocFormulaGeneral}
\begin{split}
e^{\sum_{m,n=0}\sum_{i,j}\frac{\mathcal{V}_{\text{eff},\,\Lambda=\infty}^{(m,n)}[e^{\delta \widetilde{u}^\star}\,,\,\chi]}{\delta\omega_1\delta\omega_2}} &\,\sim\, \sum_{p\,\in\,\mathbb{Z}}\biggl(e^{-\frac{N^2 (M\varphi _1)^+ (M\varphi_2)^+ (M\varphi_3)^+}{2 M (M\delta \omega_1)
(M\delta\omega_2)}+\pi \text{i} N^2 \gamma}\biggr)\,\delta_{M(\alpha+\tfrac{k}{2}),\,1+3p} \\ &\,\, \,+\,\sum_{p\,\in\, \mathbb{Z}}\biggl(e^{-\frac{N^2 (M\varphi _1)^{-} (M\varphi_2)^{-} (M\varphi_3)^-}{2 M (M\delta\omega_1)
(M\delta\omega_2)}-\pi \text{i} N^2 \gamma}\biggr) \delta_{M(\alpha+\tfrac{k}{2}),\,-1+3p}\,,
\end{split}
\end{equation}
where $\widetilde{u}^\star$ denotes the dressed eigenvalue-instanton saddle points just studied.

We reiterate that all these saddle points of $Z_{\text{BPS}}$ are also saddle points of the superconformal index. This follows from the dynamically generated balancing condition~\eqref{eq:BalancingCondition3} that in the semiclassical expansion~\eqref{eq:SemiclassicalExp2} reduces the partition function $Z_{\text{BPS}}$ to an ensemble of superconformal indices.

}}

The asymptotic actions of more generic saddles will not be computed here. 

To summarize, the on-shell action of all the configurations that we have explored has the form
\begin{equation}
\mathcal{F}_{g_c}[\omega_1,\omega_2,\varphi_1,\varphi_2]\,=\,\frac{P_{3,g_c}[\delta\omega_1,\delta\omega_2,\varphi_1,\varphi_2]}{\delta\omega_1\delta\omega_2}\,+\, \pi\text{i} s_{g_c} N^2\gamma_{g_c}
\end{equation}
at every order in perturbative expansions around $\delta \omega_a=0\,$ and $N\gg1\,$.
$P_{3,g_c}$ is a cubic polynomial in $\omega_1,\omega_2,\varphi_1,\varphi_2\,$. $\gamma_{g_c}$ is an arbitrary complex function of chemical potentials and of $g_c$ that is regular
\[\gamma_{g_c}\underset{\delta \omega_a\to0\atop\zeta=\text{fixed}}{\sim} O(\delta\omega_a^0)\]
and real-holomorphic at leading order in the small-$\delta\omega_a$ expansion. For pairs of complex conjugate solutions $\gamma_{g_c}$ is the same.

That $P_{3,g_c}$ is a cubic polynomial follows from the truncation feature of the on-shell components $\mathcal{V}^{(m,n)}_{\text{eff},\Lambda\to\infty}[U^\star;\chi]\,$, for~$(m+n)>3\,$, at the dynamically generated balancing condition $Z^M=1\,$, previously explained.

The choice of saddle point $g_c$ determines the cubic polynomial and the dressing function $\gamma_{g_c}\,$. 

At $\omega_{1}=\omega_2\,$, the dressing function $\gamma_{g_c}$ is a $c$-number times a function of the filling fraction $x$: this means that the $M=1$ dressed eigenvalue-instantons (with bare orbifold number $M=1$) are indistinguishable from the dominant orbifold saddle $M=1\,$ (in the microcanonical ensemble). In other words, from~\eqref{eq:DressedContributions} it is clear that the asymptotic behaviour of $d[J_{1,2},Q]$ in the semiclassical expansion~\eqref{eq:SemiclassicalExp2} at large enough normalized charges \[j_{1,0}+\tfrac{1}{2}q_0\,=\,j_{2,0}+\tfrac{1}{2}q_0=\frac{(J+\tfrac{1}{2}Q)}{N^2}\] is dominated by pure orbifold solutions (more concretely by those with $M=1$). That is because at $\zeta=1$, the function $\gamma$ reduces to a $c$-number, and it is then spurious as $\delta\omega_1=\delta \omega_2\to0\,$.

On the other hand, still from~\eqref{eq:DressedContributions}, it follows that the dressed eigenvalue-instanton saddles are dominant in the following normalized large-charge expansions \begin{equation}\label{eq:MicroHalfCardy}
\tfrac{j_{0,1}+j_{0,2}}{2}=j=j_L=\text{fixed}\,, \qquad \tfrac{|j_{0,1}-j_{0,2}|}{2}=j_R\,\gg 1\,.
\end{equation}
In terms of Legendre-dual chemical potentials,~\eqref{eq:MicroHalfCardy} corresponds to the expansion~\eqref{eq:HalfCardyExp}. Indeed, in that expansion the dressing contribution~\eqref{eq:gammaDEIS} competes with the core orbifold contribution. In particular, in this expansion our approach predicts that the dressed orbifold saddle points above, with core orbifold number~$M=1\,$, may dominate the microcanonical ensemble.~\footnote{For example, at fixed $\delta\omega_2\neq 0$ and $\zeta\to0$ both the core orbifold and the dressing contribution scale as $\frac{1}{\zeta}$ and compete.  }

In the context of the superconformal index, this previous observation is consistent with the empirical deviation reported in figure 1 of~\cite{Choi:2025lck} (represented as a transition of color there) as well as in~\cite{Deddo:2025jrg}. Our results indicate that such deviation may correspond to the transition in dominance between the bare orbifolds $M=1$ and the dressed orbifold solutions $M=1\,$ in expansion~\eqref{eq:MicroHalfCardy}. The verification of this last possibility lies beyond the scope of this paper.

\section{On the microcanonical perspective}\label{sec:Microcanonical}

At leading order in the expansion~\eqref{eq:SemiclassicalExp2}, the asymptotic contribution of a single saddle point $g_c$ to $d[J,Q]$ is the exponential of the Legendre transform
\begin{equation}\label{eq:CanonicalMicrocanonical}
\begin{split}
\mathcal{S}_{g_c}[J,Q]&:=\text{ext}_{\omega}\biggl(\mathcal{F}_{{g_c}}[\omega]-(\omega+2\pi \text{i}\tfrac{n_{g_c}}{M_{g_c}}) (2J+Q)
-2\pi \text{i} \alpha_{g_c}  Q\biggr)\,
\\
&\,=:\,\text{ext}_{\omega}\biggl(\mathcal{F}_{{g_c}}[\omega]-\omega (2J+Q) -\pi \text{i} s_{g_c}\biggl(\underbrace{\tfrac{2n_{g_c}}{M_{g_c}s_{g_c}} \,(2 J+Q)+\tfrac{2\alpha_{g_c}}{s_{g_c}}Q}_{\mathfrak{Q}_{g_c}
}\biggr)\biggr)\,,
\end{split}
\end{equation}
where $s_{g_c}=\pm 1$ has opposite signs for complex or time-reversal conjugated $g_c$'s. The free energy is defined as
\begin{equation}\label{eq:FgcOx}
\mathcal{F}_{{g_c}}[\omega]\,=\,-\frac{4 N^2}{27}\,\frac{(\frac{\pi\text{i} s_{g_c}}{M_{g_c}}+\omega )^3}{\omega^2}\,+\,N^2\Gamma_{g_c}\,.
\end{equation}
The extra contributions to free-energy dressing and imbalanced eigenvalue-instanton configurations are collected in the term
\begin{equation}
\Gamma_{{g_c}}[\omega]\,=\,O(\underline{x})\,+\,\pi \text{i}s_{g_c}\gamma_{g_c}\,.
\end{equation}
The integers are
\begin{equation}
n_{g_c} \in \mathbb{Z}\,, \qquad |n_{g_c}|\,\leq\, M\,.
\end{equation}
The linear combination of $J$ and $Q\,$\[\mathfrak{Q}_{g_c}\,,\] is the same for two complex conjugate $g_c$'s.
The explicit form of the free energy $\mathcal{F}_{g_c}[\omega]$ is obtained from the previous results for
\[e^{\sum_{m,n=0}\frac{V_{\text{eff},\,\Lambda=\infty}^{(m,n)}[1+\ldots\,,\,\chi]}{\delta\omega_1\delta\omega_2}}\]
after a local change of the integration variables from $\omega$ to $\delta \omega$
\begin{equation}
\delta{\omega}\,=\, \omega -2\pi \text{i}\frac{n_{g_c}}{M_{g_c}}
\end{equation}
and then dropping the $\delta$ to ease the presentation. 
$\alpha_{g_c}$ is the localized value of the potential $\alpha$ chosen by the integration contour that defines $d[J,Q]\,$: the value of $\alpha$ fixed by the delta function in expansions~\eqref{eq:AsymptLocFormulaGeneral} that intersect the saddle point solution $g_c\,$. For example, for the undressed saddle points ($\gamma_{g_c}\equiv0$) with~$M_{g_c}=1\,$ and $x=0\,$, and after dropping the trivial contribution $4\pi \text{i} n_{g_c} J\,$ (an integer multiple of $2\pi\text{i}$), as $2J \in \mathbb{Z}\,$, we obtain
\[\alpha_{g_c}\,=\,\tfrac{s_{g_c}}{2}\,\bmod\,{2}\,, \qquad \mathfrak{Q}_{g_c}\,=\, Q\,,\]
where again $s_{g_c}=\pm 1\,$, for complex conjugate $g_c$'s.

Even the $x$-dependent contributions can be collected in the form (e.g., see equation~\ref{eq:xDependence})
\begin{equation}
\mathcal{F}_{g_c}[\omega]\,=\,\frac{P_{3,g_c}[\omega]}{\omega^2}
\end{equation}
where~$P_{3,g_c}[\omega]$ is a cubic polynomial in $\omega\,$. The expectation value of the single charge $2J+Q$ is 
\begin{equation}
2J +Q=\partial_\omega \mathcal{F}_{g_c}\,.
\end{equation}
At large-$N$ the space of $g_c$'s we have studied is characterized by a potential dressing function
\begin{equation}\label{eq:regularitygammagc}
\gamma_{g_c} \underset{\delta{\omega}\to0}{\sim} \text{const} \in\mathbb{R}\,,
\end{equation}
which for generic $\omega_{1}=\omega_2=\omega$ is just a spurious $c$-number times the filling fraction contribution, e.g., $4 x(1-x)\,$; by the core orbifold number
\[M_{g_c}\,=\,1\,,\,2\,, \ldots\,;\] by the discrete variables
\begin{eqnarray}
\xi=\frac{n_{g_c}}{M_{g_c}}\,\in\, \mathbb{R}\,, \qquad \alpha_{g_c}=\pm \frac{1}{2}+ {2}\frac{p_{g_c}}{M_{g_c}} \in \mathbb{R}\,;
\end{eqnarray}
and filling fractions $x$ characterizing eigenvalue-instanton configurations. For example, for the $\mathbb{Z}_2$ eigenvalue-instanton representations
\begin{equation}
{x} \in {[0,1]}\,.
\end{equation}
As explained before, the integral over the filling fraction moduli $x$ localizes to orbifold configurations. We have shown this to be the case for $\mathbb{Z}_2$ eigenvalue-instantons, but we expect this to be the case for generic $\mathbb{Z}_{N}$ eigenvalue-instantons.

{{\label{rem:AbsdJQ}
The saddle-point prediction for BPS entropy $d[J,Q]$ in the semiclassical expansion~\eqref{eq:SemiclassicalExp2} is
\begin{equation}\label{eq:StrongerClaim}
|d[J,Q]|\,\sim\,e^{\max_{g_c}\text{Re}\biggl(\mathcal{S}_{g_c}[J,Q]\biggr)}\,.
\end{equation}
From the extremization problem~\eqref{eq:CanonicalMicrocanonical} it follows that $|d[J,Q]|$ is only a function of $2J+Q\,$. The maximization process selects the pairs of complex conjugate saddle points $g^\star \in \{g_c\}$ that dominate the real part of $\mathcal{S}_{g_c}[J,Q]$ at a given value of $2J+Q\,$.

Moreover, equation~\eqref{eq:CanonicalMicrocanonical} as well, it follows that in the semiclassical expansion~\eqref{eq:SemiclassicalExp2}
\begin{equation}\label{eq:RelationdDtilded}
\biggl|d[J\,,\,Q]\biggr| \sim \biggl|\widetilde{d}[2J+Q]\biggr|\,.
\end{equation}
The problem then is to find the saddle points $g^\star$ that maximize either the absolute value of the index $\widetilde{d}[2J+Q]\,$, or equivalently $d[J,Q]\,$, at a given value of $2J+Q\,$.

}}

{{\label{rm:71}The microcanonical partition function $d[J,Q]$ does not have sign oscillations. Thus, it cannot oscillate in the large-$N$ semiclassical expansion~\eqref{eq:SemiclassicalExp2}.}}
Thus, if individual $g_c$'s can dominate the integral representation of $d[J,Q]\,$, then they can only do so on a locus of charges defined by the following constraint
\begin{equation}\label{eq:ConditionConstraintNonL}
\text{Im}(\mathcal{S}_{g_c}[J,Q])\,=\,0\,.
\end{equation}
We will call this condition either the non-linear constraint of charges or the non-oscillation trajectory associated to the solution $g_c$ and its complex conjugate dual (simultaneously).

The non-oscillation constraint of $d[J,Q]$ also implies, in conjunction with~\eqref{eq:RelationdDtilded}, that in the semiclassical expansion~\eqref{eq:SemiclassicalExp2}, and in the corresponding co-dimension 1 locus of charges, the BPS partition function and the absolute value of the index give the same asymptotic answer
\begin{equation}\label{eq:dVsAbsdt}
d[J,Q] \,\sim\, |\widetilde{d}[2J+Q]|\,.
\end{equation}
Namely, the absolute value of the superconformal index is enough to reproduce the total number of states.

However, the natural observable to define entropy is the BPS partition function $d[J,Q]$ that counts the total number of physical BPS states, which has no large oscillations.

{{The asymptotic expansion of $d[J,Q]$ is a sum over all possible $g_c$'s
\begin{equation}d[J,Q]\,\sim\, \sum_{g_c} \chi_{g_c,J,Q}\,F_{g_c}[2J+Q]\, \exp\biggl(\pi\text{i}(s_{g_c}\mathfrak{Q}_{g_c}-\mathcal{C}_{g_c}[2J+Q])\biggr) \end{equation}}
where
\[F_{g_c}[2J+Q]\,\]
is the leading large-$N$ asymptotics of the absolute value of the contribution of $g_c$, and
\begin{equation}\label{eq:FunctionRcDef}
\mathcal{C}_{g_c}[\,\mathfrak{j}\,]\,:=\, -\frac{1}{\pi}\, \text{Im}\, \text{ext}_{\omega} \biggl(\mathcal{F}_{g_c}[\omega]-\omega \,\mathfrak{j}\,\biggr)\,\qquad \,\mathfrak{j}\,\in \mathbb{R}\,,
\end{equation}
and 
\begin{equation}
\chi_{g_c,J,Q}
\end{equation}
is the intersection number of the integration contour in~\eqref{eq:DTotal} with the Lefschetz thimbles ending in $g_c$'s. For example, the Lefschetz thimbles of the undressed orbifold solutions were studied in~\cite{Cabo-Bizet:2020ewf}. These intersection numbers can only change abruptly when the imaginary parts of two or more saddles $g_c$ coincide. Such a condition is precisely realized on the non-oscillation locus~\eqref{eq:ConditionConstraintNonL}.

The non-oscillation constraint implies that any potential oscillation of the~$\chi_{g_c,J,Q}$, if present, must comply with the positivity condition
\begin{equation}
d[J,Q]\,\sim\, F_{g^\star}[2J+Q] \,\times\,\sum_{g^\star} \chi_{g^\star,J,Q}\,\, \exp\biggl(\pi\text{i}(s_{g^\star}\mathfrak{Q}_{g^\star}-\mathcal{C}_{g^\star}[2J+Q])\biggr) \,>0\,.
\end{equation}
A detailed study of this question starting from the integral~\eqref{eq:DTotal} is beyond the scope of this paper.

}

{{Even with this lack of understanding of the sum over dominating $g^\star\,$, we can conclude that if a single $g^\star$ is to be identified with a single complex BPS cigar with real horizon area computing the logarithm of the number of field theory BPS states
\[d[J,Q]\,,\] then such an identification can only hold in the codimension 1 locus of charges $Q=Q_{g^\star}[J]$ defined by the non-oscillation condition~\eqref{eq:ConditionConstraintNonL}
\begin{equation}\label{eq:NonLinearConstraintChargesFieldTheory}
s_{g^\star}\mathfrak{Q}_{g^\star}\,=\mathcal{C}_{g^\star}[2J+Q] \,.
\end{equation} 
For example, if $g^\star$ is one of the two trivial orbifold saddle points with~$M_{g_c}=1\,$ and $x=0\,$
\[\alpha_{g_c}\,=\,\tfrac{s_{g_c}}{2}\,, \qquad \mathfrak{Q}_{g_c}\,=\, Q\,.\]
with $s_{g_c}=\pm 1\,$, then the non-oscillation constraint~\eqref{eq:NonLinearConstraintChargesFieldTheory} reduces to the well-known non-linear constraint among charges associated to the absence of naked CTCs in the supersymmetric locus of CCLP solutions~\cite{Chong:2005da,Gutowski:2004ez}\cite{Cabo-Bizet:2018ehj,Hosseini:2017mds} (to be reviewed below in equation~\eqref{eq:NonLConConcrete}). 

Our expectation is that these latter solutions will dominate the microcanonical ensemble in the expansion~\eqref{eq:SemiclassicalExp2} in the section $J_1=J_2=J\,$ with $Q$ fixed by the non-oscillation condition. At $J_1=J_2=J$ and in regions of $Q$ away from the non-oscillation locus of these saddle points, we expect that other saddle point solutions will dominate $d[J,Q]\,$. In order to check so, we need to compute the Taylor coefficients of integral~\eqref{eq:ZBPSNonPerIntro} for large enough $N$, in contradistinction to the tests in~\cite{Murthy:2020scj,Agarwal:2020zwm}, which were performed for a single superconformal index, in this case there will not be oscillations (See appendix~\ref{sec:App}).~\footnote{Orbifold contributions are suppressed. Their entropy is proportional to $\tfrac{1}{M}\,$~\cite{Cabo-Bizet:2019eaf,Aharony:2021zkr}.}

}}

\subsection{A comment on the localization of $Z_{\text{BPS}}$ to the index in supergravity}\label{sec:BPSCigars}

We end with a comment on the dual gravitational description regarding the meaning of the localization of $Z_{\text{BPS}}$ to saddle points of indices.

The equation~\eqref{eq:BPSPartitionFunction} in field theory
\begin{equation}
Z_{\text{BPS}}[\omega,\varphi]\,=\, Z[\beta=\infty,\omega,\varphi]
\end{equation}
is telling us that, given $\omega_{1,2}\,$, there are infinitely many independent limits to extremality $\beta\to\infty$ which project the counting of physical states to the BPS locus~\eqref{eq:BPSEq1}
\begin{equation}
E-2J-3/2Q\,=\,0\,.
\end{equation}
These limits are parameterized by an extra single parameter $\varphi$ or equivalently $\alpha\,$. We will call them the \emph{extended-BPS locus}.

In gravity, an analogous family of \emph{extended-BPS limits} was reported in~\cite{Cabo-Bizet:2018ehj} for the periodic identification~\eqref{eq:ThermalPeriodicity}.~\footnote{More general time-orbifold identifications are possible $t\sim t+\tfrac{\beta}{M}$\,, $M\in\mathbb{Z}$, $M\geq 1\,$~\cite{Aharony:2021zkr}. In this discussion we will focus on the choice $M=1\,$~\cite{Cabo-Bizet:2018ehj}.} These limits are characterized by the single parameter $u_{\text{there}}$ defined in Section 3.3 of~\cite{Cabo-Bizet:2018ehj}, which corresponds to $-2\text{i}\alpha_{\text{here}}\,$ at
\[\beta_{\text{there}}=\infty\,.\] 
{{The two regions
\begin{equation}\label{eq:Geometry1} u_{\text{there}}=\mp \text{i}\end{equation} are called the supersymmetric locus and correspond to the lines \(\alpha_{\text{here}} = \pm \tfrac{1}{2}\bmod 1.\)
The worldline parameter is the regulator temperature mentioned in the introduction
\[\beta_{\text{there}} \,. \]
Only at $\beta_{\text{there}}=\infty\,$, the supersymmetric locus will be called the BPS locus (again, this locus represents two points of the extended-BPS locus).

{On the other hand, the gravitational geometries in the complementary region
\[
u_{\text{there}}\, \neq\, \mp \text{i}
\]
can still be supersymmetric but only at leading and next-to-leading order around $\beta_{\text{there}} = +\infty$ ($\epsilon_{\text{there}}\,=\,0$)~\cite{Cabo-Bizet:2018ehj}
\begin{equation}\label{eq:AlmostBPS}
E_{\text{there}}-2J_{\text{there}}-\tfrac{3}{2}Q_{\text{there}}= f(u_{\text{there}})\,O\biggl(\frac{1}{\beta_{\text{there}}^2}\biggr)\,.
\end{equation}
Namely, they are not supersymmetric all along the entire flow to extremality $\beta_{\text{there}} \to \infty\,$.~\footnote{$f(u_{\text{there}})$ has zeroes at $u_{\text{there}}=\mp \text{i}\,$.}}

Thus, assuming the natural identification of charges between gravity and field theory, equation~\eqref{eq:AlmostBPS} implies that:

\begin{itemize}
\item only the BPS geometries~\eqref{eq:Geometry1} and the \emph{extended-BPS geometries} which are extremal:
\begin{equation}\label{eq:RapiditiesuBeta}
u_{\text{there}}\,\neq\, \mp \text{i}\,, \qquad \beta_{\text{there}}\,=\,\frac{1}{\epsilon_{\text{there}}}\,=\,\infty
\end{equation}
correspond to saddle points of the microscopic BPS partition function~$Z_{\text{BPS}}\,=\,Z[\beta=\infty,\omega,\varphi]\,$. Geometries beyond these, e.g., extended-BPS geometries which are not extremal, do not correspond to saddles of $Z_{\text{BPS}}$\,.
\end{itemize}

Note that the extremal extended-BPS geometries~\eqref{eq:RapiditiesuBeta}
do not have a cigar topology, 
since the time direction is not a cycle anymore at $\beta_{\text{there}} = \infty\,$. From now on, we will simply call them extended-BPS geometries or cigars (assuming implicitly that they are extremal)

In the extended-BPS locus
\begin{equation}
u_{\text{there}}\,=:\,\mp \text{i}+\widehat{u}\,\neq\, \mp \text{i}\,, \qquad \beta_{\text{there}}\,=\,\frac{1}{\epsilon_{\text{there}}}\,=\,\infty\,,
\end{equation}
we cannot identify a total differential of $u_{\text{there}}$ (at fixed $a_{\text{there}}$ and $b_{\text{there}}$), $\delta u_{\text{there}
}$, with a differential of $\alpha_{\text{here}}$ (at fixed $\omega_{a,\text{here}}$), $\delta \alpha_{\text{here}}\,$. For example, for differential variations about the BPS loci \[\widehat{u}=0\,,\] we would like to identify
\begin{equation}
\delta \widehat{u}\,\leftrightarrow\, -2 \text{i}\,\delta\alpha_{\text{here}}\,,
\end{equation}
keeping fixed $\omega_{1}$ and $\omega_2\,$.

This is achieved by implementing a specific reparameterization of the parameters $a_{\text{there}}$ and $b_{\text{there}}$ in section 3.3 of~\cite{Cabo-Bizet:2018ehj}. A reparameterization of the form
\begin{equation}\label{eq:Repars}
\begin{split}
a_{\text{there}} &\to a_0+ \delta\widetilde{a}[u_{\text{there}}]\,, \qquad  b_{\text{there}}\to b_0+ \delta\widetilde{b}[u_{\text{there}}]\,,\\
\delta \widetilde{a}[u_{there}]& =r_0+\sum_{n=1}^\infty r_n (\underbrace{u\pm \pi\text{i}}_{\delta \widehat{u}})^n\,, \qquad  \delta \widetilde{b}[u_{there}] =s_0+\sum_{n=1}^\infty s_n (u\pm \pi\text{i})^n\,,\\
r_0&=r_0[a_0,b_0]\,, \qquad s_0=s_0[a_0,b_0]\,.
\end{split}
\end{equation}
In this paper we focus on what we will call the canonical parameterization of the BPS locus, which corresponds to the choice of functions
\[r_0=s_0\equiv0.\]
The particular form within~\eqref{eq:Repars} that we are looking after, for the canonical choice of zero mode functions $r_0$ and $s_0$, is such that the angular velocities
\begin{equation}\label{eq:OmegasThere}
\begin{split}
\omega_{1,\text{there}}&\,=\,\omega_{1,\text{there}}[a_{\text{there}},b_{\text{there}},u_{\text{there}}] \\
\omega_{1,\text{there}}&\,=\,\omega_{1,\text{there}}[a_{\text{there}},b_{\text{there}},u_{\text{there}}]
\end{split}
\end{equation}
(which are the functions of three variables given by equations~(3.35) in~\cite{Cabo-Bizet:2018ehj}) when written as a function of $a_0\,$, $b_0\,$, and $u_{\text{there}}$, remain constant as we vary $u_{\text{there}}\,$, and equal to
\begin{equation}
\begin{split}
\omega_{1,\text{there}}&\,=\,\omega_{1,\text{there}}[a_{0},b_{0},\mp \text{i}]\,,\\&\,=\,\frac{2 \pi  \left(a_0-1\right) \left(b_0\mp\text{i} \sqrt{a_0
   b_0+a_0+b_0}\right)}{2 \left(a_0+b_0+1\right) \sqrt{a_0
   b_0+a_0+b_0}\,\mp\,2 \text{i} \left(a_0 b_0+a_0+b_0\right)} \\
\omega_{2,\text{there}}&\,=\,\omega_{2,\text{there}}[a_{0},b_{0},\mp \text{i}]\,\\&\,=\,\frac{2 \pi  \left(b_0-1\right) \left(a_0\mp\text{i} \sqrt{a_0
   b_0+a_0+b_0}\right)}{2 \left(a_0+b_0+1\right) \sqrt{a_0
   b_0+a_0+b_0}\mp 2 \text{i} \left(a_0 b_0+a_0+b_0\right)}
\end{split}
\end{equation}
The coefficients $r_{n\geq 1}$, $s_{n\geq 1}$ are straightforwardly solved in terms of $a_0\in (0,1)$, $b_0\in (0,1)$ in perturbations around $u_{\text{there}}=\mp \pi\text{i}$ ($\widehat{u}=0$).

This choice identifies the differentials
\[\delta \widehat{u} \quad (\text{at fixed $a_0$ and $b_0$})\] as the variation of the chemical potential dual to the $U(1)$ R-charge \[R_{\text{there}}=Q_{\text{here}}\] at fixed \[\omega_{1,\text{there}}, \qquad \omega_{2,\text{there}}\,.\] The latter are the chemical potentials dual to the linear combinations of charges \[J_{1,\text{there}}+\tfrac{1}{2} R_{\text{there}}\,,\qquad J_{2,\text{there}}+\tfrac{1}{2} R_{\text{there}}\,.\]
In the new parameterization
\begin{equation}
\omega_{1\text{there}}+\omega_{2\text{there}}- 2\varphi_{\text{there}} =\pm 2\pi \text{i}  +2\,\widehat{u}\,=2 u_{\text{there}}\,.
\end{equation}
A straightforward evaluation shows that as $a_0$ and $b_0$ are real, then the first variation of the horizon area
\begin{equation}
\begin{split}
\mathcal{S}_{g_c}&\,:=\, S_{\text{there}}[a_0,b_0,u_{\text{there}}]\\
&\,=\,\frac{1}{G_N}\,\frac{\pi ^2 \left(a_0+b_0\right) \sqrt{a_0+b_0+a_0 b_0}}{2
   \left(a_0-1\right) \left(b_0-1\right)}+O(\delta\widehat{u})
\end{split}
\end{equation}
and charges
\begin{equation}\label{eq:Charges}
\begin{split}
J_{1,\text{there}}[a_0,b_0,u_{\text{there}}]&\,=\,-\frac{1}{G_N}\frac{\pi  \left(a_0+b_0\right) \left(a_0
   \left(b_0+2\right)+b_0\right)}{4 \left(a_0-1\right){}^2
   \left(b_0-1\right)}+O(\delta\widehat{u})\,,\\ J_{2,\text{there}}[a_0,b_0,u_{\text{there}}]&=-\frac{1}{G_N}\frac{\pi  \left(a_0+b_0\right) \left(\left(a_0+2\right)
   b_0+a_0\right)}{4 \left(a_0-1\right) \left(b_0-1\right){}^2}+O(\delta\widehat{u})\\ R_{\text{there}}[a_0,b_0,u_{\text{there}}]&= \frac{1}{G_N}\frac{\pi  \left(a_0+b_0\right)}{2 \left(a_0-1\right)
   \left(b_0-1\right)}+O(\delta\widehat{u})\,,
\end{split}
\end{equation}
away from the BPS locus~\eqref{eq:Geometry1}, is real only for a trivial choice of variation
\begin{equation}
\delta \widehat{u}\,=\,0\,.
\end{equation}
For example, in the particular case $b_0=a_0$ ($J_{1,\text{there}}=J_{2,\text{there}}$) the first differential correction to the horizon area for one of the two possible sign choices in~\eqref{eq:Geometry1}  -- and for the canonical choice of zero-mode functions $r_0=s_0=0$, is
\begin{equation}
\frac{\pi ^2 a_0 (a_0+2) \left(a_0 \left(11 a_0+2 \text{i} \sqrt{a_0
   (a_0+2)}+8\right)+4 \text{i} \sqrt{a_0 (a_0+2)}-1\right)\delta \widehat{u} }{((4-5 a_0) a_0+1)^2}\,.
\end{equation}
For the charges~\eqref{eq:Charges}, instead, we obtain for the very same sign choice, respectively,
\begin{equation}\small
\begin{split}
\frac{\pi  a_0 (a_0+2) \left(-\sqrt{a_0 (a_0+2)}+a_0
   \left(a_0 \left(2 \text{i} a_0+11 \sqrt{a_0 (a_0+2)}+8 \text{i}\right)+8
   \sqrt{a_0 (a_0+2)}+8 \text{i}\right)\right) \delta \widehat{u}}{(a_0-1)^3 (5
   a_0+1)^2}\,, \\
   \frac{\pi  (a_0+2) \left(\sqrt{a_0 (a_0+2)}-\text{i} a_0 \left(a_0
   \left(2 a_0-11 \text{i} \sqrt{a_0 (a_0+2)}+8\right)-8 \text{i} \sqrt{a_0
   (a_0+2)}+8\right)\right) \delta \widehat{u}}{3 (a_0-1)^2 (5 a_0+1)^2}\,.
   \end{split}
\end{equation}
From these expressions it is easy to see that these corrections can only be simultaneously real if $\delta \widehat{u}=0\,$ (the same happens for the opposite sign choice and for the general case $a_0\neq b_0$). 

Thus, in a small enough vicinity of the BPS locus there is no complex cigar geometry -- within the extended-BPS locus, with real charges and real horizon area, other than the geometries associated to the BPS locus itself. 

This isolation also means that from all complex geometries in the extended-BPS locus, only those in the BPS locus~\eqref{eq:Geometry1} could count the total number of BPS states in the dual microscopic description $d[J,Q]$ (in the expansion~\eqref{eq:SemiclassicalExp2}). 

Such dual microscopic counting does not involve $(-1)^F$ grading, as these geometries correspond to saddle points of $Z_{\text{BPS}}\,$. This is because their defining constraint
\begin{equation}\label{eq:LinConstraintIntermezzo}
\omega_{1\text{there}}+\omega_{2\text{there}}- 2\varphi_{\text{there}} \to\pm 2\pi \text{i}\,,
\end{equation}
is equivalent to imposing periodic and anti-periodic for bosons and fermions, respectively, along the thermal cycle at $\beta<\infty\,$~\cite{Cabo-Bizet:2018ehj}. This condition corresponds to the choice of core orbifold condition $M=1$ in the field theory~\cite{Cabo-Bizet:2018ehj}.  

We understand this \emph{isolation} feature of the gravitational BPS locus as the dual realization of:

\begin{itemize}
\item the asymptotic localization of $Z_{\text{BPS}}$ to a specific saddle point in a specific superconformal index within the ensemble~\eqref{eq:AsymptLocFormulaGeneral}. In the previous example we were looking for saddles with core orbifold number~$M=1\,$. 

As mentioned before, the analogous conclusion applies for $M>1$, extending the analysis in section 3.3 of~\cite{Cabo-Bizet:2018ehj} but this time starting from the geometries dictated by the periodic orbifold prescription of~\cite{Aharony:2021zkr} at the tip of the non-extremal and non-supersymmetric cigars, then imposing $\beta\to \infty\,$ (at fixed $\omega_a$)~\cite{Aharony:2021zkr}. 
\end{itemize}

At any choice of parameterization of the BPS locus the 3 charges $J_{1,2,\text{there}}$ and $R_{\text{there}}$ are functions of 2 variables $a_0$ and $b_0$ which can be recovered from~\eqref{eq:Charges}. In the simplified case \[b_0=a_0\,\implies J_{1\text{there}}=J_{2\text{there}}={J_{\text{there}}}\,,\] they are functions of a single parameter. For example, for the canonical choice of parameterization of the BPS locus
\begin{equation}\label{eq:CanonicalParameterizationCharges}
R^\star_{\text{there}}\,=\,\frac{1}{G_N}\frac{\pi  a_0}{\left(a_0-1\right){}^2}\,,\qquad 2J^\star_{\text{there}}+R^\star_{\text{there}}\,=\,\frac{1}{G_N}\frac{\pi  a_0 \left(a_0+1\right){}^2}{\left(1-a_0\right){}^3}\,.
\end{equation}
This means that there is a non-linear constraint among charges. In the relevant conventions this constraint takes the form~\cite{Cabo-Bizet:2018ehj}
\begin{equation}\label{eq:NonLConConcrete}
p_0-p_1 p_2\,=\,0\,,
\end{equation}
where
\begin{equation}
\begin{split}
p_0&:=\frac{1}{8} \left(-\frac{2 \pi  J^\star_{1,\text{there}}
   J^\star_{2,\text{there}}}{G_N}-R_{\text{there}}^{\star\,3}\right)\,,\\ p_1&:=-\frac{3}{8} \left(\frac{2 \pi 
   \left(J^\star_{1,\text{there}}+J^\star_{2,\text{there}}\right)}{3 G_N}-2
   R_{\text{there}}^{\star\,2}\right)\,,\\p_2&:=-\frac{3}{8} \left(\frac{9}{G_N}+4 R^\star_\text{there}\right)\,.
\end{split}
\end{equation}
This is the non-CTC constraint in Lorentzian signature that we have mentioned in the Introduction~\cite{Cabo-Bizet:2018ehj}. This region corresponds to a codimension 1 region in Lorentzian signature for which in the limit to extremality there are no naked CTCs remaining.

The BPS cigar we have analyzed corresponds to the time-periodicity identification $M=1$~\cite{Cabo-Bizet:2018ehj}
\[
t_E\sim t_E + \beta\,.
\]
at the horizon tip and controls the partition function only in the codimension 1 region of charges corresponding to the non-CTC condition. Instead, the field-theory analysis is saying that BPS cigars associated to the time-periodicity identification $M>1$~\cite{Aharony:2021zkr} 
\[
t_E\sim t_E + \frac{\beta}{M}\,.
\]
can only dominate the partition function in different codimension 1 sections in the space of charges,~\footnote{Assuming that charges in field theory should be identified with charges in gravity.} however, concluding this in gravity requires a more detailed analysis there that lies beyond the scope of this paper. If this turns out to be the case then they would correspond to supersymmetric Lorentzian solutions with naked CTC.

Another related question that we have not addressed in this paper is what is the holographic dual of dressed orbifold and eigenvalue-instanton saddle points. We hope to return to this and related interesting problems in the future.

\section{Final comments}\label{sec:FinalComments}

Let us summarize the main conclusions of this work.
\begin{itemize}
\item We used the supersymmetric localization method to show that the partition function over BPS states $Z_{\text{BPS}}$ is a perturbatively protected quantity, piecewise independent of the gauge coupling. This offers a fresh perspective on the non-renormalization argument of~\cite{Grant:2008sk} using the contemporary perspective on protectedness provided by supersymmetric localization~\cite{Pestun:2007rz}. 

\item The piecewise coupling-independent $Z_{\text{BPS}}$ is the natural observable to count supersymmetric black hole microstates, as it is a positive quantity without large-$N$ sign oscillations. 

\item We have shown that in the semiclassical expansion~\eqref{eq:SemiclassicalExp2} the zero-coupling version of $Z_{\text{BPS}}$ localizes to an ensemble of superconformal indices. This means that at the leading order in the expansion~\eqref{eq:SemiclassicalExp2} the zero-coupling and strong-coupling versions of $Z_{\text{BPS}}$ generate the very same asymptotic growth in the number of states in co-dimension one loci of charges (which are related with a choice of a single delta function in, e.g.,~\eqref{eq:AsymptLocFormulaGeneral} \footnote{Our results strongly indicate that different choices of co-dimension 1 regions correspond to different choices of $M>1$ and $p$ in~\eqref{eq:AsymptLocFormulaGeneral}}). 

\ That is because the indices entering in the relevant asymptotic localization formula count only genuine BPS states. Namely, states that remain BPS all the way from weak to strong coupling and viceversa. Thus, the aforementioned asymptotic localization mechanism implements, dynamically, a projection to genuine BPS state. 

\ Consequently, to make contact with the gravitational perspectives that emerge in expansions where the dynamical localization holds, it is enough to work with the zero-coupling version of $Z_{\text{BPS}}\,$.

\item The localization of $Z_{\text{BPS}}$ to ensembles of superconformal indices also explains why using the index is enough to compute the asymptotic growth of the total number of states over a co-dimension 1 region of charges of order $N^2$ at leading order in a large-$N$ expansion (after averaging-out large oscillations).

\item Remarkably, this localization mechanism follows as well from a saddle-point analysis in a large-charge expansion at finite $N$, e.g.\,, for $N=2\,$.  We exploit this to check our conclusions. In apprendix~\ref{sec:App} we explicitly compute the total degeneracy in the free theory, $d[J,Q]\,$, at gauge rank $N=2$ at large-enough charges using the Cauchy residue formula. The results in apprendix~\ref{sec:App} confirm the predictions obtained from the saddle-point method.

\item At small enough charges the most natural observable to count microstates of the (quantum extension of) black holes is not the index but the BPS partition function $Z_{\text{BPS}}\,$, this time at strong coupling. The latter distinction is essential because at small enough charges the aforementioned asymptotic localization formula does not apply. 

\item We have identified and computed contributions of novel large-$N$ (or large charge) saddle point configurations in field theory.

\item These are saddle points of the unitary matrix integral representation of the zero-coupling $Z_{\text{BPS}}$. At large $N\,$, we have found continuous families that include dressed and undressed eigenvalue-instanton saddles. We have found some evidence that the dressed solutions come from large-$N$ limits of continuous families of finite-$N$ Bethe roots (in cases of chemical potentials where such a representation is available). Instead, the undressed solutions come from discrete solutions at finite-$N$ that become continuous only at $N=\infty\,$. In a sense $1/N$ controls the discreteness of their moduli space.

\item The eigenvalue-instanton large-$N$ saddles we have found flow to orbifold saddle points of $Z_{\text{BPS}}$ (and of indices). We expect the same to happen for the generic eigenvalue-instanton saddles that we have not studied in detail. If that turns out to be the case, then it would be natural to expect that these type of saddles do not correspond to stable gravitational configurations and instead to (quantum-geometric) flows among stable gravitational solutions generated by quantum-backreaction effects, e.g., by some kind of brane deformations as studied recently in~\cite{Choi:2025lck}, and related works~\cite{Aharony:2021zkr,Chen:2023lzq,Cabo-Bizet:2023ejm,Amariti:2024bsr}. It would be interesting to study this point in depth.

\end{itemize}

Although our focus was on four-dimensional superconformal gauge theories, with particular emphasis on $\mathcal{N}=4$ SYM, we expect our main conclusions to generalize to any setup where an instance of the BPS limit procedure~\cite{Cabo-Bizet:2018ehj} has been shown to apply. These include other setups within AdS/CFT~\cite{Cassani:2019mms} but also setups involving asymptotically flat BPS black holes in string theory~\cite{Iliesiu:2021are}~\cite{Cassani:2025iix,Boruch:2025sie, Colombo:2025yqy}.

We finalize by mentioning various related open problems:
\begin{itemize}

\item The saddle-point analysis of this paper in field theory corresponds to gravitational horizons with topology of $S^3\,$. Our conclusions will generalize to setups in which $\mathcal{N}=4$ is placed in spaces topologically different from $S^3\,$. Our methods can be used to compute the on-shell action of saddles in those cases aiming at comparing their on-shell action with the conjectured gravitational on-shell actions of~\cite{Colombo:2025yqy,Park:2025fon}.

\item Relation to the Schwarzian: the fact that on the gravitational side of the duality supersymmetry is preserved beyond the extended-BPS locus up to order of $\frac{1}{\beta_{\text{there}}}=\epsilon_{\text{there}}\,$, see equation~\eqref{eq:AlmostBPS}, strongly suggests that the same conclusion should hold on the field theory side of the duality.

In field theory we may be able to extend the supersymmetric localization computation to localize the partition function $Z[\beta,\omega,\varphi]$ but only at next-to-leading order in the small-temperature expansion dual to the near BPS locus studied in section 3.3 of~\cite{Cabo-Bizet:2018ehj}, i.e., at non-vanishing $\epsilon\,$. 

Assuming this is the case, the computation to do would be equivalent to the one already reported in~\cite{Cabo-Bizet:2024kfe}; however, conceptually, this observation of~\cite{Cabo-Bizet:2018ehj} provides an independent holographic check of the conclusions in~\cite{Cabo-Bizet:2024kfe}, specifically, of those regarding the Schwarzian correction about the 1/16-BPS sector.~\footnote{That paper also identified protected Schwarzian corrections about BPS sectors with enhanced supersymmetry.}

\item It would also be interesting to understand in simpler toy models the mechanism (which we believe to be due to a change in dominant thimbles) enforcing the transition among dominating saddles when one steps out of a zero-oscillation condition region.

\item A natural extension of our work would be to identify giant-brane expansions of the matrix integral representation of $Z_{\text{BPS}}$  (projected to genuine BPS operators). In field theory~\cite{Imamura:2021ytr,Gaiotto:2021xce,Murthy:2022ien,Lee:2022vig}\cite{Liu:2022olj,Beccaria:2023zjw,Chen:2024cvf,Ezroura:2024wmp} this problem seems reachable. In gravity it seems more challenging~\cite{Gaiotto:2021xce,Lee:2022vig,Eleftheriou:2023jxr,Beccaria:2024vfx,Gautason:2024nru,Lee:2024hef,Eleftheriou:2025lac,Deddo:2025lfm}.

\item It would be interesting to understand the meaning of dressed and undressed eigenvalue-instantons in the bulk theory. 

We plan to address some of these problems in the near future.

\end{itemize}

\section*{Acknowledgements}
The author is grateful to S. van Leuven and L. Ruggeri for useful discussions, and to S. van Leuven for correspondence.
The work of the author is supported by the INFN grant GSS (Gauge Theories,
Strings and Supergravity). Subsections~\eqref{sec:BHIntro} and~\eqref{sec:BHIntro2} partially overlap with two sections of the summary of~\cite{Cabo-Bizet:2018ehj} to appear in the \emph{Proceedings of the 2025 International Congress of Basic Science: Frontiers of Science Awards}. The initial idea for this work emerged during the preparation of that summary paper.

\appendix

\section{Asymptotic localization at $N=2\,$: a Cauchy-residue test}\label{sec:App}

In section~\ref{sec:LocPartitionF} we found that the $N=\infty$ asymptotic behavior of $Z_{\text{BPS}}$ near its essential singularities localizes to an ensemble of superconformal indices due to the dynamically generated balancing conditions~\eqref{eq:BalancingCondition3}. More generally, this asymptotic localization leads to the following conclusion:

{\theorem{\label{obs:EntropyCharges}In a co-dimension one region of charges Q, e.g., \begin{equation}Q\,=\,Q_{\text{ctr}}\equiv Q_{\text{ctr}}[2J]\,,\qquad \mathfrak{j}\,=\,2J+Q_{\text{ctr}}\end{equation}
where the exponential behaviour of $Z_{\text{BPS}}$ near one of its essential singularities dominates the Laplace transform $d[J,Q]$ (as defined in~\eqref{eq:dE}). The total number of BPS operators is predicted to asymptotically match the absolute value of the superconformal index~\footnote{Up to oscillations of the latter corresponding to interference between two time-reversal conjugate leading saddle points. By asymptotic relation we mean modulo these oscillations, which are not present in the left hand side. Such oscillations are removed by a natural averaging procedure corresponding to the selecion of a single saddle point which in quantum gravity means selecting a classical geometry.}
\begin{equation}\label{eq:AsMatchApp}
d[J,Q_{\text{ctr}}] \,\sim\,\biggl|\widetilde{d}\bigl[\,\mathfrak{j}\,\bigr]\biggr|\,.
\end{equation}

The matching is predicted to be asymptotic in any such hypothetical expansion, which does not necessarily require $N$ to be large. Indeed, we have checked that the very same saddle-point analysis and conclusions of section~\ref{sec:LocPartitionF} repeat themselves at finite values of $N$, even at $N=2$, in the relevant large-charge expansions.

}{}}

In this appendix, we take advantage of this last feature and provide a consistency check of the asymptotic relation~\eqref{eq:AsMatchApp} for $U(N)$ $\mathcal{N}=4$ SYM at small rank $N=2\,$, using explicit Cauchy-residue evaluation instead of saddle-point approximation.~\footnote{At $N=\infty$ this countercheck would practically impossible.}

We start by recalling that in certain regions of (large) charges
\begin{equation}\label{eq:LargeChargeApp}
N=2\,,\quad \mathfrak{j}\,=\,2J+Q\,\gg 1\,,
\end{equation}
\footnote{This is a different expansion than the one discussed in the main body of the text.} 
the absolute value of the microcanonical index, 
\[
|\widetilde{d}[\,\mathfrak{j}\,]|\,, \qquad 3\mathfrak{j} \in \mathbb{Z}
\]
is known to be closely approximated (up to oscillations, see the plot in figure~\ref{fig:IndexSPoint})
\begin{figure}[htbp]
  \centering
\quad\includegraphics[width=0.6
\linewidth]{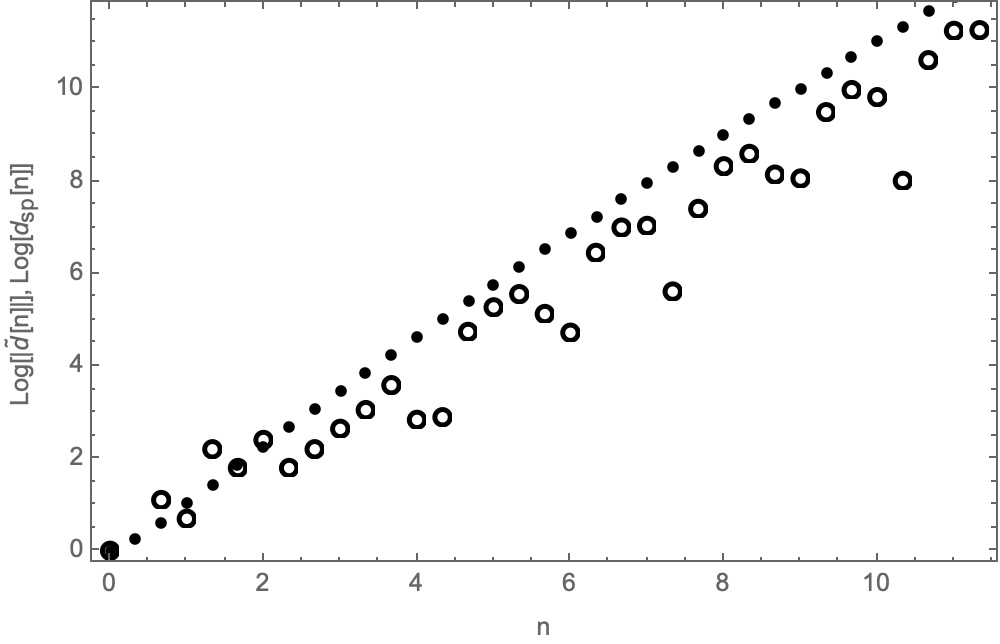}
  \caption{The circles denote the values of $\log |\widetilde{d}[\,\mathfrak{j}\,]|\,$. Their oscillations continue at larger values of $\,\mathfrak{j}\,\equiv\,n\,$. The black dots denote the values of the orbifold(s) $M=1$ saddle point $\log d_{\text{sp}}[\,\mathfrak{j}\,]\,$ defined in equation~\eqref{eq:DefdSP} at $\,\mathfrak{j}\,=n\,$.}
  \label{fig:IndexSPoint}
\end{figure}
by the asymptotic expansion of its canonical counterpart
\begin{equation}
\begin{split}
\mathcal{I}&\,=\, Z_{\text{BPS}}\,\biggl|_{t_1\,=\,t_2\,=\,t_3\,,\,p\,=\,q\,, \,\chi^{\frac{1}{2}}\,=\,-\bigl(\frac{t_1 t_2 t_3}{p q}\bigr)^{1/2}\,=\,-\,{\frac{t}{q}}\,=\,-e^{2\pi \text{i}\alpha}=\,1\,,}
\end{split}
\end{equation}
near its essential singularities, and more precisely by the trivial orbifold saddles $M=1$ with free energy~\footnote{See, for example, the plots in references \cite{Murthy:2020scj,Agarwal:2020zwm}.}
\[\mathcal{F}_{{g_c\to \pm 1}}[\omega]=-\frac{4 N^2}{27}\,\frac{({\pm\pi\text{i}}+\omega )^3}{\omega^2}\,\,, \qquad N=\text{finite}\,.\]
Moreover, as explained in observation~\ref{obs:EntropyCharges}, even at rank $N=2$ and for large charge 
\[
\mathfrak{j} \quad \text{or}\quad  2J\,\gg\,1
\] 
our analysis predicts that these complex saddles must also asymptotically count the total number of BPS states
\begin{equation}\label{eq:DJQApp}
d[J,Q]\,=\,\oint_{|q|=1} \frac{dq}{2\pi \text{i}q}\oint^{(3)}_{|{t}|=1} \frac{d{t}}{6\pi \text{i} {t}} \,Z_{\text{BPS}}\, q^{-2J} {t}^{-Q}\,,
\end{equation}
over the non-oscillation regions of coarse-grained charges~\footnote{By coarse-grained charges we mean charges that belong to a continuum which is not to be confused with the discrete spectrum of the fundamental theory $\mathcal{N}=4$ SYM. This coarse-grained continuum of charges is to be identified with the continuum of classical charges on the gravitational side~\cite{Cabo-Bizet:2024gny}.} satisfying the conditions
\begin{equation}\label{eq:ConditionConstraintNonLApp}
\frac{1}{\pi}\text{Im}(\mathcal{S}_{g_c\to \pm 1}[\,\mathfrak{j}\,,Q])\,=\,0\,\text{mod}\,2\,,
\end{equation}
where
\begin{equation}\label{eq:SgcAppendix}
    \mathcal{S}_{g_c\to\pm 1}[\,\mathfrak{j}\,,Q]\,=\,\text{ext}_{\omega}\biggl(\mathcal{F}_{{g_c\to \pm 1}}[\omega]-\omega \,\mathfrak{j}\, \mp \pi \text{i} Q\biggr)\,,
\end{equation}
and the upper (resp.\ lower) choices of signs are correlated. Equivalently, over the region of charges defined by the relation
\[
Q= Q_{\text{ctr}}[2J]\, \text{mod}\, 2\,.
\]
where the function $Q_{\text{ctr}}[2J]$ is defined by the condition $\text{Im}(\mathcal{S}_{g_c\to \pm 1}[\,\mathfrak{j}\,,Q])\,=\,0\,$. In the computations in this section we have found this function numerically with Mathematica, starting from~\eqref{eq:SgcAppendix}.   

From now on we drop the $g_c$ and remain with the labels $\pm 1\,$. The latter denote the orbifold saddles with $M=1\,$, the ones that dominate the microcanonical index at sufficiently large charge $\mathfrak{j}$ (figure~\ref{fig:IndexSPoint})
\begin{equation}\label{eq:DefdSP}
|\widetilde{d}[\,\mathfrak{j}\,]|\sim e^{\text{Re}\mathcal{S}_{\pm 1}[\,\mathfrak{j}\,]}\,=:\, d_{\text{sp}}[\,\mathfrak{j}\,]\,,
\end{equation}
Equivalently, $d_{\text{sp}}[\,\mathfrak{j}\,]$ happens to equal the exponential of the Bekenstein-Hawking entropy of the dual supersymmetric black holes with angular momenta $J_1=J_2=J$ and electric charge $Q\,$, i.e., the exponential of the area of its horizon divided by $4 G_N\,\equiv\,2\pi /N^2\,$, after particularizing $N=2\,$.

Given the above definitions, a simple calculation shows the following relation
\begin{equation}
\text{Re}\mathcal{S}_{\pm 1}[\,\mathfrak{j}\,]= \text{Re}\mathcal{S}_{-1}[\,\mathfrak{j}\,]\,.
\end{equation}
From now on we focus on the choice $-1\,$. We have removed the explicit dependence on $Q$ from these equations because by definition the real part of $\mathcal{S}_{\pm 1}$ is independent of $Q$ when written as a function of $n\,$.

Observation~\ref{obs:EntropyCharges} can be rephrased in the following asymptotic relation as $n\to \infty$
\begin{equation}
\begin{split}
d\biggl[J,Q_{\text{ctr}}\biggr]\,\sim d_{\text{sp}}[\,\mathfrak{j}\,]\,\sim\, \bigl|\widetilde{d}[\,\mathfrak{j}\,]\bigr| \,.
\end{split}
\end{equation}
Checking this via the exact evaluation of integral~\eqref{eq:DJQApp} at large $n$ and for sufficiently large values of $Q$ is a computationally demanding problem. The main challenge resides in the fact that $Q$ needs to be scaled as well with $\mathfrak{j}\,$. That means that for large $\mathfrak{j}$ we need to work with truncations of the integrand of~\eqref{eq:DJQApp} (or more precisely of~\eqref{eq:ZBPSNonPerIntro}) that include large enough powers of ${t}\,$. Adding larger such powers eventually turns out to be computationally expensive.

With our computational capabilities we were able to explore fairly large values of $2 J\approx 50$ at values of $R$-charge 
\begin{equation}\label{eq:ValuesQExplored}
Q=\biggl\{\,0\,,\frac{1}{3}\,,\,\frac{2}{3}\,,\,1\,,\, \frac{4}{3}\,\biggr\}\,.
\end{equation}
At each value of $Q$ in the list above we define a set of coarse-grained values of the spin variable $2J$ (non-integer values) denoted as $2J_{\text{ctr}}\,$. These are the possible values of $2J$ that are fixed by the non-oscillation condition
\begin{equation}
3Q=3Q_{\text{ctr}}[2 J]\,\biggl|_{J=J_{\text{ctr}}}\,\text{mod}\,6\,.
\end{equation}
There are infinitely many solutions to this equation, but the expectation is that there exists only one such solution $J=J_{\text{ctr}}$ for which a coarse-grained value of the number of BPS operators $d[J_{\text{ctr}},Q]$ approximates the absolute value of the saddle-point approximation to the index, namely:
\begin{equation}\label{eq:Eqddsp}
\begin{split}
d\bigl[J_{\text{ctr}},Q\bigr]\,\approx d_{\text{sp}}[\,\mathfrak{j}\,] \,.
\end{split}
\end{equation}
By the value of the total number of BPS operators, and its logarithm $\log d[J,Q]$ at the coarse-grained value of the spin $J=J_{\text{ctr}}$, which need not be a semi-integer, we broadly mean a (coarse-grained) interpolation that approximates the limit-function obtained in a large-$\mathfrak{j}\,$ zooming-out procedure of the discrete points obtained via explicit evaluation of~\eqref{eq:DJQApp} using the unitary matrix representation~\eqref{eq:ZBPSdefinition}. 

Plotting a large number of these points in a finite-size domain, the discrete plot for $\log d[J,Q]$ becomes quasi-continuous. In contradistinction to the index, for $\log d[J,Q]$ the limit-plot has no oscillations and becomes a smooth limit-curve which we can use to extrapolate the originally discrete plot.

Reassuringly, for the values of $Q$ we were able to study we have found not only that there is a unique coarse-grained value of $2J$ for which $d[J,Q] =d_{\text{sp}}[\,\mathfrak{j}\,]$ (see the plots in figure~\ref{fig:Plot12}),
\begin{figure}[htbp]
  \centering
\quad\includegraphics[width=0.54
\linewidth]{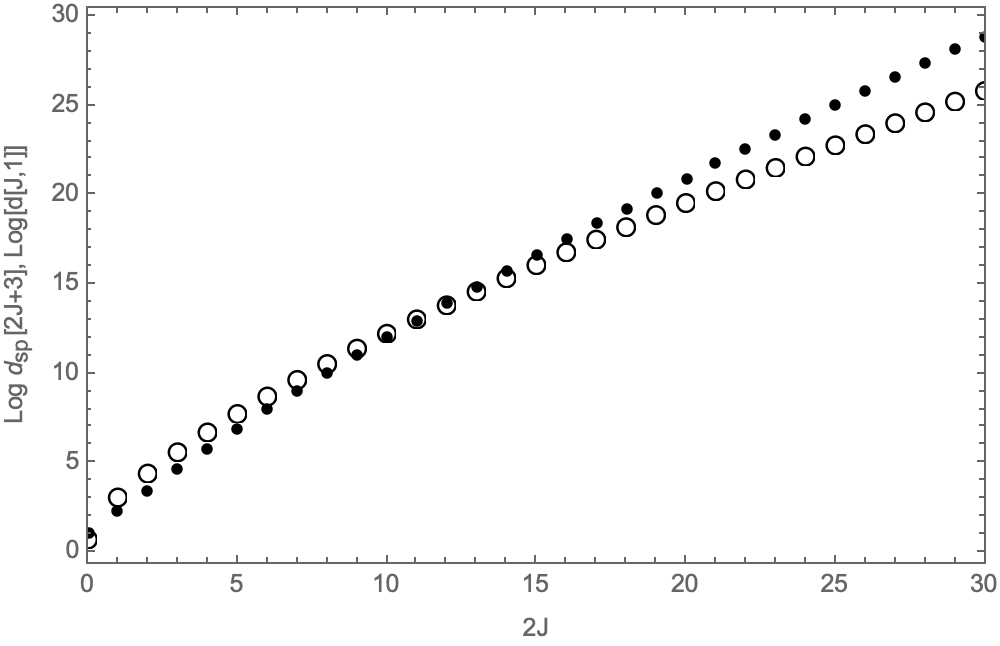}\\
\vspace{.3cm}
\includegraphics[width=0.56
\linewidth]{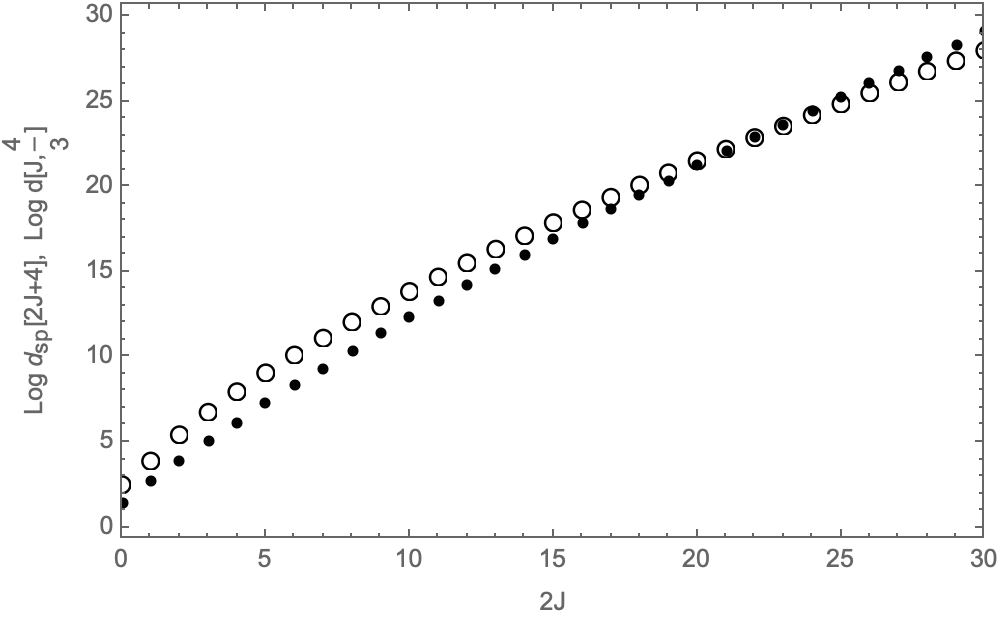} 
  \caption{The circles denote the values of $\log d[J,Q]\,$. The black dots denote the values of the saddle point approximation $\log d_{\text{sp}}[2J+Q]$. Both at fixed values $Q=1, \frac{4}{3}\,$. Note that the total number of BPS states starts being larger than the saddle point prediction but eventually the saddle point prediction starts overestimating them. We have checked that this pattern repeats itself at every value of $Q$ we were able to explore with our current computational power. Our results indicate that the intersection point among both plots define the non-linear constraint of charges. }
  \label{fig:Plot12}
\end{figure}
but also that such a value is approximately one of the $J_{\text{ctr}}$'s, as predicted by our analytic approach in the main body of the paper. Remarkably, such a unique value happens to approximate an integer as well.

Concretely, it will be convenient to focus on the object
\begin{equation}
f_{Q}[2J]:=\frac{\log d_{\text{sp}}[2J+Q] }{\log d[J, Q]}\,, \qquad  
\end{equation}
as a function of the discrete integer values of $2J$, at the fixed values of $Q$ in~\eqref{eq:ValuesQExplored}. Relation~\eqref{eq:Eqddsp} in terms of $f_Q$ reads
\[
f_{Q}[2J_{\text{ctr}}]\,\approx\, 1\,.
\]
This condition need not be true a priori. In particular, should $f_Q[2J]$ be larger or smaller than $1$ for all integer $2J$, then this would contradict observation~\ref{obs:EntropyCharges} and consequently the conclusions of our saddle-point results would be invalidated.

Consistently with our previous conclusions, we find that for the explored values of charges $Q\,$, the function $f_{Q}[2J]$ (and its coarse-grained limit function) always crosses $1$ at a unique value of $2J$, which is a function of $Q$. Our results also confirm that such a value is approximately the one predicted by the non-oscillation constraint among charges $Q$ and $J\,$ modulo periodic identifications. 

For example, the profile for $f_{Q}[2J]$ for the case \[Q=\frac{4}{3}\] is plotted in figure~\ref{fig:Plot}. 
\begin{figure}[htbp]
  \centering
\includegraphics[width=0.6
\linewidth]{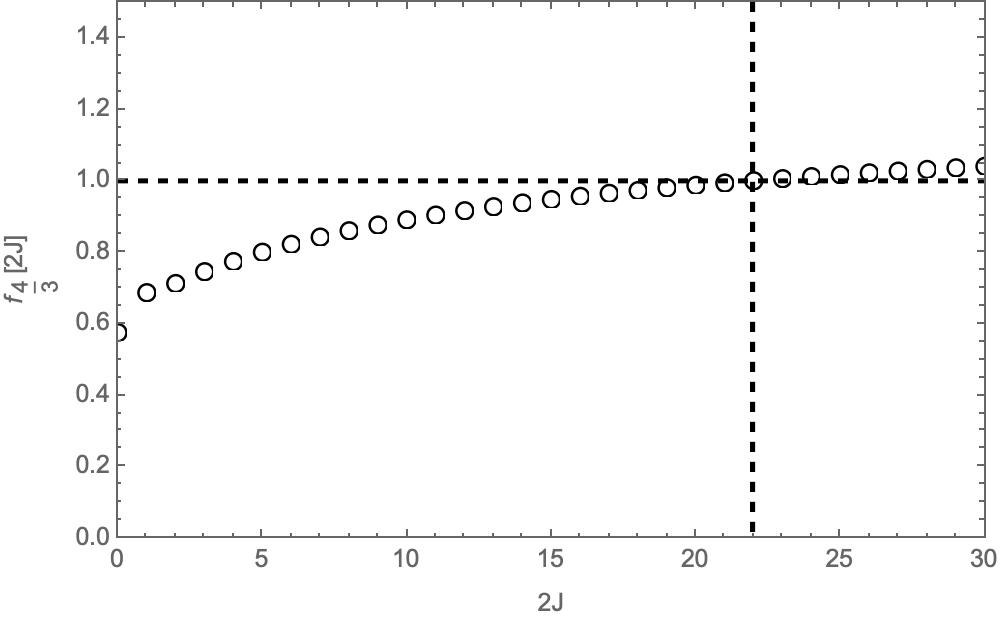} 
  \caption{The dots correspond to the discrete values of the function $f_{Q=\frac{4}{3}}[2J]$ at integer values of $2J$ corresponding to the discrete spectrum of the fundamental quantum theory. A first key feature to note is that the function $f_{Q}[J]$ grows monotonically and crosses the value $1$ at a single point. A second key feature to note is the approximate triple intersection between the dashed horizontal line denoting $1\,$, the dashed vertical line denoting $2J=2J_{\text{ctr}}[Q=\frac{4}{3}]=21.9575\ldots\,$, and a discrete point among the discrete UV values for $f_{Q}[2J]\,$. A third key feature to note is that there exists a limit curve for $f_Q[2J]\,$.  This follows from the fact $\log d[J,Q]\,$, in contradistinction with the logarithm of the absolute value of the microcanonical index $\log |\widetilde{d}[2J+Q]|\,$, has no large-charge oscillations,}
  \label{fig:Plot}
\end{figure}

Note the approximate triple intersection between the dashed horizontal line (marking the value $1$), the vertical dashed line marking the coarse-grained value
\begin{equation}
\begin{split}
2J_{\text{ctr}}&\,=\, 21.9575 \ldots\approx 22 \,, \qquad \\  3Q_{\text{ctr}}
\bigl[2J=2J_{\text{ctr}}\bigr]\,\text{mod}\,6&\,=\,4.00003\ldots\approx 3Q \,,
\end{split}
\end{equation}
and one of the discrete values of $f_{Q}[2J]\,$ (obtained from the explicit computation of $d[J,Q]$ starting from~\eqref{eq:DJQApp}). 

The same conclusion holds for every other values of $Q$ that we have explored. The trivial case being $Q=J_{\text{ctr}}=0$ for which $f_{Q=0}[0]=1$ by definition. In all of these cases we find an approximate triple intersection between the horizontal dashed line at value $1$, a vertical dashed line at $2J=2 J_{\text{ctr}}\,$, and a point in the lattice defined by the function $f_{Q}[2J]\,$ for integer $2J\,$. These results confirm the conclusions of our saddle-point analysis (as summarized in observation~\ref{obs:EntropyCharges}). 

Notice that for fixed values of $Q$, the (index) saddle-point counting $\log d_{\text{sp}}[2J+Q]$ is larger than $\log d[J,Q]$ when $2J$ is larger enough than the approximate intersection point $2J_{\text{ctr}}\,$. This is consistent with the prediction that in such a large-spin region, i.e. for a sufficiently large $2J\,$ in relation to $Q\,$, other subleading saddles of $Z_{\text{BPS}}$ (in relation to orbifold ones $M=1$) should dominate the coarse-grained asymptotic form of $d[J,Q]\,$.~\footnote{We note that it is fine that $|{\widetilde{d}}[\,\mathfrak{j}\,]|\,\geq\, d[J,\mathfrak{j}-2J]\,$. Recall that the inequality between the latter two observables comes from their relation~\eqref{eq:IndexAveragePhysStates}. Such relation does imply $|\widetilde{d}[\,\mathfrak{j}\,]|\leq |\sum_{J}(-1)^J d[J,\mathfrak{j}-2J]|$ which indeed we have checked to be satisfied using the approach in this appendix.} We leave the study of this question for the future.

These results reaffirm the analytic conclusions reached in the main body of this paper by means of saddle-point approximation: that the non-linear constraint among charges $(J,Q)$ corresponds to the locus of charges where the saddle-point approximation of the total number of BPS states $d[J,Q]$ matches the absolute value of the saddle-point approximation of the microcanonical index $\widetilde{d}[2J+Q]$ (up to oscillations of the latter).

\bibliographystyle{JHEP}

\end{document}